\documentclass[a4paper,11pt]{article}
\pdfoutput=1
\usepackage{jheppub} 
\usepackage{comment}
\usepackage{amsmath,amssymb,amsfonts,booktabs,empheq,lmodern,mathtools,nccmath,physics,slashed}
\usepackage{tikz}
\usepackage[compat=1.1.0]{tikz-feynman}
\usetikzlibrary{arrows.meta, shapes.misc,decorations.markings,decorations.pathmorphing,calc,snakes}
\usepackage{xcolor}
\usepackage{subfig}
\usepackage{pdfpages}
\usepackage{bookmark}
 
 \DeclareMathOperator*{\dpitemp}{\mathcal{D}}
 \newcommand*{\dpi}[1]{\dpitemp \left[#1\right]} 
 \newcommand*{\dm}[1]{\dfrac{\dd[3]{#1}}{(2\pi)^3}} 
 \newcommand*{\ddm}[1]{\dfrac{\dd[2]{#1}}{(2\pi)^2}} 
 \newcommand*{\dsm}[1]{\dfrac{\dd{#1}}{2\pi}} 
 \DeclareMathOperator*{\Det}{\text{Det}} 
 \DeclareMathOperator*{\Pv}{\text{PV}} 
 \DeclareMathOperator*{\sgn}{\text{sgn}} 

\preprint{TIFR/TH/23-15}

\title{\boldmath The free energy of the large-\texorpdfstring{$N$}{N} fermionic Chern--Simons theory in the `temporal' gauge}

\author[a,1]{Shiraz Minwalla,\note{minwalla@theory.tifr.res.in}}
\author[a,2]{ Souparna Nath,\note{souparna.nath@tifr.res.in}}
\author[a,3]{ Nikhil Tanwar,\note{nikhil.tanwar@tifr.res.in}}
\author[a,b,4]{ Vatsal,\note{vatsal1005@gmail.com}}

\affiliation[a]{Department of Theoretical Physics,\\ 
Tata Institute of Fundamental Research, Homi Bhabha Road, Mumbai 400005, India}
\affiliation[b]{Department of Chemistry,\\
Indian Institute of Technology Bombay, Powai, Mumbai 400076, India}

\abstract{Most of the computational evidence for the Bose--Fermi duality of fundamental fields coupled to $U(N)$ Chern--Simons theories originates in  large-$N$ calculations performed in the light-cone gauge. This gauge is ill-suited to computations in curved spacetimes, like the evaluation of the partition function on $\Sigma_g\times S^1$ for arbitrary genus $g$. In this paper, we use another gauge, the `temporal' gauge, to set up the computation of this partition function. In the large-$N$ limit, and the special case $\Sigma_g=\mathbb{R}^2$, we take the computation through to the end by setting up and solving the gap equations, generalizing tricks explored in \cite{Moshe:2014bja} to finite temperature. Our final results are in perfect agreement with earlier light-cone gauge results, providing a consistency check of both the formalism developed in this paper as well as previously performed light-cone gauge computations. In a follow-up paper, we will report on using our formalism to explicitly compute the partition function on $S^2 \times S^1$ for a finite-sized sphere.}

\begin{document}
\maketitle
\flushbottom


\section{Introduction}

Three-dimensional Chern--Simons theories coupled to matter are fascinating from several points of view. When all matter fields are massive, these theories reduce at long distances to the topological pure Chern--Simons theory. Particulate excitations of the gapped matter can be thought of as fluctuating line defects about the IR topological field theory and inherit several of their unusual properties. For instance, they generically carry non-half-integer spins \cite{Ongoing}. Their S-matrices have unusual forward scattering and crossing properties \cite{Jain:2013gza, Jain:2014nza, Gabai:2022snc, Mehta:2022lgq}. Their gauge indices participate in exchange statistics, allowing fermions and bosons in differently-coloured states to mimic each other \cite{Jain:2013gza, Jain:2014nza, Gabai:2022snc,  Mehta:2022lgq, Ongoing}. The Hilbert space of these theories on $\mathbb{R}^2$ turns out to be the projection of a Fock space down to a Wess–Zumino–Witten singlet sector \cite{Minwalla:2020ysu, Minwalla:2022sef}. These unusual features, which blur the difference between fermions and bosons, allow these theories to enjoy invariance under level-rank Bose--Fermi dualities.

Another reason for the interest in these theories is that they often produce conformal dynamics in the massless limit. Conformality is easily achieved in these theories as their quantized gauge couplings do not run. The CFTs in question are often very dynamically rich: in the famous example of the ABJM theory \cite{Aharony_2008}, they admit an AdS-dual gravitational description. Less strongly interacting versions of these theories are dual to more exotic `Vasiliev'-type gravitational theories \cite{Klebanov:2002ja, Chang:2012kt}.

Much of the current understanding of matter-Chern--Simons theories has emerged from the detailed study of a particular class of these theories over the last twelve years or so
\cite{Sezgin:2002rt, Klebanov:2002ja, Giombi:2009wh,
		Benini:2011mf, Giombi:2011kc, Aharony:2011jz, Maldacena:2011jn,
		Maldacena:2012sf, Chang:2012kt, Jain:2012qi, Aharony:2012nh,
		Yokoyama:2012fa, GurAri:2012is, Aharony:2012ns, Jain:2013py,
		Takimi:2013zca, Jain:2013gza, Yokoyama:2013pxa, Bardeen:2014paa,
		Jain:2014nza, Bardeen:2014qua, Gurucharan:2014cva, Dandekar:2014era,
		Frishman:2014cma, Moshe:2014bja, Aharony:2015pla, Inbasekar:2015tsa,
		Bedhotiya:2015uga, Gur-Ari:2015pca, Minwalla:2015sca,
		Radicevic:2015yla, Geracie:2015drf, Aharony:2015mjs,
		Yokoyama:2016sbx, Gur-Ari:2016xff, Karch:2016sxi, Murugan:2016zal,
		Seiberg:2016gmd, Giombi:2016ejx, Hsin:2016blu, Radicevic:2016wqn,
		Karch:2016aux, Giombi:2016zwa, Wadia:2016zpd, Aharony:2016jvv,
		Giombi:2017rhm, Benini:2017dus, Sezgin:2017jgm, Nosaka:2017ohr,
		Komargodski:2017keh, Giombi:2017txg, Gaiotto:2017tne,
		Jensen:2017dso, Jensen:2017xbs, Gomis:2017ixy, Inbasekar:2017ieo,
		Inbasekar:2017sqp, Cordova:2017vab, Charan:2017jyc, Benini:2017aed,
		Aitken:2017nfd, Argurio:2018uup, Jensen:2017bjo, Chattopadhyay:2018wkp,
		Turiaci:2018nua, Choudhury:2018iwf, Karch:2018mer, Aharony:2018npf,
		Yacoby:2018yvy, Aitken:2018cvh, Aharony:2018pjn, Dey:2018ykx, Skvortsov:2018uru, Argurio:2019tvw, Armoni:2019lgb,
		Chattopadhyay:2019lpr, Dey:2019ihe, Halder:2019foo, Aharony:2019mbc,
		Li:2019twz, Jain:2019fja, Inbasekar:2019wdw, Inbasekar:2019azv,
		Jensen:2019mga, Kalloor:2019xjb, Ghosh:2019sqf, Argurio:2020her, Inbasekar:2020hla,
		Jain:2020rmw, Minwalla:2020ysu, Jain:2020puw, Mishra:2020wos,
		Jain:2021wyn, Jain:2021vrv, Gandhi:2021gwn, Gabai:2022snc,Gabai:2022vri}.
These especially simple theories are $SU(N)$ or $U(N)$ Chern--Simons theories, coupled to matter in the fundamental representation, and turn out to be solvable in the large-$N$ limit.\footnote{Very roughly, this solvability may be understood as follows: Chern--Simons theories have no propagating degrees of freedom but effectively induce interactions between the fundamental matter fields. As is well known, however, that $U(N)$ charged interacting matter fields in the fundamental representation are often exactly solvable. As explained in \cite{Giombi:2011kc}, this turns out to be the case in the current context.} In a version of the light-cone gauge (see below for more details), it is possible to explicitly integrate out the gauge bosons and explicitly obtain a (nonlocal) interacting theory of the fundamental fermions. In the large-$N$ limit, it then proves possible to explicitly solve the gap equations for the fermions \cite{Giombi:2011kc}. The general procedure is similar to that of 't Hooft in his classic solution of large-$N$ 2D Yang--Mills with fundamental matter \cite{tHooft:1974pnl}. Unlike the 2D example of \cite{tHooft:1974pnl}, however, the gap equations obtained in the current context turn out to admit a simple analytic solution. This fact allows for the explicit computation of the spectrum of `single-sum' operators \cite{Giombi:2011kc} in these theories, their correlation functions \cite{Aharony:2012ns, GurAri:2012is, Bedhotiya:2015uga, Aharony:2019mbc, Gandhi:2021gwn, Jain:2022ajd}, the thermal partition function of these theories \cite{Giombi:2011kc, Jain:2012qi, Yokoyama:2012fa, Aharony:2012nh, Jain:2013py, Yokoyama:2013pxa, Choudhury:2018iwf, Dey:2018ykx, Dey:2019ihe,Halder:2019foo, Minwalla:2020ysu, Minwalla:2022sef}, their S-matrices \cite{Jain:2013gza, Jain:2014nza, Gabai:2022snc, Mehta:2022lgq}, and renormalization group flows \cite{Jain:2013gza, Aharony:2019mbc}. Almost all the results mentioned above were obtained from explicit calculations performed in the light-cone gauge, first introduced in \cite{Giombi:2011kc}.\footnote{In addition,  symmetry-based structural results obtained in \cite{Maldacena:2011jn, Maldacena:2012sf} played a key role in the computation of correlation functions.}

While the light-cone gauge is well suited to the computation of observables in flat (or nearly flat) spacetimes, this gauge is ill-suited to curved space. Interesting observables like the partition function of matter-Chern--Simons theories on $\Sigma_g \times S^1$ (where $\Sigma_g$ is a genus $g$ manifold)\footnote{This observable is of particular interest in the conformal limit and for the case $\Sigma_g=S^2$  because the state-operator correspondence establishes a map between the states of a three-dimensional conformal field theory on $S^2$ and local operators. For conformal field theories, consequently, the partition function on $S^2 \times S^1$ completely characterizes the spectrum of local operators in the theory.} are difficult to obtain in this gauge. It seems clear that progress on this front would be facilitated by moving to a different gauge.

In this paper, we work in the `temporal' gauge and develop a formalism that allows for efficient computation of the large-$N$ free energy of matter-Chern--Simons theories in $\Sigma_g \times S^1$. The `temporal' gauge we work with in this paper, more precisely, involves setting the gauge field to be independent of time (i.e., to obey $\partial_3 A_3=0$; here $3$ is the direction along the Euclidean time $S^1$).\footnote{See e.g. \cite{Li:1987hx} and \cite{Bars:1977ud} for the analyses of the two-dimensional 't Hooft model in non-light-cone gauges.} The same gauge was used in the pioneering early work of Blau and Thompson \cite{Blau:1993tv} in their study of the pure Chern--Simons partition function on $\Sigma_g \times S^1$. 

We demonstrate that this gauge is computationally tractable, by explicitly working out the special case in which $\Sigma_g$ is taken to be a very large sphere (in the rest of this paper we refer to this as the partition function on $\mathbb{R}^2 \times S^1$). We do this by setting up the relevant Schwinger--Dyson equations and explicitly solving these equations in this special case. The tricks we employ to solve our Schwinger--Dyson equations are a generalization to finite temperature of the manipulations used by Moshe-Moshe and Zinn-Justin \cite{Moshe:2014bja} in their analysis of fermionic Chern--Simons theories on $\mathbb{R}^3$ in a similar gauge (simply $A_3=0$). Our final result for the $\mathbb{R}^2 \times S^1$ partition function turn out to agree perfectly with an earlier computation of the same quantity in the light-cone gauge (this is so even though all intermediate results of the two computations look very different from each other). 

The important point is that (unlike in the case of the light-cone gauge), the explicit `temporal' gauge  $\mathbb{R}^2 \times S^1$ computation described above, generalizes in a relatively simple manner to the computation of the partition function on $S^2 \times S^1$, where the $S^2$ has finite radius.\footnote{The explicit computations of the partition function on $\Sigma_g \times S^1$ should also be possible, but involves dealing with one qualitatively new element, namely the 
holonomies on spatial cycles of $\Sigma_g$.} We have explicitly carried out this computation and will report on it in a subsequent paper, soon to appear \cite{minwalla202X:CSMonS2}.\footnote{This computation has been performed previously in the case that the $S^2$ has a parametrically large radius (of order $\sqrt{N}$). However, it has never been performed on smaller spheres previously.}

Apart from setting the stage for the computation of interesting new observables on compact spaces (like the computation we will report in \cite{minwalla202X:CSMonS2}), we view the agreement of our `temporal' gauge computation with earlier light-cone gauge results on $\mathbb{R}^2$ as an important confirmation of the consistency of those results. Earlier `light-cone gauge' results were obtained by setting $A_z=0$ (here $z$ is $x_1+ ix_2$) without requiring $A_{\bar z}$ to vanish. In other words, these computations involve a rotation of the contour of integration (in field space) away from the usual contour on which $A_z=A_{{\bar z}}$. While this procedure yields results that seem physically reasonable (and are in perfect agreement with conjectured Bose--Fermi dualities), the logical basis for this gauge has never been clearly spelt out (especially at the non-perturbative level). In contrast, the `temporal' gauge studied in this paper can be developed in a completely systematic manner without any logical jumps (see section \ref{sup} below). In our opinion, the fact that the carefully obtained results (for the $\mathbb{R}^2 \times S^1$ partition function in this paper) agree with earlier results from the light-cone gauge lends significant additional confidence to all the numerous previous computations performed in that gauge.

The rest of this paper is organized as follows. In section \ref{sup}, we set up the path integral that computes the partition function of the regular fermion theory on $\Sigma_g \times S^1$ at finite $N$ and finite volume. Working in the `temporal' gauge, we integrate out almost all gauge bosons. The remaining path integral is that of a two-dimensional Abelian $U(1)^N$ gauge theory, coupled to the gauge holonomies plus an infinite number\footnote{While the gauge bosons and holonomies are two-dimensional fields, the fermions are three-dimensional in nature. The Kaluza--Klein procedure can be used to convert these into an infinite number of two-dimensional fields.} of two-dimensional fermions. This exact effective description is completely local on $\Sigma_g$, a fact that has its origin in both the non-propagating nature of the Chern--Simons coupled gauge fields and our particular choice of gauge.

In section \ref{subsec:large_N&large_V}, we specialize this exact but unwieldy effective description to the large-$N$ limit. In this paper, we also specialize the base manifold $\Sigma_g$ to a sphere of very large volume (we will report on results for a sphere of arbitrary radius in our upcoming work \cite{minwalla202X:CSMonS2}). We demonstrate that in this limit, the thermal free energy of interest may be obtained by solving a couple of integral equations (called the gap equations) and plugging the solutions of these equations into a relatively simple expression \eqref{eq:Final_thermal_action}. In section \ref{sge}, the technical heart of this paper, we analyze and solve the gap equations using manipulations inspired by those presented in \cite{Moshe:2014bja}. In section \ref{sec:free_energy_FT}, we plug our solution to the gap equations back into the expression for the free energy, and demonstrate that the final results obtained from this procedure are identical to those obtained from the earlier light-cone gauge computations, even though intermediate results in the two gauges are very different. Finally, in section \ref{disc}, we present a discussion of our results and interesting potential generalizations. In several appendices, we present the technical details that support the analysis in the main text.


\section{Setting up the problem} \label{sup}
In this paper, we study the Type 1 $U(N)_k$  3D Chern–Simons theory,\footnote{See appendix A of \cite{Minwalla:2020ysu} for a detailed explanation of the nomenclature.} with fermionic matter. Our theory is defined by the Euclidean path integral,
\begin{equation}\label{pathint}
    Z = \int \dpi{A_\mu} \dpi{\psi} \dpi{\bar \psi} e^{-S_E},
\end{equation}
and the dimensional regularization scheme. The Euclidean action $S_E$ is given by,
\begin{equation}\label{eq:CSm_action}
    S_E = \dfrac{i\kappa}{4\pi}\int_{\Sigma\times S^1}\Tr{A\wedge \dd{A}-\dfrac{2i}{3}A\wedge A\wedge A}+\int_{\Sigma\times S^1}\bar{\psi}(\slashed{D}+M)\psi.
\end{equation}
In this section, we work at finite $N$, and study the path integral of this theory on $\Sigma \times S^1$ where $\Sigma$ is an arbitrary Riemann surface.\footnote{The generalization to the case where the $S^1$ is fibred over $\Sigma$ may well prove to be straightforward.} This path integral \eqref{pathint} can be interpreted as the finite temperature partition function of our theory on manifold $\Sigma$, at temperature $T=\beta^{-1}$, where $\beta$ is the circumference of the $S^1$.


\subsection{Reduction to a two-dimensional Abelian gauge theory}\label{subsec:2D_Abelian_gauge_theory}

In this section we follow the discussion of \cite{Blau:1993tv}, \cite{Jain:2013py} and \cite{Minwalla:2022sef} to choose a convenient gauge that reduces the path integral \eqref{pathint} to that of a two-dimensional Abelian $U(1)^N$ gauge theory, interacting {\it locally} (on $\Sigma$) with the infinite number of fermionic fields obtained from the Kaluza--Klein reduction of the matter fermions. The gauge we adopt is,
\begin{equation}\label{eq:gc}
\partial_3 A_3=0,
\end{equation} 
where $x_3$ is the coordinate along the $S^1$. The condition \eqref{eq:gc} is imposed at each point ${\vec x}$ on $\Sigma$, and effectively sets $A_3$ to be (pointwise) constant along the $S^1$, such that,
\begin{align}
    A_3 = A_3({\vec x}).
\end{align}
The gauge \eqref{eq:gc} is the nearest we can come to imposing the temporal gauge condition $A_3=0$. The obstruction to imposing the strict temporal gauge comes from the two-dimensional adjoint-valued holonomy field, $U({\vec x})$, around $S^1$. In the gauge \eqref{eq:gc}, $U({\vec x})$ is given by,\footnote{Working with $U({\vec x})$ rather than $A_3$ helps us account for large gauge transformations: configurations that have the same $U({\vec x})$ but different $A_3(\vec{x})$ are actually gauge equivalent.}
\begin{equation} \label{eq:holonomyfield} 
U({\vec x}) = e^{i \beta A_3(\vec{x})}.
\end{equation} 
We emphasize that $U({\vec x})$ is only a function of the coordinates ${\vec x}$ on $\Sigma$, and is thus, effectively, a two-dimensional field.

The condition \eqref{eq:gc} leaves a two-dimensional $U(N)$ gauge freedom unfixed. Following \cite{Blau:1993tv, Jain:2013py, Minwalla:2022sef}, we further fix the gauge down to the (two-dimensional) gauge group $U(1)^N$ by imposing the condition that $U(\vec{x})$ be diagonal everywhere.

We denote the Faddeev–Popov determinant associated with the choice of gauge-fixing \eqref{eq:gc}, together with the diagonalization of $U({\vec x})$, by $\Delta_{\rm FP}(A_3)$. Roughly speaking, this determinant is a product of the Vandermonde factors (one for each point in space), which effectively converts the path integral over $A_3(\vec{x})$ to a path integral over $U(\vec{x})$, with the Haar measure,\footnote{At large $N$, on $S^2$, the Haar measure is given by \eqref{eq:haar_measure}.} for every point $\vec{x}$. We will return to this gauge-fixing factor below. 

Our gauge choice \eqref{eq:gc} leaves the two-dimensional $U(1)^N$ gauge transformations (those that depend only on $\Sigma$) unfixed. In order to perform practical computations, we will also need to adopt a gauge choice to fix this remaining gauge-invariance. Of course, this gauge can be picked in any convenient manner: we reserve a discussion of this choice to later (see  \ref{subsec:2D_gauge-fixing}).


\subsection{Summation over the fluxes}

In addition to the above-mentioned, there is another subtlety that is worth remarking. Recall that a key step in our gauge-fixing was the diagonalization of the field $A_3(\vec{x})$. This gauge-fixing condition is ambiguous when two or more eigenvalues of the holonomy matrix are equal. As has been described in \cite{Blau:1993tv} (and in a greater detail in subsection 4.1 of \cite{Jain:2013py}), anywhere on $\mathbb{R}^2$, two eigenvalues of $U({\vec x})$ become equal on a codimension one surface in the field space. Codimension one surfaces are domain walls that break up the field space into regions. As explained in \cite{Jain:2013py}, effectively, different regions in the field space have different background $U(1)$ fluxes for the $U(1)^N$ gauge group. Essentially, we are instructed to sum our path integral over all flux sectors for each of the $U(1)^N$ groups. We denote the flux in the $\alpha^{\rm th}$ $U(1)$ component by $2 \pi n_{\alpha}$ (where $n_{\alpha}$ is an integer). It follows that the $U(1)$ factors in a two-dimensional gauge theory are effectively compact, explaining the summation over fluxes for the computation of the path integral. 


\subsection{Integrating out the 3D spatial gauge fields}\label{igf} 

In addition to the holonomy fields $U({\vec x})$, we need to perform the 
path integral over the spatial gauge fields (and also over the fermions, but we reserve that discussion for later).

Spatial gauge fields can be divided into:
\begin{itemize}
    \item[(1)] Diagonal zero modes (on the $S^1$), i.e. the gauge fields that are both time (i.e. the $3^{\text{rd}}$ coordinate) independent as well as diagonal. These are the gauge fields of the two-dimensional $U(1)^N$ gauge theory. As we will see below, these fields appear linearly in the Lagrangian. At this stage, we will not integrate out these fields but keep them around.
    \item[(2)] The remaining gauge fields, viz. the gauge fields that have nonzero momenta around the $S^1$ circle, the gauge fields that are off-diagonal, and the gauge fields that satisfy both. These fields appear quadratically (and linearly) in the Lagrangian. Moreover, their kinetic term involves only time (but no spatial) derivatives. We now proceed to integrate out these fields. The result will be a quartic contribution to the effective action for the fermions. This new term is nonlocal on the $S^1$, but is perfectly local on $\Sigma$.
\end{itemize}
The part of the action that involves the gauge fields listed in item (2) above takes the form,
\begin{align}\label{nzndact}
    S_{{\rm int}} = \dfrac{i\kappa}{4 \pi} &\int_{\Sigma \times S^1} \, \sum_{\substack{\alpha, \sigma \\ }} \Big( (A_2)_{\alpha}^{~\sigma} \partial_3 (A_{1})_{\sigma}^{~\alpha} - (A_1)_\sigma^{~\alpha} \partial_3 (A_2)_\alpha^{~\sigma}  - 2iA_3^{\sigma} (A_1)_\sigma^{~\alpha} (A_2)_\alpha^{~\sigma} + 2iA_3^{\alpha} (A_2)_{\alpha}^{~\sigma} (A_1)_\sigma^{~\alpha} \Big) \nonumber \\
    &-i \int_{\Sigma \times S^1} \, \sum_{\substack{ \alpha, \sigma \\ }} \bar{\psi}^{\alpha} \Big( \gamma^1 (A_1)_\alpha^{~\sigma} + \gamma^2 (A_2)_\alpha^{~\sigma} \Big) \psi_{\sigma}.
\end{align}
In equation \eqref{nzndact}, the lower (Greek) indices on the fermions and the fields $A_1$ and $A_2$ are fundamental indices while all the upper (Greek) indices on these fields are antifundamental indices. On the other hand, the indices on the field $A_3$ are neither fundamental nor antifundamental- as we explain in appendix \ref{notation}, the symbol $A_3^\sigma$ denotes $(A_3)^{~\sigma}_\sigma$, i.e. the $\sigma^{\rm th}$ diagonal component of the $A_3$ matrix.

As foreshadowed above, \eqref{nzndact} is quadratic in $A_1$ and $A_2$ and involves only temporal but no spatial derivatives of $A_1$ and $A_2$. It is useful to Kaluza--Klein-decompose our fields on the temporal $S^1$:
\begin{align}
    (A_{\dot{\mu}})_{\alpha}^{~\sigma}(x) & = \dfrac{1}{\beta}\sum_{l \in \mathbb{Z}} e^{2 \pi i l x_3 / \beta} (A_{\dot{\mu} ,l})_{\alpha}^{~\sigma}(\vec{x}), \label{kkdecomp1} \\
    \psi_{\alpha}(x) &= \dfrac{1}{\beta} \sum_{m \in \mathbb{Z} + \frac{1}{2}} e^{ 2 \pi i m x_3 /\beta} \psi_{\alpha, m}(\vec{x}),\label{kkdecomp2}\\
    \bar{\psi}^{\alpha}_{m}(\vec{x}) &= \big( \psi_{\alpha, -m}(\vec{x}) \big)^{*}.\label{kkdecomp3}
\end{align}
Inserting \eqref{kkdecomp1}, \eqref{kkdecomp2}, and \eqref{kkdecomp3} into \eqref{nzndact}, we obtain the action,
\begin{align} \label{actn}
    S_{{\rm int}} =& \dfrac{i\kappa}{2 \pi \beta} \int_{\Sigma} \, \sum_{\substack{l \in \mathbb{Z} \\ \alpha, \sigma \\ \alpha \neq \sigma \text{ at } l=0}} i \, (A_{2,-l})_{\alpha}^{~\sigma}(\vec{x}) \left( \dfrac{2 \pi l}{\beta}  - A_3^{\sigma}(\vec{x}) + A_3^{\alpha}(\vec{x})  \right) (A_{1,l})_{\sigma}^{~\alpha}(\vec{x})  \nonumber \\
    &  - \dfrac{i}{\beta^2} \int_{\Sigma} \sum_{\substack{ l \in \mathbb{Z} \\ m \in \mathbb{Z} + \frac{1}{2} \\ \alpha, \sigma \\ \alpha \neq \sigma \text{ at } l=0 }} \bar{\psi}^{\alpha}_{-m}(\vec{x}) \Big( \gamma^1 (A_{1,-l})_{\alpha}^{~\sigma}(\vec{x}) + \gamma^2 (A_{2,-l})_{\alpha}^{~\sigma}(\vec{x}) \Big) \psi_{\sigma, m+l}(\vec{x}).
\end{align}
We can now integrate out all the spatial gauge fields (listed in item (2) above) in \eqref{actn}. This will have two effects. First, the procedure of completing the square will yield a quartic contribution (in the fermionic fields) to the effective action given by, 
\begin{align}\label{eq:fer_int}
    S_{{\rm int}} = -\dfrac{2 \pi}{\kappa \beta^3} \int_{\Sigma}  \sum_{\substack{ l \in \mathbb{Z} \\ m,n \in \mathbb{Z} + \frac{1}{2} \\ \alpha, \sigma \\ \alpha \neq \sigma \text{ at } l=0 }} \dfrac{1}{\dfrac{2 \pi l}{\beta}  - A_3^{\alpha}(\vec{x}) + A_3^{\sigma}(\vec{x})} ~ \bar{\psi}^{\alpha}_{-m}(\vec{x}) \gamma^1  \psi_{\sigma, m-l}(\vec{x}) \bar{\psi}^{\sigma}_{-n}(\vec{x}) \gamma^2 \psi_{\alpha, n+l}(\vec{x}).
\end{align}
Second, we will obtain a contribution of $\displaystyle \dfrac{1}{\sqrt{\Det_V}}$, where $\Det_V$ is the determinant of the quadratic operator on $A_{\dot{\mu}}$ (where $\dot{\mu} =1,2\,$) in the first line of \eqref{actn}. 


\subsection{Evaluating the ratio of the Faddeev–Popov and the gauge field determinants}

It is convenient to club the contribution from the determinant $\Det_V$ with that of the Faddeev–Popov determinant $\Delta_{\rm FP}(A_3)$ (described in subsection \ref{subsec:2D_Abelian_gauge_theory}), i.e. to study the ratio,
\begin{equation}\label{ratofdet}
    \dfrac{\Delta_{\rm FP}(A_3)}{\sqrt{\Det_V}}.
\end{equation}
Each of the determinants that appears in \eqref{ratofdet} is highly divergent.\footnote{Recall that the operators under study have no spatial derivatives, so the determinants in question are simply a product of determinants, one for each point on the two-dimensional spatial manifold: as this number of points is an uncountable infinity, the determinants under study are both highly divergent.} However, the numerator and denominator of \eqref{ratofdet} almost completely cancel against each other. It was demonstrated in \cite{Blau:1993tv} that the ratio \eqref{ratofdet} is given by the finite local (on $\Sigma$) expression,
\begin{align}\label{deeint}
    \prod_{\alpha < \sigma} \exp\left(\dfrac{1}{8\pi}\int_\Sigma R\, \ln \left(\left(1-e^{i\beta (A_3^\alpha-A_3^\sigma)}\right)\left(1-e^{-i\beta (A_3^\alpha-A_3^\sigma)}\right)\right) + \dfrac{1}{2\pi} \int_{\Sigma} A_3^\alpha (F_{12})_{\alpha}^{~\alpha} \ln \left(\dfrac{1-e^{i\beta (A_3^\alpha-A_3^\sigma)}}{1-e^{-i\beta (A_3^\alpha-A_3^\sigma)}}\right) \right),
\end{align}
where $\Sigma$ is the spatial manifold, $R$ is the Ricci scalar of $\Sigma$, and $(F_{12})_{\alpha}^{~\alpha}$ is the $\alpha \alpha^{\rm th}$ component of the $F_{12}$ field strength tensor. The result of \cite{Blau:1993tv} are obtained in a heat kernel regularization scheme (similar to the regularization scheme obtained by turning on a small Yang--Mills coupling) and will be modified in our scheme, in a manner we explain below. 

The argument of the exponential in \eqref{deeint} has two terms. The contribution of the first of these terms can be succinctly written in the form,
\begin{align} \label{succint}
    \exp\left(\dfrac{1}{8\pi}\int_\Sigma R\, \ln V(\vec{x}) \right),
\end{align}
where $V(\vec{x})$ is the local Vandermonde factor,
\begin{align}\label{eq:local_Vandermonde}
    V(\vec{x}) = \prod_{\alpha < \sigma} \sin^2 \left( \dfrac{ \beta (A_3^\alpha(\vec{x})  - A_3^\sigma(\vec{x})) }{2} \right).
\end{align}
The second term in \eqref{deeint}, namely,
\begin{align}
     \exp\left( \sum_{\alpha<\sigma} \dfrac{1}{2\pi} \int_{\Sigma} A_3^\alpha (F_{12})_{\alpha}^{~\alpha} \ln \left(\dfrac{1-e^{i\beta (A_3^\alpha-A_3^\sigma)}}{1-e^{-i\beta (A_3^\alpha-A_3^\sigma)}}\right)  \right), \label{deeint_2}
\end{align}
was studied in detail in \cite{Blau:1993tv}, and more recently in appendix J of \cite{Minwalla:2022sef}. In appendix J of \cite{Minwalla:2022sef}, it was demonstrated that \eqref{deeint_2} evaluates to, 
\begin{equation}\label{thete}
    (-1)^{(N-1)  \sum \limits_{\alpha} n_{\alpha} }\,,
\end{equation}
where $N$ is the rank of the matrix under study, and $n_{\alpha}$ is the flux in the $\alpha^{\rm th}$ $U(1)$ factor. In addition, the term \eqref{deeint_2} affects the renormalization of the level of the Chern--Simons term in the bare action (the effect of this renormalization is to add $N\sgn(k)=N\sgn(\kappa)$ to the bare level in the action).

As we have mentioned above, the result \eqref{deeint} was evaluated by \cite{Blau:1993tv} in a Yang--Mills-type regularization scheme. Working in the dimensional regularization scheme (adopted in this paper) yields a similar result, except that the renormalization of the Chern--Simons level described above is absent (see e.g. appendix A.1.2 of \cite{Minwalla:2020ysu} for a closely related discussion).

In summary, it follows that the ratio of determinants \eqref{ratofdet} studied in this subsection evaluates to the manifestly local and simple expression,
\begin{align}\label{succintsum}
    (-1)^{(N-1) \sum \limits_{\alpha} n_{\alpha} } \times
    \exp\left(\dfrac{1}{8\pi}\int_\Sigma R \ln V(\vec{x}) \right).
\end{align}
Notice that the term,
\begin{align}
    (-1)^{(N-1) \sum \limits_{\alpha} n_{\alpha} },
\end{align}
vanishes identically in the $SU(N)$ theory, but is nontrivial in the 
$U(N)$ theory studied in this paper. 


\subsection{The fermion kinetic term}

The second term in the action \eqref{eq:CSm_action} is the covariantized  kinetic term for fermions. In the second line of \eqref{nzndact}, we have already accounted for one piece of this kinetic term, the part that involves the interaction with the gauge fields listed in item (2) of subsection \ref{igf}. It was this gauge-fermion interaction that gave rise to the quartic fermion interaction term \eqref{eq:fer_int} upon integrating out the gauge bosons in \eqref{actn}. 

The rest of the fermion kinetic term takes the form,
\begin{equation} \label{fermionke}
 S_{{\rm kin}} =\dfrac{1}{\beta} \int_{\Sigma} \, \sum_{\substack{m \in \mathbb{Z} + \frac{1}{2} \\ \alpha}} \bar{\psi}^{\alpha}_{-m}(\vec{x}) \left\{ (\tilde{\slashed{D}})_\alpha^{\:\:\alpha} + M +i \gamma^3 \left( \dfrac{2 \pi m}{\beta} - A_3^{\alpha}(\vec{x}) \right) \right\} \,\psi_{\alpha,m}(\vec{x}),
 \end{equation} 
where $\tilde{\slashed{D}}$ is $\gamma^{\dot{\mu}} D_{\dot{\mu}}$ ($\dot{\mu}$ runs over spatial indices), $D_{\dot{\mu}}$ is the $U(1)^N$ two-dimensional covariant derivative (denoted by $\tilde{D}$), and $\displaystyle (D_{\dot{\mu}})_\alpha^{\:\:\alpha}$ is the diagonal component in the gauge index:
\begin{align}\label{eq:cov-der-gauge-spin}
    (D_{\dot{\mu}})_\alpha^{\:\:\alpha} = \partial_{\dot{\mu}} - \dfrac{i}{2} \omega_{\dot{\mu}}^{a b} \sigma_{a b} - i (A_{\dot{\mu}})_{\alpha}{}^{\alpha}, \quad \gamma^{\dot{\mu}} = e^{\dot{\mu}}_{a} \, \gamma^a, \quad \sigma_{ab} = \dfrac{i}{4} [\gamma_a,\gamma_b].
\end{align}
Here $\displaystyle (D_{\dot{\mu}})_\alpha^{\:\:\alpha}$ contains both the constant fluxes and the dynamical $(A_{1})_{\alpha}^{~\alpha}$, $(A_{2})_{\alpha}^{~\alpha}$ fields, $e_a$ are the vielbeins on the spatial manifold $\Sigma$, and $\omega_{\dot{\mu}}^{a b}$ is the spin connection on $\Sigma$.\footnote{We will soon make a gauge choice for the $(A_{1})_{\alpha}^{~\alpha}$, $(A_{2})_{\alpha}^{~\alpha}$ fields (see subsection \ref{subsec:2D_gauge-fixing}) and perform the path integral over the residual dynamical fields (see subsection \ref{subsec:specializing_to_S2}). We denote the remaining part of \eqref{eq:cov-der-gauge-spin} by,
\begin{equation}
    (D_{\dot{\mu}}^{'})_\alpha^{\:\:\alpha} = \partial_{\dot{\mu}} - \dfrac{i}{2} \omega_{\dot{\mu}}^{a b} \sigma_{a b} - i (A_{\dot{\mu}}^{\rm background})_{\alpha}{}^{\alpha}\;.
\end{equation}}


\subsection{Interaction between the \texorpdfstring{$U(1)^N$}{U(1)\^N} gauge fields and the holonomy fields}

We have already integrated out the gauge fields listed in item (2), subsection 
\ref{igf}. The self-interactions of the remaining (two-dimensional, $U(1)^N$) gauge bosons are governed by the action,
\begin{equation}\label{inchernsimons1}
    S_{{\rm Abelian}}= \dfrac{i\kappa \beta}{2 \pi} \sum_{\alpha} \int_{\Sigma} A_3^\alpha (F_{12})_{\alpha}^{~\alpha},
\end{equation}
where $(F_{12})_{\alpha}^{~\alpha}$ contains both the constant fluxes and the dynamical $(A_{1})_{\alpha}^{~\alpha}$, $(A_{2})_{\alpha}^{~\alpha}$ fields.\footnote{Since $(A_{\dot{\mu}})_\alpha{^\alpha}= (A_{\dot{\mu}}^{\rm{background}})_\alpha{^\alpha} + (A_{\dot{\mu}}^{\rm{dynamical}})_\alpha{^\alpha}$, $(F_{12})_\alpha{^\alpha}$ can be split into two pieces: \begin{equation}\label{eq:F12_split} 
    (F_{12})_\alpha{^\alpha} = (F_{12}^{\rm{background}})_\alpha{^\alpha} + (F_{12}^{\rm{dynamical}})_\alpha{^\alpha}.
\end{equation}}
We get \eqref{inchernsimons1} by substituting the $U(1)^N$ part of the gauge fields into \eqref{eq:CSm_action}.


\subsection{Summary of the 2D \texorpdfstring{$U(1)^N$}{U(1)\^N} path integral}

In summary, it follows that the partition function \eqref{pathint} is equivalently given by the expression,
\begin{equation}\label{pfwc}
    Z = \left( \prod_{\alpha=1}^N \sum_{n_\alpha =-\infty}^\infty \right) (-1)^{(N-1) \sum \limits_{\alpha} n_{\alpha} } \int \prod_{\alpha=1}^N \left(  \dpi{A_3^{\alpha}(\vec{x})} \dpi{(A_{\Dot{\mu}})_{\alpha}^{~\alpha}(\vec{x})} 
    \prod_{m \in \mathbb{Z} + \frac{1}{2}} \dpi{\bar{\psi}_m^\alpha(\vec{x})} \dpi{\psi_{\alpha,m}(\vec{x})} \right) e^{-S_{U(1)^N}},
\end{equation}
where $\Dot{\mu} =1,2$, and the action $S_{U(1)^N}$ has $U(1)^N$ gauge symmetry. It is given by the sum of \eqref{inchernsimons1}, \eqref{succint}, \eqref{fermionke}, and \eqref{eq:fer_int}, as follows,
\begin{align}\label{exact}
    S_{U(1)^N} =& \dfrac{i\kappa \beta}{2 \pi} \sum_{\alpha} \int_{\Sigma} A_3^\alpha (F_{12})_{\alpha}^{~\alpha} - \dfrac{1}{8\pi}\int_\Sigma R \ln V(\vec{x}) \nonumber\\
    &+ \dfrac{1}{\beta} \int_\Sigma \, \sum_{\substack{m \in \mathbb{Z} + \frac{1}{2} \\ \alpha}} \bar{\psi}^{\alpha}_{-m}(\vec{x}) \left\{ (\tilde{\slashed{D}})_\alpha^{\:\:\alpha} + M +i \gamma^3 \left( \dfrac{2 \pi m}{\beta} - A_3^{\alpha}(\vec{x}) \right) \right\} \,\psi_{\alpha,m}(\vec{x}) \nonumber\\
    &  -\dfrac{2 \pi}{\kappa \beta^3} \int_\Sigma \sum_{\substack{ l \in \mathbb{Z} \\ m,n \in \mathbb{Z} + \frac{1}{2} \\ \alpha, \sigma \\ \alpha \neq \sigma \text{ at } l=0 }} \dfrac{1}{\dfrac{2 \pi l}{\beta}  - A_3^{\alpha}(\vec{x}) + A_3^{\sigma}(\vec{x})} ~ \bar{\psi}^{\alpha}_{-m}(\vec{x}) \gamma^1  \psi_{\sigma, m-l}(\vec{x}) \bar{\psi}^{\sigma}_{-n}(\vec{x}) \gamma^2 \psi_{\alpha, n+l}(\vec{x}).
\end{align}
The term \eqref{deeint_2} comes as the product \eqref{thete} in the path integral \eqref{pfwc}. Recall that $n_\alpha$ is the flux in the $\alpha^\text{th}$ $U(1)$ (where $\alpha = 1, \ldots, N$).  The infinite sum over fluxes $n_\alpha$ tells us that the two-dimensional $U(1)^N$ gauge bosons, which appear in \eqref{pfwc}, are effectively compact. 

Effectively, \eqref{pfwc} describes two-dimensional $U(1)^N$ gauge fields interacting with $N$ neutral scalar fields $A_3^\alpha$, together with an infinite number of fermionic fields $\psi_{\alpha, n}(\vec{x})$, via the complicated yet completely local (in two dimensions) action \eqref{exact}. As far as we can tell, the expression \eqref{pfwc} is exact.


\subsection{Gauge-fixing the two-dimensional \texorpdfstring{$U(1)^N$}{U(1)\^N} gauge-invariance}\label{subsec:2D_gauge-fixing}

As noted towards the end of subsection \ref{subsec:2D_Abelian_gauge_theory}, in order to proceed with actual computations, it is useful to fix the remaining unfixed gauge-invariance, i.e. the two-dimensional $U(1)^N$ gauge-invariance. One convenient gauge-fixing condition for this purpose is the Coulomb gauge condition for each of the $U(1)$ factors. This condition is imposed via the equations,
\begin{equation}\label{eq:deliai}
    \nabla\cdot ({\vec A}^{\rm{dynamical}})_{\sigma}^{~\sigma}=0, \qquad \text{for }\,\sigma= 1,\dots,N.
\end{equation}
The gauge conditions \eqref{eq:deliai} are solved by setting,
\begin{equation}\label{gcmo}
    ({A}^{\rm{dynamical}}_{\dot{\mu}})_{\sigma}^{~\sigma}= \sum_{j=1}^{N(g)} \alpha_j^{\sigma} a_{\dot{\mu},j} +  \epsilon_{\dot{\mu} \dot{\nu}} \partial_{\dot{\nu}} \chi^{\sigma}, \qquad \dot{\mu}, \dot{\nu} = 1,2,
\end{equation}
where $\{a_j\}$ is a basis for the nontrivial flat one-forms on the genus $g$ Riemann surface $\Sigma$ (the number of such one-forms is $N(g)= 2g$).\footnote {Note that for a genus-$g$ manifold with $p$ punctures, the formula \eqref{gcmo} above remains the same, with the number of such one-forms $N(g,p) = 2g + p-1$. Now, the flat-one forms, of course, need to be computed for this manifold with punctures.} The Faddeev–Popov determinant corresponding to this (Abelian) gauge-fixing is field-independent, hence unimportant. Consequently, in this gauge, the path integral over $(A^{\rm{dynamical}}_{\dot{\mu}})_{\sigma}^{~\sigma}$ turns into an integral over the $2 N g$ $\{\alpha_j^{\sigma}\}$ plus a path integral over the $N$ $\{\chi^{\sigma}\}$. 


\subsection{Specializing to \texorpdfstring{$\Sigma=S^2$}{Σ=S\^2}}\label{subsec:specializing_to_S2}

Up to this point, our analysis has been general and exact. Hereon, we begin to specialize our analysis. In this subsection, we specialize to the case $g=0$, i.e. to the special case of a round sphere. In this case, the spatial holonomies $\{\alpha_j\}$ in \eqref{gcmo} are all zero. It follows that our two-dimensional $U(1)^N$ gauge fields are characterized by the constant fluxes (whose integral is $2\pi n_\alpha$) together with $\chi^\alpha$, the fluctuating (topologically trivial) parts of these $U(1)$ gauge fields. Recall, of course, that the indices $\alpha$ range from $1$ to $N$.

The part of the action \eqref{exact} that depends on $\{\chi^{\alpha}\}$ is,
\begin{equation}\label{zetact}
    S_{{\rm dynamical}} = i \int_{S^2} \sum_{\substack{\alpha}}  \chi^{\alpha} \left( -\dfrac{\kappa\beta}{2 \pi} \nabla^2 A_3^{\alpha}  + \dfrac{1}{\beta} \sum_{\substack{m \in \mathbb{Z} + \frac{1}{2}}}\epsilon_{\dot{\mu}\dot{\nu}} \partial_{\dot{\nu}} (\bar{\psi}^{\alpha}_{-m} \gamma^{\dot{\mu}} \psi_{\alpha,m} )\right).
\end{equation}
As the variables $\chi^{\alpha}$ appear linearly in the action, they play the role of Lagrange multipliers. Integrating out the $\chi^{\alpha}$ fields imposes the following constraint equations,
\begin{align} \label{athreei}
    \nabla^2 A_3^{\alpha} = \dfrac{2 \pi}{\kappa \beta^2} \epsilon_{\dot{\mu}\dot{\nu}} \sum_{\substack{m \in \mathbb{Z} + \frac{1}{2}}} \partial_{\dot{\nu}} (\bar{\psi}^{\alpha}_{-m} \gamma^{\dot{\mu}} \psi_{\alpha,m} ).
\end{align}
These constraints are exact as they appear as a $\delta$-function within the path integral. This $\delta$-function can now be used to perform 
the path integral over the fields $A_3^{\alpha}$. 

Note that the expression on the RHS of \eqref{athreei} is a total derivative on 
$S^2$. It follows that the integral on the RHS of \eqref{athreei} vanishes on $S^2$, so \eqref{athreei} always admits a solution, given by,
\begin{equation} \label{athreein} 
    A_3^{\alpha}= \dfrac{\lambda_{\alpha}}{\beta} + \dfrac{1}{\nabla^2} \left( \dfrac{2 \pi}{\kappa \beta^2} \epsilon_{\dot{\mu}\dot{\nu}} \sum_{\substack{m \in \mathbb{Z} + \frac{1}{2}}} \partial_{\dot{\nu}} (\bar{\psi}^{\alpha}_{-m} \gamma^{\dot{\mu}} \psi_{\alpha,m} ) \right),
\end{equation} 
where $\lambda_{\alpha}$ is a constant. It follows that the path integral over $\{A_3^{\alpha}\}$ reduces to an ordinary integral over the $N$ numbers $\{\lambda_{\alpha}\}$. At this stage, it remains to perform the path integral 
over $\psi_\alpha$, the integral over the $N$ numbers $\{\lambda_{\alpha}\}$, and the summation over the fluxes $\{n_{\alpha}\}$. The action that we are instructed to use for this purpose is \eqref{exact}, with $A_3^\alpha$ given by \eqref{athreein}. Note, in particular, that the first term in the first line of \eqref{exact} reduces to,\footnote{This term arises from the constant background flux piece of $F_{12}$, i.e., the first term in \eqref{eq:F12_split}.}
\begin{equation}\label{inchernsimonsflux}
    S_{{\rm constant-flux}}= \dfrac{i\kappa \beta}{ V_2} \sum_{\alpha} n_\alpha  \int_{S^2} \sqrt{g_2}  A_3^\alpha,
\end{equation}  
where $g_2$ is the determinant of the metric on the sphere, and $V_2$ is the volume of the two-sphere.\footnote{In order to obtain the detailed form of this term, we use the fact that the constant flux on a sphere is proportional to its volume form.} However, since the second term on the RHS of \eqref{athreein} has no $l=0$ piece (since it is a total derivative), its contribution to \eqref{inchernsimonsflux} vanishes, and \eqref{inchernsimonsflux} simplifies to, 
\begin{equation}\label{inchernsimonsfluxn}
S_{{\rm constant-flux}}= i \kappa  \sum_{\alpha} n_\alpha  \lambda_\alpha.
\end{equation} 
Thus, \eqref{pfwc} and \eqref{exact} simplify to,
\begin{equation}\label{pfwcn}
    Z = \left( \prod_{\alpha=1}^N \sum_{n_\alpha =-\infty}^\infty \right) (-1)^{(N-1) \sum \limits_{\alpha} n_{\alpha} }\int \prod_{\alpha=1}^N \left( \dpi{ A_3^{\alpha}(\vec{x})} \prod_{m \in \mathbb{Z} + \frac{1}{2}} \dpi{\bar{\psi}_m^\alpha(\vec{x})} \dpi{\psi_{\alpha,m}(\vec{x})}\right) e^{-S_{{\rm exact}}},
\end{equation}
with,
\begin{align}\label{exactn}
    S_{{\rm exact}} =& ~ i \kappa \sum_{\alpha} n_\alpha  \lambda_\alpha - \dfrac{1}{8\pi}\int_{S^2} R \, \ln V(\vec{x}) \nonumber\\
    &+ \dfrac{1}{\beta} \int_{S^2} \, \sum_{\substack{m \in \mathbb{Z} + \frac{1}{2} \\ \alpha}} \bar{\psi}^{\alpha}_{-m}(\vec{x}) \left\{ \tilde{\slashed{D}^{'}} + M +i \gamma^3 \left( \dfrac{2 \pi m}{\beta} - A_3^{\alpha}(\vec{x}) \right) \right\} \,\psi_{\alpha,m}(\vec{x}) \nonumber\\
    &  -\dfrac{2 \pi}{\kappa \beta^3} \int_{S^2} \sum_{\substack{ l \in \mathbb{Z} \\ m,n \in \mathbb{Z} + \frac{1}{2} \\ \alpha, \sigma \\ \alpha \neq \sigma \text{ at } l=0 }} \dfrac{1}{\dfrac{2 \pi l}{\beta}  - A_3^{\alpha}(\vec{x}) + A_3^{\sigma}(\vec{x})} ~ \bar{\psi}^{\alpha}_{-m}(\vec{x}) \gamma^1  \psi_{\sigma, m-l}(\vec{x}) \bar{\psi}^{\sigma}_{-n}(\vec{x}) \gamma^2 \psi_{\alpha, n+l}(\vec{x}),
\end{align}
where $A_3^\alpha$ is given by \eqref{athreein}, $V(\vec{x})$ is given (in terms of $A_3^\alpha$) by \eqref{eq:local_Vandermonde}, and $\tilde{\slashed{D}^{'}}$ is the two-dimensional covariant derivative with the contribution from the (dynamical) spatial gauge fields integrated out.


\section{The gap equations at large \texorpdfstring{$N$}{N} and large volume}\label{subsec:large_N&large_V}

The reader may be forgiven for suspecting that the path integral \eqref{pfwcn} (with the action \eqref{exactn}) is impossibly complicated. In this section, we will demonstrate that its evaluation is, in fact, surprisingly simple in the limit,
\begin{itemize}
    \item $N \to \infty$,
    \item $V_2 \to \infty$,
\end{itemize} 
where $V_2$ is the volume of the base $S^2$.


\subsection{Simplifications at large \texorpdfstring{$N$}{N}}\label{subsec:largeN_gc}

 In order to evaluate the path integral \eqref{pfwcn}, we can proceed as follows. We first integrate out the fermionic fields $\psi(\vec{x})$ for fixed values of the gauge eigenvalues, to obtain an effective action for the eigenvalues $\{\lambda_\alpha\}$. In the second step, we perform the integral over $\{\lambda_{\alpha}\}$.
 
 In this paper, we will demonstrate that the effective action as a function of $\{\lambda_{\alpha}\}$, obtained at the end of the first step described above, agrees exactly with the corresponding effective action obtained using the light-cone gauge in earlier work. It follows that the second step is identical to that in earlier work (see \cite{Jain:2013py}), and will not be reconsidered here. In this paper, we focus only on the first step, namely, integrating out 
 the fermionic fields to obtain an effective action as a function of the 
 gauge holonomy eigenvalues $\{\lambda_\alpha\}$.

As usual, taking the 't Hooft limit, $N\to\infty$, $\kappa \to \infty$ with 
$\dfrac{N}{\kappa}=\lambda$ held fixed, turns the evaluation of the path integral \eqref{pfwcn} into a problem of determination of a classical saddle point. This works as follows. Both terms in the first line of \eqref{exactn} are of the order $N^2$.\footnote{The underlying reason for this is that both these terms have their origin in the gauge part of the action \eqref{eq:CSm_action}.} The first of these terms is a function of $\{\lambda_\alpha\}$ only (independent of the fermionic fields), and is, thus, just a spectator during our evaluation. Inserting \eqref{athreein} into the second term of \eqref{exactn}, and expanding in $1/N$ we find,
\begin{equation}\label{onebnexp}
    - \dfrac{1}{8\pi}\int_{\Sigma= S^2} R \, \ln V(\vec{x})
    =-\ln V + {\cal O}(1),
\end{equation} 
where $V$ is the Vandermonde defined by,
\begin{align}
     V = \prod_{\alpha < \sigma} \sin^2 \left( \dfrac{ \lambda_\alpha  - \lambda_\sigma }{2} \right). 
\end{align}
Here we have used the fact that $\displaystyle \dfrac{1}{8\pi} \int_\Sigma R=1$, when $\Sigma=S^2$. Notice that $-\ln V$, the first term on the RHS in \eqref{onebnexp}, is a sum over $N^2$ terms, and is therefore of the order $N^2$. Using \eqref{athreein}, the first correction to this result is of the order $N$. However, this correction is proportional to a total derivative, thus integrating to zero (where we use the fact that we are working on a round sphere). For this reason, the first correction in \eqref{onebnexp} is only of the order unity, and can thus be ignored at the leading order in $1/N$. 

In summary, it follows that up to corrections of the order unity (which we ignore), the first line of \eqref{exactn} evaluates to,
\begin{equation}\label{firstline}
    i \kappa \sum_{\alpha} n_\alpha  \lambda_\alpha -\ln V.
\end{equation}
As \eqref{firstline} is independent of the fermionic fields, it simply goes for the ride unchanged in most of the rest of our computation. In the rest of 
this paper, we will focus on evaluating the path integral over the fermions governed by the action $S_{{\rm f}}$, listed in the second and third lines of \eqref{exactn}, i.e.,
\begin{equation}\label{pfwcnn}
    e^{-S_{\rm eff}\left(\{\lambda_\alpha\}, \{n_\alpha\}\right)}= \prod_{\alpha=1}^N \int 
    \prod_{m \in \mathbb{Z} + \frac{1}{2}} \dpi{\bar{\psi}_m^\alpha(\vec{x})} \dpi{\psi_{\alpha,m}(\vec{x})}\, e^{-S_{{\rm f}}},
\end{equation}
where $S_{\rm eff}\left(\{\lambda_\alpha\}, \{n_\alpha\}\right)$ is the renormalized effective action we get after integrating out the fermionic fields, and,
\begin{align}\label{exactnn}
    S_{{\rm f}} = & \dfrac{1}{\beta} \int_{S^2} \, \sum_{\substack{m \in \mathbb{Z} + \frac{1}{2} \\ \alpha}} \bar{\psi}^{\alpha}_{-m}(\vec{x}) \left\{ \tilde{\slashed{D}^{'}} + M +i \gamma^3 \left( \dfrac{2 \pi m}{\beta} - \dfrac{\lambda_{\alpha}}{\beta} \right) \right\} \,\psi_{\alpha,m}(\vec{x}) \nonumber\\
    &  -\dfrac{2 \pi}{\kappa \beta^3} \int_{S^2} \sum_{\substack{ l \in \mathbb{Z} \\ m,n \in \mathbb{Z} + \frac{1}{2} \\ \alpha, \sigma \\ \alpha \neq \sigma \text{ at } l=0 }} \dfrac{1}{\dfrac{2 \pi l}{\beta}  - \dfrac{\lambda_{\alpha}}{\beta} + \dfrac{\lambda_{\sigma}}{\beta}} ~ \bar{\psi}^{\alpha}_{-m}(\vec{x}) \gamma^1  \psi_{\sigma, m-l}(\vec{x}) \bar{\psi}^{\sigma}_{-n}(\vec{x}) \gamma^2 \psi_{\alpha, n+l}(\vec{x}).
\end{align}
The spatial covariant derivative $\tilde{{D}^{'}}$ in \eqref{exactnn} is taken in the background of the background spatial gauge fields corresponding to the constant fluxes in the $U(1)^N$. These fluxes obey,
\begin{align}
    \displaystyle \dfrac{1}{2\pi} \int_{S^2} (F_{12})_{\alpha}^{~\alpha}=n_\alpha.
\end{align}
Once we have determined $S_{\rm eff}\left(\{\lambda_\alpha\}, \{n_\alpha\}\right)$, we can obtain the full thermal free energy by evaluating,
\begin{equation}\label{finans}
    \begin{split}
        Z&= \left( \prod_{\alpha=1}^N \sum_{n_\alpha =-\infty}^\infty \right)  (-1)^{(N-1) \sum \limits_{\alpha} n_{\alpha} } \int \left( \prod_{\alpha=1}^N \dd{\lambda_\alpha} \right) ~ V ~\exp \left(- i\kappa  \sum_{\alpha} n_\alpha  \lambda_\alpha  -S_{\rm eff}\left(\{\lambda_\alpha\}, \{n_\alpha\}\right) \right) \\
        &=\left( \prod_{\alpha=1}^N \sum_{n_\alpha =-\infty}^\infty \right)  (-1)^{(N-1) \sum \limits_{\alpha} n_{\alpha} } \int \prod_{\alpha=1}^N  \dd{U}~\exp \left(- i\kappa  \sum_{\alpha} n_\alpha  \lambda_\alpha  -S_{\rm eff}\left(\{\lambda_\alpha\}, \{n_\alpha\}\right) \right), \\
    \end{split}
\end{equation}
where,
\begin{equation}\label{eq:haar_measure}
    \dd{U} = \left( \prod_{\alpha=1}^N \dd{\lambda_\alpha} 
    \right) ~ V,
\end{equation} 
denotes the Haar measure for the integral over the unitary holonomy matrix. As mentioned previously, the focus of this paper is on the evaluation of $S_{\rm eff}\left(\{\lambda_\alpha\}, \{n_\alpha\}\right)$.


\subsection{The large-\texorpdfstring{$V_2$}{V\_2} limit} \label{subsec:large_V_2}

In addition to the large-$N$ limit, if we also take the large-$V_2$ limit now, the evaluation of the path integral \eqref{pfwcnn} simplifies further, as follows: The local value of the $\alpha\alpha^{\rm th}$ component of the field strength tensor is,
\begin{align}\label{eq:flux-F12}
    (F_{12}^{\rm{background}})_{\alpha}^{~\alpha} = \dfrac{ 2 \pi n_\alpha}{V_2} \omega,
\end{align} 
where $\omega$ is the volume form on $S^2$. In the large-$V_2$ limit (and at any finite value of $n_\alpha$, for each $\alpha$), this field strength vanishes. In this limit, it thus follows that the action $S_{{\rm f}}$, listed in \eqref{exactnn}, is independent of $\{n_\alpha\}$.\footnote{If $V$ is taken to $\infty$ before $N$ is taken to $\infty$, then this is the case for every value of $\{n_\alpha\}$, no matter how large.}

As a consequence, the effective action $S_{\rm eff}$, defined in \eqref{pfwcnn}, is independent of $\{n_\alpha\}$, and depends only on $\{\lambda_\alpha\}$.\footnote{A more precise statement goes as follows. The free energy can be expanded in a power series in $1/V_2$. The leading term in this expansion is proportional to $V_2$ (as might be expected from extensivity). The flux corrections to $S_{\rm eff}$ do not contribute to this term, but only to the subleading corrections (in particular to the terms that are ${\cal O}(V_2^0$) and smaller).} This fact simplifies the computation of $S_{\rm eff}$.\footnote{This observation also allows one to immediately perform the sum over $n_\alpha$ in \eqref{finans}. As explained in appendix J of \cite{Minwalla:2022sef}, this sum quantizes the eigenvalues in precisely the manner expected from the Type I $U(N)$ Verlinde formula.}

The limit $N \to \infty$, taken in the previous subsection \ref{subsec:largeN_gc}, was essential for reducing the computation of the path integral to the evaluation of a classical saddle point equation. This feature seems essential for computability. In contrast, while the limit $V_2 \to \infty$ has simplified the evaluation of $S_{\rm eff}$ (in particular, ensuring that this quantity is independent of $\{n_\alpha\}$), this simplification does not seem crucial for computability. Though we only study the limit $V_2 \to \infty$ in this paper, a part of the motivation for this study was to prepare the ground for the evaluation of the free energy at finite $V_2$. We hope to return to this interesting problem in the near future.


\subsection{Rewriting the four-fermion interaction as a bilinear of the gauge-singlet `Wilson lines'}

It now remains to perform the path integral over the fermions. This path integral may be evaluated employing the method described in section 2 of 
\cite{Giombi:2011kc}.

As we are working in the limit $V_2 \to \infty$, we are effectively on the space $\mathbb{R}^2$. In order to deal with propagating fermions in this space, it is convenient to move to the momentum-space on $\mathbb{R}^2$. The fermionic contribution to our action is given by, 
\begin{align}\label{eq:eff_act}
    &S_{\text{f}}  = \dfrac{1}{\beta}\sum_{m\in\mathbb{Z} + \frac{1}{2}}\int \ddm{p} \,\bar{\psi}^\alpha(-P_m) (i \slashed{P}_{m\alpha} + M)\, \psi_\alpha(P_m) \nonumber \\
    &\resizebox{1.15\hsize}{!}{\text{\small $\displaystyle{-\dfrac{2 \pi}{\beta^3\kappa} \sum_{\substack{ l\in\mathbb{Z} \\ m,n\in\mathbb{Z} + \frac{1}{2} \\ \alpha \neq \sigma \text{ at } m=n }}\int \ddm{p} \ddm{q} \ddm{r} ~\dfrac{ \bar{\psi}^{\alpha}\left(\dfrac{P_{2l}}{2}-Q_m\right) \gamma^1 \,\psi_{\sigma}\left(-\dfrac{P_{2l}}{2}+R_n\right)\,\bar{\psi}^{\sigma}\left(-\dfrac{P_{2l}}{2}-R_n\right) \gamma^2 \,\psi_{\alpha}\left(\dfrac{P_{2l}}{2}+Q_m\right) }{Q_{m\alpha,3}-R_{n\sigma,3}} }$},}
\end{align}
where $\displaystyle P_m = \left(p_1,p_2,\dfrac{2\pi m}{\beta}\right)$, $\displaystyle P_{n\alpha}=\left(p_1,p_2,\dfrac{2\pi n - \lambda_{\alpha}}{\beta}\right)$, and $\displaystyle P_{n\alpha,3}=\left( \dfrac{2\pi n - \lambda_{\alpha}}{\beta} \right)$.

In the large-$N$ limit, \eqref{eq:eff_act} can be rewritten as,
\begin{align}\label{eq:action_in_B}
    S_{\text{f}}  =& \dfrac{1}{\beta}\sum_{m\in\mathbb{Z} + \frac{1}{2}}\int \ddm{p} \,\bar{\psi}^\alpha(-P_m) (i \slashed{P}_{m\alpha} + M)\, \psi_\alpha(P_m) \nonumber\\
    &+ \dfrac{2 \pi N^2}{\beta\kappa}\sum_{l\in\mathbb{Z}} \int \ddm{p} \dsm{q_3} \dsm{r_3}\dfrac{1}{q_3-r_3} \Tr \left( Y(P_l,q_3) \gamma^1 Y(-P_l,r_3) \gamma^2 \right),
\end{align}
where we have defined, 
\begin{equation} \label{Ydef}
    Y(P_l,q_3) = \dfrac{2\pi}{N\beta}\sum_{m\in\mathbb{Z} + \frac{1}{2}}\int \ddm{q} \,\psi_{\alpha}\left(\dfrac{P_{2l}}{2}+Q_m\right)\, \bar{\psi}^{\alpha}\left(\dfrac{P_{2l}}{2}-Q_m\right)\delta(q_3-Q_{m\alpha,3}).
\end{equation}
$Y$ is a $2\times 2$ matrix in the spinor space, and is a color singlet. The function $Y$ takes two arguments: the first argument, $P_l$ (in \eqref{Ydef}), is a full three-momentum (see under \eqref{eq:eff_act}, or refer to appendix \ref{notation} for notation); and the second argument, $q_3$ (in \eqref{Ydef}), is only the third component of a three-momentum (this is a consequence of the integral over $q_1$, $q_2$ in the definition of $Y$). 

$Y(P_l,q_3)$ is a singlet field. In fact, it may be thought of as the Fourier transform of,
\begin{align}
    W(t,\tau) = \dfrac{1}{N} \int d^2\vec{x} \,\, \psi_{\alpha} \left(t + \dfrac{\tau}{2}, \vec{x} \right) e^{- i \lambda_{\alpha}\tau/\beta} \, \bar{\psi}^{\alpha} \left(t - \dfrac{\tau}{2}, \vec{x} \right).
\end{align}
The gauge-invariant quantity $W$ is an open Wilson line, in the fundamental representation, that ends on a $\psi$ field at one end and a ${\bar \psi}$ field at the other. The variable $P_l$ in $Y(P_l,q_3)$ is the Fourier mode dual to the centre of mass, $t$, of $W(t, \tau)$, while $q_3$ is the Fourier mode that couples to the relative separation $\tau$ of the Wilson line. The ends of the  Wilson line $W(t, \tau)$ are separated only in the temporal direction. This explains why the relative momentum in $Y(P_l,q_3)$ lies only in the third direction. 

Note that \eqref{eq:action_in_B} is a valid rewriting of \eqref{eq:eff_act} in the large-$N$ limit only. This is because of the exclusions $\alpha\neq \sigma$, when $m=n$, in the summation on the second line of \eqref{eq:eff_act}.\footnote{Due to this removal, the second line of the expression \eqref{eq:eff_act} is not really the product of two independently-defined gauge-invariant terms, but reduces to this form in the large-$N$ limit.} This exclusion is not reproduced in the expression \eqref{eq:action_in_B}. As the exclusion involves only $N$ (out of the total $N^2$) terms, \eqref{eq:eff_act} equals \eqref{eq:action_in_B} at the leading order in the large-$N$ limit.

Upon expanding $Y$ in a complete basis of $2 \times 2$ matrices as,
\begin{align}
    Y = Y_{1} \gamma^1 + Y_{2} \gamma^2 + Y_{3} \gamma^3 + Y_{I} I, 
\end{align}
it is easy to verify that (\ref{eq:action_in_B}) reduces to,
\begin{align} \label{acty}
    \resizebox{1.1\hsize}{!}{$\displaystyle S_{\text{f}}  = \dfrac{1}{\beta} \sum_{m\in\mathbb{Z} + \frac{1}{2}}\int \ddm{p} \,\bar{\psi}^\alpha(-P_m) (i \slashed{P}_{m\alpha} + M)\, \psi_\alpha(P_m) + \dfrac{8 \pi iN^2}{\beta\kappa}\sum_{l\in\mathbb{Z}} \int \ddm{p} \dsm{q_3} \dsm{r_3}\dfrac{1}{q_3-r_3} Y_3(P_l,q_3) Y_I(-P_l,r_3).$}
\end{align}
Note that \eqref{acty} depends only $Y_I$ and $Y_3$; the quantities $Y_{1}$ and $Y_{2}$ drop out of \eqref{acty}. 


\subsection{Simplification via the Hubbard–Stratonovich trick}

As $Y$ in \eqref{Ydef} is quadratic in the fermions, the term proportional to $Y_3 Y_I$ in \eqref{acty} is quartic in the fermions. We now use the Hubbard–Stratonovich trick to rewrite \eqref{acty} in terms of an expression that is quadratic in fermions. To do this we will introduce a new $2 \times 2$ matrix-valued singlet field, $\Sigma_T(P_l,q_3)$, which we expand in terms of its components as,
\begin{align}\label{eq:decom_of_Sigma}
    \Sigma_T(P_l,q_3) = \Sigma_{T,3}(P_l,q_3) \gamma^3 + \Sigma_{T,I}(P_l,q_3) I.
\end{align}
We now insert unity rewritten as,
\begin{align}\label{identin}
     1= \displaystyle\dfrac{\int  \dpi{\Sigma_{T,3}} \dpi{\Sigma_{T,I}} \, \exp{-E }}{\int \dpi{\Sigma_{T,3}} \dpi{\Sigma_{T,I}} \, \exp{-E'}},
\end{align}
into \eqref{pfwcnn}. Here $E$ is defined by,
\begin{align}\label{defe}
    E = \dfrac{\kappa}{2 \pi i \beta} &\int \ddm{p} \dsm{q_3} \dsm{q_3'} \sum_{l \in \mathbb{Z}} \Bigg( \left( \Sigma_{T,3}(P_l,q_3) - \dfrac{4 \pi N i}{\kappa}  \int \dsm{r_3} Y_{I}(P_l,r_3) \dfrac{1}{(r_3-q_3)} \right)\Bigg. \nonumber \\
    \Bigg. &  \times  G^{-1}(q_3 - q'_3) \times \left( \Sigma_{T,I}(-P_l,q'_3) - \dfrac{4 \pi N i}{\kappa} \int \dsm{r_3'} Y_{3}(-P_l,r_3') \dfrac{1}{(q'_3-r'_3)} \right) \Bigg),
\end{align}
and $E'$ is defined by,
\begin{align}\label{defep}
    E' = \dfrac{\kappa}{2 \pi i \beta} &\int \ddm{p} \dsm{q_3} \dsm{q_3'} \sum_{l \in \mathbb{Z}} \Bigg(  \Sigma_{T,3}(P_l,q_3)  G^{-1}(q_3 - q'_3)  \Sigma_{T,I}(-P_l,q'_3) \Bigg).
\end{align}
\eqref{identin} is unity because the integral in the numerator is equal to the integral in the denominator by a change of variables. As the integral in the denominator is independent of the fermionic fields, it is just a number and we will omit it in all subsequent formulae. 

The quantity $G^{-1}(p_3-q_3)$ that appears in the expressions for $E$ and $E'$
is defined by,
\begin{align}
    \int \dsm{q_3}\, G^{-1}(p_3-q_3) \dfrac{1}{(q_3-r_3)} = \int \dsm{q_3} \dfrac{1}{(p_3-q_3)} \, G^{-1}(q_3-r_3)= 2 \pi \delta(p_3-r_3).
\end{align}
Also note that $G^{-1}$ is odd in its argument. We will never need the explicit form of $G^{-1}$.

After inserting \eqref{identin} into \eqref{pfwcnn}, we obtain,
\begin{align}\label{eq:path_in_Lag_fields}
     e^{-S_{\rm eff}(\{\lambda_\alpha\})}= \int \dpi{\psi} \dpi{ \Bar{\psi} } \dpi{\Sigma_{T,3}} \dpi{\Sigma_{T,I}} \, \exp{-S_{\text{f}}-E }.
\end{align}

We have chosen the form of the action $E$ \eqref{defe} to ensure that the terms proportional to $Y_I Y_3$ in $E$ precisely cancel those in \eqref{acty}. Consequently, $S_{\text{f}}+E$ is quadratic in fermion fields and is given by, 
\begin{align}\label{eq:eff_action_SplusE}
    S_{\text{f}}+E = &\dfrac{1}{\beta}\sum_{m\in\mathbb{Z} + \frac{1}{2}}\int \ddm{p} \,\bar{\psi}^\alpha(-P_m) (i \slashed{P}_{m\alpha} + M)\, \psi_\alpha(P_m)\nonumber \\ 
    &+ \dfrac{2 \pi}{\beta^2} \int \ddm{p} \dm{q}\, \sum_{\substack{l \in \mathbb{Z}\\ m \in \mathbb{Z} + \frac{1}{2}} } \bar{\psi}^{\alpha}\left(\dfrac{P_{2l}}{2} -Q_m \right) \Sigma_T(-P_l,q_3)\, \psi_\alpha\left( \dfrac{P_l}{2} +Q_m \right) \delta(q_3 - Q_{m \alpha,3}) \nonumber  \\
    & +  \dfrac{\kappa}{2\pi \beta i } \int \ddm{p} \dsm{q_3} \dsm{q_3'} \sum_{l \in \mathbb{Z}} \, \Sigma_{T,3}(P_l,q_3) G^{-1}(q_3-q'_3) \Sigma_{T,I}(-P_l,q'_3).
\end{align}
Equivalently,
\begin{align}\label{eq:action_quad_in_fields}
     S_{\text{f}} + E =&\resizebox{1.05\hsize}{!}{$\displaystyle  \dfrac{2 \pi}{\beta^2} \int \ddm{p} \dm{q} \sum_{\substack{l \in \mathbb{Z}\\ m \in \mathbb{Z} + \frac{1}{2}} }\, \bar{\psi}^{\alpha}\left(\dfrac{P_{2l}}{2} -Q_m \right) \left((i \slashed{q} + M) (2 \pi)^2 \delta(\vec{p}) \beta \delta_{l,0} + \Sigma_T(P_l,q_3)\right)\psi_\alpha\left( \dfrac{P_l}{2} +Q_m \right) \delta(q_3 - Q_{m \alpha,3})$} \nonumber \\
     &- \dfrac{\kappa}{8 \pi \beta} \int \ddm{p}  \dsm{q_3} \dsm{q_3'} \sum_{l \in \mathbb{Z}}\, G^{-1}(q_3-q'_3) \Tr{\Sigma_T(P_l,q_3) \gamma^1 \Sigma_T(-P_l,q'_3) \gamma^2} .
\end{align}
As is clear from \eqref{eq:action_quad_in_fields}, the field $\Sigma_T$ is effectively the self-energy of the fermionic fields in the `temporal' gauge under study. Upon integrating out the fermions we find an effective action, of the order $N$, for the singlet field $\Sigma_T$. In the large-$N$ limit, fluctuations of $\Sigma_T$ are suppressed, and thus $\Sigma_T$ can be replaced by its value at the saddle point. As yet we do not know what this saddle point value is, but we assume that it is translationally invariant. In other words, we assume that $\Sigma_T$ takes the form,
\begin{align}\label{eq:trans_inv_sig_config}
    \Sigma_T(P_l,q_3) = (2 \pi)^2 \,\delta^2(\vec{p})\, \beta \,\delta_{l,0} \,\Sigma_T(q_3).
\end{align}
The action now takes the form,
\begin{align}\label{eq:simplified_action}
    S_{\text{f}} +E = \dfrac{1}{\beta} \int \dm{p} \sum_{m \in \mathbb{Z} + \frac{1}{2}} \bar{\psi}^{\alpha}\left(-P_m \right) (i \slashed{p} + M+ \Sigma_T(p_3))\psi_\alpha\left(P_m \right)(2 \pi) \delta(p_3 - P_{m \alpha,3})&  \nonumber \\
     - \dfrac{\kappa V_2 \beta}{8 \pi} \int \dsm{q_3} \dsm{q_3'}\, G^{-1}(q_3-q'_3) \Tr{\Sigma_T(q_3) \gamma^1 \Sigma_T(q'_3) \gamma^2}& ,
\end{align}
where $V_2$ is the volume of the space. To obtain the last term in \eqref{eq:simplified_action}, we have used the fact that,
\begin{align}
    \left((2 \pi)^2 \delta^2(\vec{p}) \beta \delta_{l,0}\right)^2 = V_2 \beta^2 (2 \pi)^2 \delta^2(\vec{p}) \delta_{l,0} \,.
\end{align}
Integrating out the fermion fields in (\ref{eq:simplified_action}) yields,
\begin{align}\label{eq:final_action}
    S_{\Sigma_{T}} = - V_2 \int \dm{q} &\sum_{\substack{\alpha \\ m \in \mathbb{Z} + \frac{1}{2}}} \Tr{\ln \left( i \slashed{q} + M + \Sigma_T(q_3)  \right)} (2 \pi) \delta(q_3-Q_{m \alpha,3}) \nonumber\\
    &- \dfrac{\kappa V_2 \beta}{8 \pi} \int  \dsm{q_3} \dsm{q_3'}\, G^{-1}(q_3-q'_3) \Tr{\Sigma_T(q_3) \gamma^1 \Sigma_T(q'_3) \gamma^2},
\end{align}
such that \eqref{eq:path_in_Lag_fields} reads,
\begin{align}\label{eq:path_in_sigma_T}
    e^{-S_{\rm eff}(\{\lambda_\alpha\})} = \int\dpi{\Sigma_T} \,\exp{-S_{\Sigma_T}}.
\end{align}
As we had anticipated above, all the terms in the action (\ref{eq:final_action}) are proportional to $N$. At the leading order in the large-$N$ expansion, it follows that $S_{\rm eff}(\{\lambda_\alpha\})$ of \eqref{eq:path_in_sigma_T} may be evaluated by simply minimizing (\ref{eq:final_action}) with respect to $\Sigma_T$.


\subsection{The gap equation}

The variational equation we encounter in the minimization process mentioned above is,
\begin{align} \label{vargap}
    \resizebox{1.15\hsize}{!}{$\displaystyle 0 = \int \Tr\left\{\dsm{q_3}\, \delta \Sigma_T(q_3) \left( \sum_{\substack{\alpha \\ m \in \mathbb{Z} + \frac{1}{2}}} \int \ddm{q} \dfrac{2 \pi \delta(q_3-Q_{m \alpha,3})}{i \slashed{q} + M + \Sigma_T(q_3)} +\dfrac{\kappa \beta}{8 \pi} \int \dsm{q_3'}\, G^{-1}(q_3-q'_3) \left( \gamma^1 \Sigma_T(q'_3) \gamma^2 - \gamma^2 \Sigma_T(q'_3) \gamma^1 \right) \right)\right\}.$}
\end{align}
Defining $H(C)$ for an arbitrary $2 \times 2$ matrix $C$ by,
\begin{equation}\label{Hdef} 
    H(C) \equiv \gamma^1 C \gamma^2 - \gamma^2 C \gamma^1,
\end{equation} 
if we expand the matrix $C$ in a complete basis,
\begin{equation}\label{eq:arbit_mat_decomp}
    C = C_I I + C_1 \gamma^1 + C_2 \gamma^2 + C_3 \gamma^3,
\end{equation}
we find that,
\begin{align}
    H(C) = 2i (C_I \gamma^3 - C_3 I). \label{eq:mat_val_func_H}
\end{align}
Note that $H(C)$ is independent of $C_1$ and $C_2$. In terms of the function $H(C)$, \eqref{vargap} becomes,
\begin{align}\label{eq:variational_H}
    \int \Tr\left\{\dsm{q_3}\, \delta \Sigma_T(q_3) \left( \sum_{\substack{\alpha \\ m \in \mathbb{Z} + \frac{1}{2}}} \int \ddm{q} \dfrac{2 \pi \delta(q_3-Q_{m \alpha,3})}{i \slashed{q} + M + \Sigma_T(q_3)} + \dfrac{\kappa \beta}{8 \pi} \int \dsm{q_3'}\, G^{-1}(q_3-q'_3) H(\Sigma_T(q'_3)) \right)\right\}=0.
\end{align}
This equation takes  the form,
\begin{align}
    \int \Tr{\dsm{q_3}\, \delta \Sigma_T(q_3) B(q_3)}=0,
\end{align}
where,
\begin{align}
    B(q_3) = \sum_{\substack{\alpha \\ m \in \mathbb{Z} + \frac{1}{2}}} \int \ddm{q} \dfrac{2 \pi \delta(q_3-Q_{m \alpha,3})}{i \slashed{q} + M + \Sigma_T(q_3)} + \dfrac{\kappa \beta}{8 \pi} \int \dsm{q_3'}\, G^{-1}(q_3-q'_3) H(\Sigma_T(q'_3)) .
\end{align}
As $\delta \Sigma(q_3)$ is any matrix of the form (\ref{eq:decom_of_Sigma}), it follows that,
\begin{align}
    B_3(q_3)=B_I(q_3)=0,
\end{align}
which is the same as,
\begin{align}
    H(B(q_3))=0.
\end{align}
Using the fact that,
\begin{align}
    H(H(\Sigma_T(q'_3))) = 4 \,\Sigma_T(q'_3),
\end{align}
and integrating both sides of (\ref{eq:variational_H}) against the kernel $\dfrac{1}{(p_3-q_3)}$ and using the defining property of $G^{-1}$, it follows from $H(B(q_3))=0$ that,
\begin{align} \label{gap1}
    \Sigma_T(p_3) = - \dfrac{2 \pi}{\kappa \beta} \sum_{\substack{\alpha \\ m \in \mathbb{Z} + \frac{1}{2}}} \int \dm{q} \dfrac{1}{p_3-q_3} H\left(\dfrac{(2 \pi )\delta(q_3-Q_{m \alpha,3})}{i\slashed{q}+M+\Sigma_T(q_3)}\right).
\end{align}
Multiplying both sides of \eqref{gap1} by $\delta(p_3 - P_{n \sigma,3})$, and integrating over $p_3$, we get,
\begin{align} \label{gap2}
    \Sigma_T(P_{n \sigma,3}) = -\dfrac{2 \pi}{\kappa \beta} \sum_{\substack{ m \in \mathbb{Z} + \frac{1}{2} \\ \alpha \\ \alpha \neq \sigma \text{ at } m=n  }} \int \ddm{q} \dfrac{1}{P_{n \sigma,3}-Q_{m \alpha,3}} \, H\left(\dfrac{1}{i\slashed{Q}_{m \alpha}+M+\Sigma_T(Q_{m \alpha,3})}\right).
\end{align}
Throughout the rest of this paper, we will refer to \eqref{gap2} as the gap equation. In the next section, we will find the explicit solution to this equation.


\subsection{\texorpdfstring{$S_{\rm eff}$}{S\_eff} in terms of \texorpdfstring{$\Sigma_T$}{Σ\_T}}

Using the definition of $H$ \eqref{Hdef}, \eqref{eq:final_action} can be evaluated to,
\begin{align}\label{eq:final_action_1}
    S_{\Sigma_T} &= - V_2 \int \ddm{q} \sum_{\substack{\alpha \\ m \in \mathbb{Z} + \frac{1}{2}}} \Tr{\ln \left(  i \slashed{Q}_{m \alpha} + M + \Sigma_T(Q_{m \alpha,3}) \right)}\nonumber\\ 
    & \qquad\qquad\qquad +\dfrac{\kappa V_2 \beta}{16 \pi} \int \dfrac{dq_3}{(2 \pi)} \dfrac{dq_3'}{(2 \pi)} G^{-1}(q_3-q'_3) \Tr{\Sigma_T(q'_3) H(\Sigma_T(q_{3})) }.
\end{align}
Once the gap equation has been solved, the free energy may be obtained by evaluating the Euclidean action (\ref{eq:final_action_1}) at the saddle point to give the thermal effective action,\footnote{$S_{\rm eff}$ is free of any UV-divergences, whereas $S_T$ is divergent. The relation between $S_{\rm eff}$ and $S_T$ is discussed in subsection \ref{sec:UV_div}.}
\begin{align}\label{eq:Final_thermal_action}
    S_T &= \resizebox{1.05\hsize}{!}{$\displaystyle - V_2 \int \ddm{q} \sum_{\substack{\alpha \\ m \in \mathbb{Z} + \frac{1}{2}}} \Tr{\ln \left(  i \slashed{Q}_{m \alpha} + M + \Sigma_T(Q_{m \alpha,3}) \right) +\dfrac{1}{8} H\left( \dfrac{1}{i \slashed{Q}_{m \alpha} + M + \Sigma_T(Q_{m \alpha,3})} \right) H(\Sigma_T(Q_{m \alpha,3}))}$} \nonumber \\
    &= - V_2 \int \ddm{q} \sum_{\substack{\alpha \\ m \in \mathbb{Z} + \frac{1}{2}}} \Tr{\ln \left(  i \slashed{Q}_{m \alpha} + M + \Sigma_T(Q_{m \alpha,3}) \right) -\dfrac{1}{2} \Sigma_T(Q_{m \alpha,3}) \dfrac{1}{i \slashed{Q}_{m \alpha} + M + \Sigma_T(Q_{m \alpha,3})}},
\end{align}
where the first equality follows from the equation of motion, and the second equality follows from the fact that for an arbitrary matrix $A$,
\begin{align}
    \Tr{H(A)\,H(\Sigma_T(Q_{m \alpha,3}))} = -4 \Tr{A \,\Sigma_T(Q_{m \alpha,3})}.
\end{align}


\subsection{Diagrammatic rederivation of the gap equation}

As explained in \cite{Giombi:2011kc}, the gap equation \eqref{gap2} is simply the diagrammatic equation,
\begin{align}\label{feyndiag}
    \hbox{
    \begin{tikzpicture}
    \begin{feynman}[every blob= {/tikz/fill=white,/tikz/inner sep=2pt}, horizontal=(f1) to (f2)]
/tikz/inner sep=2pt
        \vertex (b) at (2.0,0) {=};
        \vertex (f1) at (-1.5,0);
        \vertex[blob] (a) at (0,0) {1PI};
        \vertex (f2) at (1.5,0);
        \diagram*{
            (f1) -- [fermion] (a) -- [fermion] (f2),
        };
    \end{feynman}
    \end{tikzpicture}
    }
    \hbox{
    \begin{tikzpicture}
        \begin{feynman} [horizontal=(f1) to (f3)]
            \vertex (f1) at (-2.4,0);
            \vertex (a) at (-1.2,0);
            \vertex[blob] (f2) at (0,0) {};
            \vertex (b) at (1.2,0);
            \vertex (f3) at (2.4,0);
            \diagram*{
            (f1) -- [fermion] (a) -- [fermion] (f2) -- [fermion] (b) -- [fermion] (f3),
            (a) -- [gluon, half left] (b)
            };
            \vertex (c) at (2.9,0) {,};
        \end{feynman}
    \end{tikzpicture}
    }
\end{align}
where solid lines with a blob in between denote the exact fermion propagator, and wiggly (spring-like) lines denote the gauge boson propagator. With the momentum 3-tuple $\displaystyle P_{m}$ defined in \eqref{eq:mom_matsubara},
\begin{equation}
    \expval{A^{\alpha\sigma}_{\dot{\mu}}(P_m) A^{\rho\eta}_{\dot{\nu}}(-Q_n)} = (2\pi)^2 \, \delta^2(\vec{p} - \vec{q}) \, \beta \, \delta_{m,n} G_{\dot{\mu} \dot{\nu}}^{\alpha\sigma}(P_m)\delta^{\alpha\eta}\delta^{\sigma\rho},
\end{equation}
is the bare gluon propagator, with,
\begin{equation}\label{gbprop}
     G_{\dot{\mu} \dot{\nu}}^{\alpha\sigma}(P_m) = \dfrac{2\pi}{\kappa} \epsilon_{\dot{\mu}\dot{\nu}} \dfrac{1}{P_{m,3}  - \dfrac{\lambda_\alpha}{ \beta } + \dfrac{\lambda_\sigma}{\beta} }\,,
\end{equation}
in the `temporal' gauge ($\partial_3 A_3=0$), with $\dot{\mu}, \, \dot{\nu} =1,2$, and $\epsilon_{12}=1$.

As the exact fermion propagator is given in terms of the exact fermion self-energy by,
\begin{equation}\label{eq:fermi_prop}
    \expval{\psi_\alpha(P_m)\bar{\psi}^\sigma(-Q_n)} = \delta^\sigma_\alpha \dfrac{1}{i\slashed{P}_{m \alpha}+M+\Sigma_{T}(P_{m\alpha})} \times (2\pi)^2 \delta^2(\vec{p}-\vec{q}) \, \beta \, \delta_{m,n} \,\,,
\end{equation}
where the gauge-invariant momentum 3-tuple $\displaystyle P_{m \alpha}$ is defined in \eqref{gimom}, it follows (exactly as in sections 2.1 and 2.2 of \cite{Giombi:2011kc}) that \eqref{feyndiag} becomes,\footnote{In other words, $\Sigma_T$ is minus the sum of 1PI self-energy correction graphs.} 
\begin{equation}\label{eq:gap1}
    \Sigma_T(P_{m \alpha}) = - \dfrac{1}{\beta} \sum_{\substack{\sigma \\ n \in \mathbb{Z} + \frac{1}{2}}} \, \int\ddm{q} \left( \gamma^{\dot{\mu}} \dfrac{1}{i\slashed{Q}_{n \sigma}+M+\Sigma_T(Q_{n \sigma})} \gamma^{\dot{\nu}}\right) G_{\dot{\mu} \dot{\nu}}^{\alpha\sigma}(P_m-Q_n),
\end{equation}
where, once again, the symbols ${\dot \mu}$ 
and ${\dot \nu}$ run over the spatial indices $1$ and $2$. Inserting \eqref{gbprop} into \eqref{eq:gap1}, we find the more explicit gap equation,
\begin{equation}\label{eq:gap_eq_FT}
    \Sigma_T(P_{m\alpha}) = - \dfrac{2 \pi}{\beta\kappa}\sum_{\substack{\sigma \\ j \in \mathbb{Z} + \frac{1}{2} \\ \sigma \neq \alpha \text{ at } j=m  }}\int\ddm{q}\, H\left(\dfrac{1}{i\slashed{Q}_{j\sigma}+M+\Sigma_T(Q_{j\sigma})}\right) \dfrac{1}{P_{m\alpha,3}-Q_{j\sigma,3}}.
\end{equation}
The saddle point (thermal effective) action \eqref{eq:Final_thermal_action} can also be rederived using diagrammatic techniques. The derivation proceeds exactly along the lines of section 2.1 of \cite{Giombi:2011kc}, and we will not repeat it here. 

As we explain in appendix \ref{schwinger}, it is possible to rederive the gap equation \eqref{eq:gap_eq_FT} using Schwinger--Dyson equations and large-$N$ factorization.


\subsection{Variables in \texorpdfstring{$\Sigma_T$}{Σ\_T}}

Notice that the RHS of \eqref{eq:mat_val_func_H} is proportional to (a sum of) $\gamma^3$ and $I$. From \eqref{eq:gap_eq_FT}, it follows that $\Sigma_T$ would share this property. Furthermore, as the RHS of \eqref{eq:gap_eq_FT} is a function of $P_{m\alpha,3}$ only, $\Sigma_T$ too can only depend on this component of the momenta. It follows that our self-energy takes the form,
\begin{equation}\label{eq:sigma_decomp}
    \Sigma_T(P_{m\alpha,3}) = i \gamma^3 {\Sigma}_{T,3}(P_{m\alpha,3}) + \left(\Sigma_{T,I}(P_{m\alpha,3})-M\right)I.
\end{equation}
In order to solve the gap equation \eqref{eq:gap_eq_FT}, we thus have to determine the two functions, $\Sigma_{T,3}(P_{m\alpha,3})$ and $\Sigma_{T,I}(P_{m\alpha,3})$, as yet unknown in \eqref{eq:sigma_decomp}.


\section{Exact solution of the finite temperature gap equation}\label{sge}

In this section, the technical heart of this paper, we determine the exact solution to the gap equation \eqref{eq:gap_eq_FT}. This section is organized as follows. In subsection \ref{efge}, we present a more explicit form of the gap equation \eqref{eq:gap_eq_FT} (see \eqref{gapeqdet1} and \eqref{gapeqdet2}). In subsection \ref{subsec:ansatz_sol}, we present and motivate an ansatz for the solution of these gap equations (our ansatz effectively linearizes these equations). In subsection \ref{subsec:gap_solve_outline}, we present our solution to the gap equation, and explain in broad outline, how this solution is obtained. The solution presented in subsection \ref{subsec:gap_solve_outline} is determined using several mathematical identities (derived in appendix \ref{app:omega_iden}). In subsection \ref{naive}, we give the reader a flavour of these identities by presenting the detailed derivation in a simple example. In subsection \ref{subsec:sol_gpe}, we explain how the identities of appendix \ref{app:omega_iden} are used to solve the gap equations in an iterative expansion in the coupling constant. 


\subsection{Explicit form of the gap equations} \label{efge}

Plugging \eqref{eq:sigma_decomp} into \eqref{eq:gap_eq_FT} we can obtain a more spelt out form of the gap equations as follows. Observing,
\begin{equation}\label{eq:propagator}
    \dfrac{1}{i(\slashed{P}_{m\alpha}+\slashed{\Sigma}_T(P_{m\alpha,3}))+\Sigma_{T,I}(P_{m\alpha,3})} = \dfrac{-i(\slashed{P}_{m\alpha}+\slashed{\Sigma}_T(P_{m\alpha,3}))+\Sigma_{T,I}(P_{m\alpha,3})}{(P_{m\alpha}+\Sigma_T(P_{m\alpha,3}))^2+\Sigma_{T,I}(P_{m\alpha,3})^2},
\end{equation}
where $(P_{m\alpha}+\Sigma_T(P_{m\alpha,3}))^2=(P_{m\alpha,\dot{\mu} }+\Sigma_{T,\dot{\mu}}(P_{m\alpha,3}))(P_{m\alpha}^{\dot{\mu}}+\Sigma_T^{\dot{\mu}}(P_{m\alpha,3}))$, and using (\ref{eq:sigma_decomp}) with (\ref{eq:mat_val_func_H}), we rewrite \eqref{eq:gap_eq_FT} as,\footnote{ $\Sigma_{T,1}(P_{m\alpha,3}) = \Sigma_{T,2}(P_{m\alpha,3}) = 0$ }
\begin{equation}
    \Sigma_T(P_{m\alpha,3}) = - \dfrac{4 \pi}{\beta\kappa}\sum_{\substack{\sigma \\ j \in \mathbb{Z} + \frac{1}{2}}}\int\ddm{q} \dfrac{i\Sigma_{T,I}(Q_{j\sigma,3})\gamma^3 -(Q_{j\sigma,3}+\Sigma_{T,3}(Q_{j\sigma,3}))I}{(Q_{j\sigma}+\Sigma_T(Q_{j\sigma,3}))^2+\Sigma_{T,I}(Q_{j\sigma,3})^2} \dfrac{1}{P_{m\alpha,3}-Q_{j\sigma,3}}.
\end{equation}
Using (\ref{eq:sigma_decomp}), substituting for $\Sigma_T(P_{m\alpha,3})$ on the LHS above and equating coefficients of linearly independent matrices, it follows that,
\begin{align} 
    \Sigma_{T,3}(P_{m\alpha,3}) &= - \dfrac{4 \pi}{\beta\kappa}\sum_{\substack{\sigma \\ j \in \mathbb{Z} + \frac{1}{2}}}\int\ddm{q} \dfrac{\Sigma_{T,I}(Q_{j\sigma,3})}{(Q_{j\sigma}+\Sigma_T(Q_{j\sigma,3}))^2+\Sigma_{T,I}(Q_{j\sigma,3})^2} \dfrac{1}{P_{m\alpha,3}-Q_{j\sigma,3}},\label{gapeqdet1} \\
    \Sigma_{T,I}(P_{m\alpha,3}) - M &= \dfrac{4 \pi}{\beta\kappa}\sum_{\substack{\sigma \\ j \in \mathbb{Z} + \frac{1}{2}}}\int\ddm{q} \dfrac{Q_{j\sigma,3}+\Sigma_{T,3}(Q_{j\sigma,3})}{(Q_{j\sigma}+\Sigma_T(Q_{j\sigma,3}))^2+\Sigma_{T,I}(Q_{j\sigma,3})^2} \dfrac{1}{P_{m\alpha,3}-Q_{j\sigma,3}}.\label{gapeqdet2}
\end{align}


\subsection{An ansatz for the solution to the gap equations}\label{subsec:ansatz_sol}

\eqref{gapeqdet1} and \eqref{gapeqdet2} constitute a complicated set of two coupled nonlinear integral equations. We will solve these equations using a physically motivated guess. Note that the integrand on the RHS of each of the two equations, \eqref{gapeqdet1} and \eqref{gapeqdet2}, involves the same denominator. Given that the thermal quasiparticles in vector large-$N$ theories have a lifetime of the order $N$, and that this lifetime goes to infinity in the large-$N$ limit, it is natural to guess that the denominators flagged above are simply those for a free fermion with an, as yet unknown, renormalized thermal mass. Our belief in this guess is strengthened by the fact that this has, indeed, turned out to be the case in all the computations performed, to date, in the light-cone gauge (see e.g. \cite{Giombi:2011kc, Jain:2013py, Dey:2018ykx}). 

Motivated by the considerations described above, we assume that our solution will turn out to obey the identity,
\begin{align}\label{eq:gap_eq_constraint}
    Q_{j\sigma}^2+2Q_{j\sigma,3}\Sigma_{T,3}(Q_{j\sigma,3})+\Sigma_{T,3}(Q_{j\sigma,3})^2+\Sigma_{T,I}(Q_{j\sigma,3})^2 = Q_{j\sigma}^2 + M_T^2,
\end{align}
for some (as yet unknown) thermal mass $M_T$. Later in this paper, we will 
demonstrate that our final solution indeed obeys \eqref{eq:gap_eq_constraint} for a suitable choice of the variable $M_T$.

The guess \eqref{eq:gap_eq_constraint} greatly simplifies the gap equations \eqref{gapeqdet1} and \eqref{gapeqdet2}; in particular, these equations now become linear integral equations. The form of these linear integral equations is simplified by making the definitions,
\begin{equation}\label{newdefs} 
    \Sigma_{T,\pm}(P_{m\alpha,3}) = \Sigma_{T,I}(P_{m\alpha,3}) \pm i\Sigma_{T,3}(P_{m\alpha,3}),
\end{equation} 
which end up decoupling the two integral equations, yielding, 
\begin{align} \label{gedecoup} 
    \Sigma_{T,\pm}(P_{m\alpha,3}) = M+ \dfrac{4 \pi}{\beta\kappa}\sum_{\substack{j \in \mathbb{Z} + \frac{1}{2} \\ \sigma \\ \sigma \neq \alpha\text{ at }j=m}} \int\ddm{q} \dfrac{1}{P_{m\alpha,3}-Q_{j\sigma,3}}  \dfrac{Q_{j\sigma,3} \mp i\Sigma_{T,\pm}(Q_{j\sigma,3})}{(Q_{j\sigma})^2 + M_T^2}.
\end{align}
Note that the summation over $\sigma$ in \eqref{gedecoup} does not include contributions from $\sigma=\alpha$, when $j=m$, because the gauge boson that would have given rise to such a contribution is one of the two-dimensional $U(1)^N$ gauge fields.\footnote{Recall that integrating out these special gauge fields, in the large-$N$ limit did not induce a new interaction between our fermionic fields $\psi$, but instead, simply set the holonomy field to be a constant.} For this reason the contribution of $\sigma=\alpha$, when $j=m$, is absent in \eqref{gedecoup}. Although we do not always explicitly include this fact in our notation from now on, contributions of this nature are always omitted from all sums. In practice, all summations are defined so that any term with a vanishing denominator is removed from the summation.


\subsection{Outline of the solution to the gap equations}\label{subsec:gap_solve_outline}

As we describe in appendix \ref{app:sol_gap_eq}, the equations \eqref{gedecoup} are easily solved by iteration. Let us define, 
\begin{equation}\label{defoc}
     \Sigma_{T,\pm}(P_{m\alpha,3})= \sum_{n=0}^\infty \left( \dfrac{4 \pi}{\kappa} \right)^n \Sigma^{(n)}_{T,\pm}(P_{m\alpha,3}).
\end{equation}
While we have expanded \eqref{defoc} in a power series in $\frac{1}{\kappa}$, the expansion coefficients $\Sigma^{(n)}_{T,\pm}$ will, of course, turn out to scale like $N^n$ at the leading order in the large-$N$ limit under study. As a consequence, the expansion \eqref{defoc} is effectively an expansion in a power series in $\lambda= \frac{N}{\kappa}$, as might have been anticipated on general grounds. 

By explicitly iterating \eqref{gedecoup}, it is easy to show that, 
\begin{align} 
    \Sigma^{(0)}_{T,\pm}(P_{m\alpha,3}) &= M, \label{nthorder_0}\\
    \Sigma^{(n)}_{T,\pm}(P_{m\alpha,3}) &= \left( \mp \dfrac{i}{\beta} \right)^{n-1} \dfrac{1}{\beta}  \sum_{\substack{\sigma_1\\ j_1 \in \mathbb{Z} + \frac{1}{2}}} \int\ddm{q_1} \dfrac{1}{P_{m\alpha,3}-(Q_1)_{j_1\sigma_1,3}} \dfrac{1}{((Q_1)_{j_1\sigma_1})^2 + M_T^2}   \nonumber \\
    & \qquad\qquad\qquad\times \prod_{i=2}^{n} \left( \int \ddm{q_{i}} \sum_{\substack{\sigma_{i} \\ j_{i} \in \mathbb{Z} + \frac{1}{2}} } \dfrac{1}{(Q_{i-1})_{j_{i-1} \sigma_{i-1},3} - (Q_{i})_{j_{i} \sigma_{i},3} } \dfrac{1}{(Q_{i})_{j_{i} \sigma_{i}}^2 + M_T^2} \right) \nonumber \\
    &\qquad\qquad\qquad \times \Big( (Q_n)_{j_n \sigma_n,3} \mp i M  \Big), \qquad \text{for}\quad n\geq 1. \label{nthorder}
\end{align}
Quite remarkably, it is possible to use a finite temperature version of the identities described in \cite{Moshe:2014bja}, to actually evaluate all the integrals/summations in \eqref{nthorder} at the leading order in the large-$N$ limit. In appendix \ref{app:sol_gap_eq}, we demonstrate that the expressions in \eqref{nthorder} evaluate to,
\begin{equation}\label{nthorderans}
    \resizebox{1.15\hsize}{!}{$\displaystyle\Sigma^{(n)}_{T,\pm}(P_{m\alpha,3})= \dfrac{M}{n!} \big( \mp i\Omega_T(M_T,P_{m \alpha,3})  \big)^{n} - \dfrac{\Phi_T(M_T)}{(n-1)!} \big( \mp i\Omega_T(M_T,P_{m \alpha,3})  \big)^{n-1} \pm \dfrac{i P_{m \alpha,3}}{n!} \big( \mp i\Omega_T(M_T, P_{m \alpha,3} )  \big)^{n}, \quad n \geq 1,$}
\end{equation}
where we have defined,
\begin{align}
    \Omega_T(M_T,P_{m\alpha,3}) &= \dfrac{1}{\beta}\sum_{\substack{ j \in \mathbb{Z} + \frac{1}{2} \\ \sigma \\ \sigma \neq \alpha \text{ at } j=m}}\int\ddm{q} \dfrac{1}{(Q_{j \sigma})^2+M_T^2} \dfrac{1}{(P_{m\alpha,3}-Q_{j \sigma,3})},\label{eq:integral_omega_FT}\\
    \Phi_T(M_T) &= \dfrac{1}{\beta}\sum_{\substack{\sigma \\ j \in \mathbb{Z} + \frac{1}{2}}}\int\ddm{q} \dfrac{1}{(Q_{j\sigma})^2+M_T^2}. \label{eq:integral_phi_FT}
\end{align}
Plugging \eqref{nthorder_0} and \eqref{nthorderans} into \eqref{defoc}, and performing the summation yields the summed answers,
\begin{equation}\label{sumans}
    \Sigma_{T,\pm}(P_{m\alpha,3}) = \left( M - \dfrac{4 \pi}{\kappa} \Phi_{T}(M_T) \pm i P_{m \alpha,3}  \right)  \exp( \mp \dfrac{4 \pi i}{\kappa} \Omega_T(M_T,P_{m \alpha,3}) ) \mp i P_{m \alpha,3}.
\end{equation} 

\eqref{sumans}, our almost final answer, still has to pass through a very significant check. Now it must be true that \eqref{eq:gap_eq_constraint}, which we had assumed at the beginning of our workout, is indeed obeyed, for all values of momenta. Remarkably enough, this indeed turns out to be the case, provided the thermal mass $M_T$ obeys the mass gap equation, 
\begin{align}
     M_T^2 &= \left( M - \dfrac{4 \pi}{\kappa} \Phi_{T}(M_T)  \right)^2,
\end{align}
or, taking the positive root,
\begin{align}\label{ge}
    M_T &= M - \dfrac{4 \pi}{\kappa} \Phi_{T}(M_T).
\end{align}
The function $\Phi_T(M_T)$, that appears in \eqref{ge}, is relatively elementary, and has been evaluated explicitly in appendix \ref{app:phi_eval}. In the zero temperature limit, $\Phi_T(M_T)$ becomes,
\begin{align}
    \Phi_0(M_0)= -\dfrac{NM_0}{4\pi}.
\end{align}
As a consequence, the zero temperature gap equation takes the completely elementary form,
\begin{align}\label{gezero}
    M_0 &= M + \lambda M_0,
\end{align}
where $\lambda = \dfrac{N}{\kappa}$ is the 't Hooft coupling, and $M$ and $M_0$ are the bare mass and the renormalized mass at zero temperature, respectively. 

It follows that \eqref{sumans}, with $M_T$ constrained to obey \eqref{ge}, is the final solution to the gap equations \eqref{gedecoup}. 

The general solution to the finite temperature gap equation can also be expressed in the form,
\begin{align}
    \Sigma_{T,I}(P_{m\alpha,3}) &= P_{m\alpha,3} \sin({\dfrac{4 \pi}{\kappa}\Omega_T(M_T,P_{m\alpha,3})})+ M_T\cos({\dfrac{4 \pi}{\kappa}\Omega_T(M_T,P_{m\alpha,3})}),\label{eq:gap_eq_gen_sol_FT_I}\\
    \Sigma_{T,3}(P_{m\alpha,3}) &= P_{m\alpha,3} \left(\cos({\dfrac{4 \pi}{\kappa}\Omega_T(M_T,P_{m\alpha,3})})-1\right) - M_T\sin({\dfrac{4 \pi}{\kappa}\Omega_T(M_T,P_{m\alpha,3})}).\label{eq:gap_eq_gen_sol_FT_3}
\end{align}
\eqref{sumans} (equivalently \eqref{eq:gap_eq_gen_sol_FT_3} and \eqref{eq:gap_eq_gen_sol_FT_I}), together with \eqref{ge}, \eqref{eq:integral_omega_FT}, and \eqref{eq:integral_phi_FT}, are our final answers for the self-energy. 

Our final answer is less than completely explicit for two reasons:
\begin{itemize}
    \item[(1)] Our results are given in terms of the function $\Omega_T(M_T, P_{m \alpha,3})$, which is defined as an integral (see \eqref{eq:integral_omega_FT}) that we have not been able to evaluate in a closed-form. Moreover, it turns out that $\Omega_T(M_T,P_{m\alpha,3})$ has a logarithmic divergence (or a divergence proportional to $\dfrac{1}{\epsilon}$ in our dimensional regularization scheme; see appendix \ref{app:Omegadiv}). For these reasons, $\Omega_T(M_T, P_{m \alpha,3})$ is only implicitly defined.\footnote{In contrast, the other function, $\Phi_T(M_T)$, which enters our solution \eqref{sumans} (through \eqref{ge}) is defined by an integral (see \eqref{eq:integral_phi_FT}) that is very easy to evaluate explicitly. Indeed, it turns out that, 
    \begin{equation}\label{phit}
        \Phi_T(M_T) = -\dfrac{1}{4 \pi \beta} \sum_{\sigma} \ln \left| 2 (\cosh(\beta M_T) + \cos( \lambda_{\sigma})) \right|.
    \end{equation}
    See appendix \ref{app:phi_eval} for a derivation.}
    \item[(2)] Our final expression for the self-energy is presented in terms of the thermal mass $M_T$, which is defined by the algebraic equation \eqref{ge}. While it is clear that \eqref{ge} has a unique solution (and while this solution is very easily found, numerically, on Wolfram Mathematica), we do not know of a closed-form solution for $M_T$, as a function of $T$. 
\end{itemize}

The mass gap equation \eqref{ge} is identical to the mass gap equation obtained in the light-cone gauge (see equation (3.8) of \cite{Minwalla:2020ysu}), and so the lack of a closed-form solution of this equation is a feature that the `temporal' gauge and light-cone gauge share in common (indeed, this fact is gauge-invariant: as the thermal mass determines the poles in the thermal propagator, it is gauge-invariant). On the other hand, the unevaluated integral, $\Omega_T(M_T, P_{m \alpha,3})$, in our final answer had no analogue in the light-cone gauge. Of course, the fermion propagator is not, in itself, gauge-invariant. Note that the function $\Omega_T(M_T, P_{m \alpha,3})$, unlike the function $\Phi_T(M_T)$, does not enter the equation for the (physical) finite temperature pole mass $M_T$. We will also demonstrate below that when we use the results \eqref{sumans} to evaluate the free energy, the results will be presented in terms of integrals of $\Omega_T(M_T, P_{m \alpha,3})$, which we will be able to evaluate explicitly without ever using the explicit form of $\Omega_T(M_T, P_{m \alpha,3})$. In particular, the logarithmic divergence in the quantity $\Omega_T(M_T, P_{m \alpha,3})$ does not appear in our final answer for the (physical) finite temperature free energy. Our final answers for the free energy are completely explicit and in perfect agreement with those obtained in the light-cone gauge. For all these reasons, we conjecture that the (rather ugly) function $\Omega_T(M_T, P_{m \alpha,3})$ is an unphysical gauge-artefact, in the sense that its explicit form will never be needed for the evaluation of any gauge-invariant quantity, and that the logarithmic divergence in $\Omega_T(M_T, P_{m \alpha,3})$ will not appear (so to speak, will cancel) in every physical quantity. It would be interesting to prove this conjecture, or at least to check it via the computation of more intricate gauge-invariant observables like correlators (see sections 6 and 8 of \cite{Moshe:2014bja}), or 
S-matrices, or the free energy at finite volume. We leave this exercise to future work. 


\subsection{Illustration of the integral identities} \label{naive} 

In the sketch of our solution to the gap equation, presented above, we relegated the key technical step, namely the transition from the iterated form of the solution \eqref{nthorder} to the explicit form of this solution \eqref{nthorderans}, to appendix \ref{app:sol_gap_eq}. In appendix \ref{app:sol_gap_eq}, we follow \cite{Moshe:2014bja} to explicitly evaluate the integrals in \eqref{nthorderans} at every value of $n$. In order to give the reader a taste of our methods (which are a mild adaptation of the methods of \cite{Moshe:2014bja}), in the rest of this section, we present a detailed evaluation of $\Sigma^{(n)}_{T,\pm}(P_{m\alpha,3})$ for the simplest nontrivial case $n=2$. It follows from \eqref{nthorder} that,
\begin{align} 
    \resizebox{1.1\hsize}{!}{$\displaystyle \Sigma^{(2)}_{T,\pm}(P_{m\alpha,3}) = \dfrac{\mp i}{\beta^2}  \sum_{\substack{\sigma_1,  \sigma_2\\ j_1, j_2 \in \mathbb{Z} + \frac{1}{2}}} \int\ddm{q_1} \ddm{q_2} \dfrac{(Q_2)_{j_2 \sigma_2,3} \mp i M }{\Big(P_{m\alpha,3}-(Q_1)_{j_1\sigma_1,3}\Big) \left((Q_1)^2_{j_1\sigma_1} + M_T^2 \right)  \Big((Q_{1})_{j_{1} \sigma_{1},3} - (Q_{2})_{j_{2} \sigma_{2},3} \Big) 
    \left( (Q_{2})_{j_{2} \sigma_{2}}^2 + M_T^2\right) }. \label{ntwo}$}
\end{align}
It was explained in \cite{Moshe:2014bja} that a set of simple manipulations allow one to greatly simplify integrals of the form \eqref{ntwo}. The manipulations in question are precise only in the large-$N$ limit: this is a consequence of a fact that we have emphasized above, namely that the summation range in \eqref{ntwo} includes an exclusion. Concretely, all terms with $\sigma_1=\alpha$ at $j_1=m$ are removed from the summation, and the same is also true for terms with $\sigma_2=\sigma_1$ at $j_2=j_1$. In the interest of 
clarity, we first explain our manipulations ignoring this subtlety, and then return to account for this subtlety. 

The expression for $\Sigma^{(2)}_{T,\pm}(P_{m\alpha,3})$  presented above in \eqref{ntwo} can be broken up into a sum of two terms, the first of which receives contributions only from the $\mp iM$ in the numerator, and the second of which receives contributions only from the $(Q_2)_{j_2 \sigma_2,3}$ in the numerator. Correspondingly we have,
\begin{equation}\label{signt}
    \Sigma^{(2)}_{T,\pm}(P_{m\alpha,3})=  M \left( \dfrac{\mp i}{\beta}\right)^2 I_1+ \left( \dfrac{\mp i}{\beta}\right)^2 I_2.
\end{equation} 
Let us start with $I_1$, that is,
\begin{align} \label{expione}
    \resizebox{1.1\hsize}{!}{$\displaystyle I_1 =   \sum_{\substack{ j_1,j_2 \in \mathbb{Z} + \frac{1}{2} \\ \sigma_1,\sigma_2 }} \int\ddm{q_1} \ddm{q_2} \dfrac{1}{\Big(P_{m\alpha,3}-(Q_1)_{j_1\sigma_1,3}\Big) \left((Q_1)^2_{j_1\sigma_1} + M_T^2 \right)  \Big((Q_{1})_{j_{1} \sigma_{1},3} - (Q_{2})_{j_{2} \sigma_{2},3} \Big) 
    \left( (Q_{2})_{j_{2} \sigma_{2}}^2 + M_T^2\right)}.$}
\end{align}
The expression on the RHS of \eqref{expione} involves summations (and integrals) over the dummy variables $(j_1, \sigma_1, q_1) $ and $(j_2, \sigma_2, q_2)$. Of course, we get the same final answer if we interchange the $1$ and $2$ labels on all the dummy variables. We use the term `symmetrization' (in $1$ and $2$) for this operation. It follows that the LHS of \eqref{expione} equals the sum of the RHS and the symmetrization of the RHS, divided by two. Note, however, that the integrand in the RHS of \eqref{expione} includes the factor $\displaystyle \dfrac{1}{ (Q_{1})_{j_{1} \sigma_{1},3} - (Q_{2})_{j_{2} \sigma_{2},3} }$, which is antisymmetric under symmetrization. It follows that the procedure described above gives us,
\begin{align}\label{simpexp_init}
     \text{\small $2I_1$} &\resizebox{1.05\hsize}{!}{$\displaystyle = \sum_{\substack{ j_1,j_2 \\ \sigma_1, \sigma_2 }}  \int \ddm{q_1} \ddm{q_2} \left(  \dfrac{1}{P_{m\alpha,3}-(Q_1)_{j_1\sigma_1,3}} - \dfrac{1}{P_{m\alpha,3}-(Q_2)_{j_2\sigma_2,3}}  \right) \dfrac{1}{(Q_{1})_{j_{1} \sigma_{1},3} - (Q_{2})_{j_{2} \sigma_{2},3} } \, \dfrac{1}{(Q_1)_{j_1\sigma_1}^2 + M_T^2} \, \dfrac{1}{(Q_2)_{j_2\sigma_2}^2 + M_T^2} $} \nonumber\\
     &\resizebox{1.05\hsize}{!}{$\displaystyle = \sum_{\substack{ j_1,j_2  \\ \sigma_1, \sigma_2  }}  \int \ddm{q_1} \ddm{q_2}  \dfrac{(Q_{1})_{j_{1} \sigma_{1},3} - (Q_{2})_{j_{2} \sigma_{2},3} }{ \Big( P_{m\alpha,3}-(Q_1)_{j_1\sigma_1,3} \Big) \Big( P_{m\alpha,3}-(Q_2)_{j_2\sigma_2,3} \Big)} \dfrac{1}{(Q_{1})_{j_{1} \sigma_{1},3} - (Q_{2})_{j_{2} \sigma_{2},3} } \, \dfrac{1}{(Q_1)_{j_1\sigma_1}^2 + M_T^2} \, \dfrac{1}{(Q_2)_{j_2\sigma_2}^2 + M_T^2}$},
\end{align}
which simplifies to,
\begin{align}\label{simpexp}
     2I_1 &= \sum_{\substack{ j_1,j_2  \\ \sigma_1, \sigma_2  }}  \int \ddm{q_1} \ddm{q_2}  \dfrac{1 }{ P_{m\alpha,3}-(Q_1)_{j_1\sigma_1,3} } \, \dfrac{1}{(Q_1)_{j_1\sigma_1}^2 + M_T^2} \dfrac{1}{ P_{m\alpha,3}-(Q_2)_{j_2\sigma_2,3} } \, \dfrac{1}{(Q_2)_{j_2\sigma_2}^2 + M_T^2} \nonumber\\
     &= \beta^2  \left( \dfrac{1}{\beta}  \sum_{\substack{ \sigma \\ j \in \mathbb{Z} + \frac{1}{2} }} \int \ddm{q} \dfrac{1}{P_{m\alpha,3}-Q_{j\sigma,3}} \dfrac{1}{(Q_{j\sigma})^2 + M_T^2} \right)^2\nonumber\\
     &= \beta^2 \Omega^2_T(M_T,P_{m\alpha,3}).
\end{align}
In going from the last line of \eqref{simpexp_init} to the first line of \eqref{simpexp}, we have cancelled the numerator with the equivalent term in the denominator, and grouped the remaining terms in the denominator such that all the terms with the label $`1$', and separately, with the label $`2$', are brought together. This manoeuvre then allows us to recognize the answer as a perfect square.

We now turn to $I_2$. Even before symmetrizing, it is convenient to rearrange the expression for $I_2$ as follows,
\begin{align}\label{fsline_pre}
    I_2 &= \resizebox{1.05\hsize}{!}{$\displaystyle \sum_{\substack{j_1,j_2 \in \mathbb{Z} + \frac{1}{2} \\ \sigma_1, \sigma_2  }} \int \ddm{q_1} \ddm{q_2} \dfrac{ \pm i (Q_{2})_{j_{2} \sigma_{2},3} }{\Big( P_{m\alpha,3}-(Q_1)_{j_1\sigma_1,3} \Big)  \left( (Q_1)_{j_1\sigma_1}^2 + M_T^2 \right)  \Big( (Q_{1})_{j_{1} \sigma_{1},3} - (Q_{2})_{j_{2} \sigma_{2},3} \Big)  \left( (Q_{2})_{j_{2} \sigma_{2}}^2 + M_T^2 \right)  }$}  \nonumber\\
    &= \resizebox{1.05\hsize}{!}{$\displaystyle \sum_{\substack{j_1,j_2 \in \mathbb{Z} + \frac{1}{2} \\ \sigma_1, \sigma_2  }} \int \ddm{q_1} \ddm{q_2} \dfrac{ \pm i \left[ (Q_{2})_{j_{2} \sigma_{2},3} -(Q_{1})_{j_{1} \sigma_{1},3} + (Q_{1})_{j_{1} \sigma_{1},3} \right] }{\Big( P_{m\alpha,3}-(Q_1)_{j_1\sigma_1,3} \Big)  \left( (Q_1)_{j_1\sigma_1}^2 + M_T^2 \right)  \Big( (Q_{1})_{j_{1} \sigma_{1},3} - (Q_{2})_{j_{2} \sigma_{2},3} \Big)  \left( (Q_{2})_{j_{2} \sigma_{2}}^2 + M_T^2 \right)  }$} \nonumber\\
    &= \mp i  \sum_{\substack{\sigma_1 \\ j_1 \in \mathbb{Z} + \frac{1}{2}}} \int\ddm{q_1} \dfrac{1}{P_{m\alpha,3}-(Q_1)_{j_1\sigma_1,3}} \dfrac{1}{(Q_1)_{j_1\sigma_1}^2 + M_T^2}    \int \ddm{q_{2}} \sum_{\substack{\sigma_{2} \\ j_{2} \in \mathbb{Z} + \frac{1}{2}} } \dfrac{1}{(Q_{2})_{j_{2} \sigma_{2}}^2 + M_T^2} \nonumber \\
    & \;\;\;\;\;  \resizebox{1.05\hsize}{!}{$\displaystyle\pm i  \sum_{\substack{j_1,j_2 \in \mathbb{Z} + \frac{1}{2} \\ \sigma_1, \sigma_2  }} \int \ddm{q_1} \ddm{q_2} \dfrac{ (Q_{1})_{j_{1} \sigma_{1},3} }{\Big( P_{m\alpha,3}-(Q_1)_{j_1\sigma_1,3} \Big)  \left( (Q_1)_{j_1\sigma_1}^2 + M_T^2 \right)  \Big( (Q_{1})_{j_{1} \sigma_{1},3} - (Q_{2})_{j_{2} \sigma_{2},3} \Big)  \left( (Q_{2})_{j_{2} \sigma_{2}}^2 + M_T^2 \right)  },$}
\end{align}
or,
\begin{align} \label{fsline}
    I_2 &= \beta^2 \Phi_T(M_T) \Big(\mp i \Omega_T(M_T,P_{m \alpha,3}) \Big) \nonumber\\
    & \;\;\;\;\;  \resizebox{1.05\hsize}{!}{$\displaystyle\pm i   \sum_{\substack{j_1,j_2 \in \mathbb{Z} + \frac{1}{2} \\ \sigma_1, \sigma_2  }} \int \ddm{q_1} \ddm{q_2} \dfrac{ (Q_{1})_{j_{1} \sigma_{1},3} }{\Big( P_{m\alpha,3}-(Q_1)_{j_1\sigma_1,3} \Big)  \left( (Q_1)_{j_1\sigma_1}^2 + M_T^2 \right)  \Big( (Q_{1})_{j_{1} \sigma_{1},3} - (Q_{2})_{j_{2} \sigma_{2},3} \Big)  \left( (Q_{2})_{j_{2} \sigma_{2}}^2 + M_T^2 \right)  }$} \nonumber\\
    &= \beta^2 \, \Phi_T(M_T) \Big(\mp i \Omega_T(M_T,P_{m \alpha,3}) \Big) + \tilde{I}_2.
\end{align}
In going from \eqref{fsline_pre} to \eqref{fsline}, we have used the definitions of the functions $\Phi_T(M_T)$ and $\Omega_T(M_T,P_{m \alpha,3})$, given in \eqref{eq:integral_phi_FT} and \eqref{eq:integral_omega_FT}, respectively. The function ${\tilde I}_2$ is defined by the second expression in the first equality of \eqref{fsline}. 

The computation of ${\tilde I}_2$ is very similar to that of $I_1$. To proceed with the computation, we first symmetrize the formula for ${\tilde I}_2$ (in the manner described above), to obtain,
\begin{align}
    \resizebox{1.1\hsize}{!}{$\displaystyle 2 \tilde{I}_2 = \sum_{ \substack{ j_1,j_2 \\ \sigma_1, \sigma_2  } }  \int \ddm{q_1} \ddm{q_2} \left(  \dfrac{(Q_{1})_{j_{1} \sigma_{1},3} }{P_{m\alpha,3}-(Q_1)_{j_1\sigma_1,3}} - \dfrac{(Q_{2})_{j_{2} \sigma_{2},3} }{P_{m\alpha,3}-(Q_2)_{j_2\sigma_2,3}}  \right)  \dfrac{\pm i}{(Q_{1})_{j_{1} \sigma_{1},3} - (Q_{2})_{j_{2} \sigma_{2},3} } \, \dfrac{1}{(Q_1)_{j_1\sigma_1}^2 + M_T^2} \, \dfrac{1}{(Q_2)_{j_2\sigma_2}^2 + M_T^2},$}
\end{align}
which simplifies to,
\begin{align}\label{itwotilde}
    \text{\small $2 \tilde{I}_2$} &= \resizebox{1.05\hsize}{!}{$\displaystyle \sum_{ \substack{ j_1,j_2 \\ \sigma_1, \sigma_2 } }  \int \ddm{q_1} \ddm{q_2} \dfrac{ \pm i P_{m\alpha,3} \Big((Q_1)_{j_1\sigma_1,3} -(Q_{2})_{j_{2} \sigma_{2},3} \Big) }{\Big( P_{m\alpha,3}-(Q_1)_{j_1\sigma_1,3} \Big) \Big( P_{m\alpha,3}-(Q_2)_{j_2\sigma_2,3} \Big)} \dfrac{1}{(Q_{1})_{j_{1} \sigma_{1},3} - (Q_{2})_{j_{2} \sigma_{2},3} } \, \dfrac{1}{(Q_1)_{j_1\sigma_1}^2 + M_T^2} \, \dfrac{1}{(Q_2)_{j_2\sigma_2}^2 + M_T^2}$} \nonumber\\
    &=\pm i P_{m\alpha,3} \sum_{\substack{\sigma_1, \sigma_2\\ j_1,j_2 } }  \int \ddm{q_1} \ddm{q_2}  \dfrac{1}{P_{m\alpha,3}-(Q_1)_{j_1\sigma_1,3}}  \dfrac{1}{(Q_1)_{j_1\sigma_1}^2 + M_T^2} \dfrac{1}{P_{m\alpha,3}-(Q_2)_{j_2\sigma_2,3}}  \dfrac{1}{(Q_2)_{j_2\sigma_2}^2 + M_T^2}\nonumber\\
    &= \pm i P_{m\alpha,3}  \left( \, \sum_{\substack{\sigma \\ j \in \mathbb{Z} + \frac{1}{2}}}  \int \ddm{q}  \dfrac{1}{P_{m\alpha,3}-Q_{j\sigma,3}}  \dfrac{1}{(Q_{j\sigma})^2 + M_T^2} \right)^2 \nonumber\\
    &=\pm i P_{m\alpha,3} \,\, \beta^2 \, \Big( \Omega_T(M_T,P_{m \alpha,3}) \Big)^2.
\end{align}
Plugging the result for $\tilde{I}_2$ from \eqref{itwotilde} in \eqref{fsline}, we get,
\begin{align} \label{itwomain}
    I_2 = \beta^2 \, \Phi_T(M_T) \Big(\mp i \Omega_T(M_T,P_{m \alpha,3}) \Big) \pm \dfrac{ i P_{m\alpha,3}}{2} \beta^2 \, \Big( \Omega_T(M_T,P_{m \alpha,3}) \Big)^2.
\end{align}
Collecting both the integrals \eqref{simpexp} and \eqref{itwomain}, from \eqref{signt} we get,
\begin{align}\label{eq:sigma_pm_naive_result}
    \resizebox{1.1\hsize}{!}{$\displaystyle\Sigma^{(2)}_{T,\pm}(P_{m\alpha,3}) = \dfrac{M}{2} \big( \mp i \Omega_T(M_T,P_{m\alpha,3}) \big)^2 - \Phi_T(M_T) \big(\mp i \Omega_T(M_T,P_{m \alpha,3}) \big) \pm \dfrac{ i P_{m\alpha,3}}{2} \big( \mp i \Omega_T(M_T,P_{m \alpha,3}) \big)^2.$}
\end{align}

As mentioned above, for the analysis of this subsection, we have ignored the exclusions (as described towards the end of subsection \ref{subsec:ansatz_sol}) from the summations that appear in \eqref{ntwo} (read below \eqref{ntwo}). As the exclusions occur only when two colour indices become equal, the reader might suspect that a careful accounting for these exclusions only modifies our answers at the first subleading order in $1/N$. In appendix \ref{careful}, we demonstrate that this is indeed the case. As the interest of this paper is in the large-$N$ limit, we will blithely proceed with our analysis using naive identities of the form derived in this subsection, leaving the analysis of $1/N$ corrections to future work.


\subsubsection{Exclusions in the zero temperature limit}\label{subsubsec:zt_exclusion}

We have explained above that the `exclusions', which complicate the identities derived in this section, are negligible in the large-$N$ limit. Let us recall why this turned out to be the case. As we have explained under \eqref{gedecoup}, the terms that are excluded from all summations are those corresponding to the exchange of the two-dimensional $U(1)^N$ gauge bosons. The fact that exclusions affect the answer only at order ${\cal O}(1/N)$ follows from the fact that these $N$ gauge fields are only $(1/N)^{\rm th}$ the field content of the $N^2$ gauge bosons. 

The reason only two-dimensional $U(1)^N$ exchange (rather than than the full two-dimensional $U(N)$ exchange) is excluded from summations has its roots in the fact that the $U(N)$ gauge-invariance of the underlying UV theory was broken down to $U(1)^N$ by the holonomy eigenvalues $\{\lambda_\alpha\}$. These eigenvalues ensure that the propagator of the $n=0$ Kaluza--Klein modes of the spatial components of the gauge field with gauge index structure  $A_\alpha^{~\sigma}$ is proportional to,
\begin{equation}\label{probod}
    \frac{\beta}{\lambda_\alpha-\lambda_\sigma}.
\end{equation} 
At any nonzero temperature, this propagator diverges when $\alpha=\sigma$ (and so the corresponding terms are excluded from sums like \eqref{gedecoup}) but is finite when $\alpha\neq \sigma$ (so the corresponding terms are not excluded from summations like \eqref{gedecoup}). This discussion explains why the exclusions affect the identities derived in this subsection only at the subleading order in $1/N$ at any finite temperature. 

At least naively, the situation appears to be different in the strict zero temperature limit. In this limit, the propagator listed in \eqref{probod} diverges for every choice of the indices $\alpha$ and $\sigma$ (this is a consequence of the factor of $\beta$ in the numerator of \eqref{probod}).\footnote{This fact may have been anticipated from the following physical observation. At absolute zero temperature, time is non-compact, there are no temporal holonomies, and the breaking of $U(N) \to U(1)^N$ should disappear.} At least naively, it thus appears that the limits $T \to 0$ and $N \to \infty$ may not commute. All the results of this paper are certainly valid if we first take $N \to \infty$ and then take the limit $T \to 0$. It is not clear to us whether (and if so why) the formulae presented in this paper also apply if the limits are taken in the opposite order, i.e. if we first take $T \to 0$ and then take $N \to \infty$. The interested reader will find some more discussion of this point in appendix \ref{zertemp}.


\subsection{Outline of the derivation of the solution to the gap equations}\label{subsec:sol_gpe}

In the previous subsection, we described the identities needed to evaluate $\Sigma^{(2)}_{T,\pm}(P_{m\alpha,3})$. Following \cite{Moshe:2014bja}, we use similar methods in appendix \ref{app:omega_iden}, to derive the identities needed to evaluate $\Sigma^{(n)}_{T,\pm}(P_{m\alpha,3})$, listed in \eqref{nthorder}, for all values of $n$. The identities we need have been derived in detail in appendix \ref{app:omega_iden}; in the next few paragraphs we describe very briefly, the solution to the gap equations using them, leaving all details to appendix \ref{app:sol_gap_eq}. Let us first consider the part of the RHS of \eqref{nthorder} that is proportional to $M$. In order to evaluate this expression, we need to perform summations and integrals over the $n$ different flavours of dummy variables, with indices $1, \ldots, n$. Let us first focus on the summations/integrals over the variables with indices $n$ and $n-1$. The summation over these two variables involves exactly the expression given in \eqref{expione}, but with $P_{m \alpha, 3} \rightarrow (Q_{n-2})_{j_{n-2} \sigma_{n-2},3}$. The result of these summations/integrals is thus given by (see \eqref{simpexp}),
\begin{align}
    \dfrac{\beta^2}{2}  \Omega_T^2\left(M_T, (Q_{n-2})_{j_{n-2} \sigma_{n-2},3}\right).
\end{align}
We now turn to performing the summation/integral over the variables with index $n-2$. This computation is now easily performed using the identity (see \eqref{eq:omega_identity_1}),
\begin{align}\label{ident-maintext12}
    \dfrac{1}{\beta} \int\ddm{q}  \sum_{\substack{\sigma \\ l \in \mathbb{Z} + \frac{1}{2}}} \dfrac{1}{Q_{l \sigma}^2+M_T^2} \dfrac{1}{P_{m \alpha,3}-Q_{l \sigma,3}} \Omega_T(M_T,Q_{l \sigma,3})^{(n-1)} = \dfrac{1}{n}\Omega_T(M_T,P_{m \alpha,3})^{n},
\end{align}
derived in appendix \ref{app:omega_iden}. For the immediate case at hand we need this identity for $n=3$. It follows that the result of this summation/integral is given by,
\begin{align}
    \dfrac{\beta^3}{3!}  \Omega_T^3(M_T, (Q_{n-3})_{j_{n-3} \sigma_{n-3},3}).
\end{align}
We now iterate this procedure. The summation/integral over the variables with 
labels $n-4$ is once again evaluated using \eqref{ident-maintext12}, yielding,
\begin{align}
    \dfrac{\beta^4}{4!}  \Omega_T^4(M_T, (Q_{n-4})_{j_{n-4} \sigma_{n-4},3} ),
\end{align}
and so on. The final result is thus, 
\begin{align}
    \dfrac{\beta^n}{n!}  \Omega_T^n(M_T, P_{ m \alpha, 3}),
\end{align}
giving the first term on the RHS of \eqref{nthorderans}. 

The other terms in \eqref{nthorderans} are obtained in a similar manner. The summation/integral over the variables with indices $n$ and $n-1$ gives \eqref{itwomain} with $P_{m \alpha, 3} \rightarrow (Q_{n-2})_{j_{n-2} \sigma_{n-2},3} $. We now use the identity \eqref{ident-maintext12} to iterate the first term on the RHS of \eqref{itwomain}, and a new identity (see \eqref{eq:omega_identity_2}),
\begin{align}\label{eq:mainomega_identity_2}
    \dfrac{1}{\beta} \int\ddm{q}  \sum_{\substack{\sigma \\ l \in \mathbb{Z} + \frac{1}{2}}} \dfrac{Q_{l\sigma,3}}{Q_{l \sigma}^2+M_T^2} \dfrac{1}{P_{m \alpha,3}-Q_{l \sigma,3}} \Omega_T(M_T,Q_{l \sigma,3})^{(n-1)} = \dfrac{P_{m \alpha,3}}{n}\Omega_T(M_T,P_{m \alpha,3})^{n}.
\end{align}
to iterate the second term on the RHS of \eqref{itwomain}. The result of this procedure yields the last two terms on the RHS of \eqref{nthorderans}.

The identities \eqref{ident-maintext12} and \eqref{eq:mainomega_identity_2}
are themselves derived (in appendix \ref{app:omega_iden}) using the `symmetrization' procedure explained under \eqref{expione}. Consider, for instance, the identity \eqref{ident-maintext12}. In order to derive this formula, we first substitute with the definition of $\Omega_T(M_T,Q_{l \sigma,3})$ \eqref{eq:integral_omega_FT} on the LHS of \eqref{ident-maintext12}. The LHS of \eqref{ident-maintext12} thus has $n$ dummy summation/integration variables (one of these was explicit in \eqref{ident-maintext12} before the substitution while the remaining $n-1$ come from substituting with the definition of $\Omega_T(M_T,Q_{l \sigma,3})$). We then completely symmetrize over all $n$ sets of dummy variables. The expression was already symmetric in the last $n-1$ sets of indices, and so we obtain $1/n$ times a summation over $n$ terms. The summation that thus appears (on the LHS of \eqref{ident-maintext12}) takes the the form of the RHS of \eqref{eq:identity1}, and so equals the simple expression of the form given on the LHS of \eqref{eq:identity1}. It now follows that the summation/integral over the $n$ sets of variables separates into $n$ identical products of summations/integrals, yielding the RHS of \eqref{ident-maintext12}. Very similar manipulations, using \eqref{eq:identity2}, give \eqref{eq:mainomega_identity_2}.


\section{Free energy as a function of temperature}\label{sec:free_energy_FT}

In this section, we will plug our explicit results for the fermion self-energy \eqref{sumans} into our thermal effective action (see \eqref{eq:Final_thermal_action}),
\begin{equation}\label{eq:action_CS_m_FT_2}
    S_T= - V_2 \sum_{\substack{\alpha \\ n \in \mathbb{Z} + \frac{1}{2}}} \int \ddm{q} \Tr{\ln \left(  i \slashed{Q}_{n\alpha} + M + \Sigma_T(Q_{n\alpha,3}) \right) -\dfrac{1}{2} \Sigma_T(Q_{n\alpha,3}) \dfrac{1}{i \slashed{Q}_{n\alpha} + M + \Sigma_T(Q_{n\alpha,3})}},
\end{equation}
to find an explicit expression for $S_T$, the `holonomy-dependent thermal free energy'. 


\subsection{Dealing with UV divergences}\label{sec:UV_div}

As was explained in \cite{Giombi:2011kc} (and as we will explain below), the integral that defines $S_T$ is divergent. However, all divergences that appear in this expression turn out to be proportional to $\beta$. In other words, the divergences appear only in the vacuum energy, and can be removed by the subtraction of appropriate (temperature-independent) counter-terms. In order to avoid discussing these counter-terms explicitly, instead of $S_T$, we evaluate,
\begin{align}\label{eq:S_eff_S_T-S_0}
    S_{\rm eff} = S_T - S_0,
\end{align}
where,
\begin{equation}\label{snot}
    S_0= \beta \lim_{T' \to 0} T'\, S_{T'}.
\end{equation} 
Clearly, $S_T-S_0$ is the free energy of our system defined such that its vacuum has zero energy. This quantity, which contains all the physical information relevant to thermodynamics, will turn out to be completely UV-finite (as expected on physical grounds).  

For later use, we note (the obvious fact) that the summation over the Kaluza--Klein modes, which defines $S_0$ can be replaced by an integral (this is a consequence of the limit $T' \to 0$ in \eqref{snot}). As a consequence of this, the expression for $S_0$ is independent of holonomies.\footnote{This fact is easy to understand on physical grounds. In the limit $T'\to 0$, the thermal circle is decompactified, and we no longer have a gauge-invariant holonomy variable.} In equations, \eqref{snot} thus becomes,
\begin{align}\label{eq:S_0}
    S_0 &= \resizebox{1.05\hsize}{!}{$\displaystyle -V_2 \beta \lim \limits_{\beta' \to \infty} \dfrac{1}{\beta'} \sum_{\substack{\alpha \\ n \in \mathbb{Z} + \frac{1}{2}}} \int \ddm{q} \Tr{\ln \left(  i \slashed{Q}_{n\alpha} + M + \Sigma_T(Q_{n\alpha,3}) \right) -\dfrac{1}{2} \Sigma_T(Q_{n\alpha,3}) \dfrac{1}{i \slashed{Q}_{n\alpha} + M + \Sigma_T(Q_{n\alpha,3})}}$} \nonumber \\
     &= - V_2 \beta \sum_\alpha \int \dm{q} \Tr{\ln \left(  i \slashed{q} + M + \Sigma_0(q_3) \right) -\dfrac{1}{2} \Sigma_0(q_3) \dfrac{1}{i \slashed{q} + M + \Sigma_0(q_3)}} \nonumber\\
    &= - NV_2\beta \int \dm{q} \Tr{\ln \left(i \slashed{q} + M + \Sigma_0(q_{3}) \right) -\dfrac{1}{2} \Sigma_0(q_{3}) \dfrac{1}{i \slashed{q} + M + \Sigma_0(q_{3})}}.
\end{align}


\subsection{Explicit evaluation of the free energy}

Using \eqref{eq:gap_eq_constraint}, it is easy to convince oneself that (recall that the trace is over the Dirac matrices),
\begin{align} \label{detexp}
    \Tr\Big\{\ln \left(i\slashed{Q}_{n\alpha} + M + \Sigma_T(Q_{n\alpha,3}) \right)\Big\} = \ln{\left((Q_{n\alpha})^2+M_T^2\right)}.
\end{align}
Moreover, 
\begin{align}
    \resizebox{1.1\hsize}{!}{$\displaystyle\dfrac{1}{2}\Tr{\dfrac{\Sigma_T(Q_{n\alpha,3})}{i \slashed{Q}_{n\alpha} + M + \Sigma_T(Q_{n\alpha,3})}} = \dfrac{\left((Q_{n\alpha,3})^2+M_T^2\right) \left(1-\cos({\dfrac{4 \pi}{\kappa}\Omega_T(M_T,Q_{n\alpha,3})})\right)-\dfrac{4\pi}{\kappa}\Phi_T(M_T)\Sigma_{T,I}(Q_{n\alpha,3})}{(Q_{n\alpha})^2+M_T^2},$}
\end{align}
where we have used the mass gap equation (\ref{ge}), the decomposition of $\Sigma_T(P_{m\alpha,3})$ in a complete basis of $2\times 2$ matrices,
\begin{align}
    \Sigma_T(P_{m\alpha,3}) = i\slashed{\Sigma}_T(P_{m\alpha,3}) + (\Sigma_{T,I}(P_{m\alpha,3})-M)I,
\end{align}
and the identity (from \eqref{eq:gap_eq_constraint}),
\begin{align}
    2Q_{j\alpha,3}\Sigma_{T,3}(Q_{j\alpha,3})+\left(\Sigma_{T,3}(Q_{j\alpha,3})\right)^2+\left(\Sigma_{T,I}(Q_{j\alpha,3})\right)^2 = M_T^2.
\end{align}
It follows that the second integral on the RHS of \eqref{eq:action_CS_m_FT_2} takes the form,
\begin{align}\label{eq:S_T_2nd_term_intermed}
    &V_2 \int \ddm{q} \sum_{\substack{\alpha \\ n \in \mathbb{Z} + \frac{1}{2}}} \dfrac{1}{2}\Tr{\dfrac{\Sigma_T(Q_{n\alpha,3})}{i \slashed{Q}_{n\alpha} + M + \Sigma_T(Q_{n\alpha,3})}} \nonumber \\
    &= V_2 \int \ddm{q} \sum_{\substack{\alpha \\ n \in \mathbb{Z} + \frac{1}{2}}} \dfrac{\left((Q_{n\alpha,3})^2+M_T^2\right) \left(1-\cos({\dfrac{4 \pi}{\kappa}\Omega_T(M_T,Q_{n\alpha,3})})\right)-\dfrac{4\pi}{\kappa}\Phi_T(M_T)\Sigma_{T,I}(Q_{n\alpha,3})}{(Q_{n\alpha})^2+M_T^2} \nonumber \\
    &= I_1 + I_2,
\end{align}
where $I_1$ and $I_2$ are, respectively,
\begin{align}
     I_1 &= V_2 \int \ddm{q} \sum_{\substack{\alpha \\ n \in \mathbb{Z} + \frac{1}{2}}} \dfrac{\left((Q_{n\alpha,3})^2+M_T^2\right) \left(1-\cos({\dfrac{4 \pi}{\kappa}\Omega_T(M_T,Q_{n\alpha,3})})\right)}{(Q_{n\alpha})^2+M_T^2}, \\
     I_2 &= - \dfrac{4 \pi V_2 }{\kappa} \Phi_T(M_T) \int \ddm{q} \sum_{\substack{\alpha \\ n \in \mathbb{Z} + \frac{1}{2}}} \dfrac{\Sigma_{T,I}(Q_{n\alpha,3})}{(Q_{n\alpha})^2+M_T^2}.
\end{align}
We will evaluate $I_1$ and $I_2$ separately.

Lets first take a look at $I_1$:
\begin{align} \label{ione}
    I_1 &= V_2 \int \ddm{q} \sum_{\substack{\alpha \\ n \in \mathbb{Z} + \frac{1}{2}}} \dfrac{\left((Q_{n\alpha,3})^2+M_T^2\right) \left(1-\cos({\dfrac{4 \pi}{\kappa}\Omega_T(M_T,Q_{n\alpha,3})})\right)}{(Q_{n\alpha})^2+M_T^2} \nonumber \\
    &= V_2 \int \ddm{q} \sum_{\substack{\alpha \\ n \in \mathbb{Z} + \frac{1}{2}}} \sum_{m=1}^{\infty} \dfrac{(-1)^{m+1}}{(2m)!} \left( \dfrac{(Q_{n \alpha,3})^2}{(Q_{n \alpha})^2 + M_T^2}  \left( \dfrac{4 \pi \Omega_T(M_T,Q_{n \alpha,3})}{\kappa} \right)^{2 m}\right. \nonumber \\
    & \qquad\qquad\qquad\qquad\qquad\qquad\qquad\quad \left. + \dfrac{M_T^2}{(Q_{n \alpha})^2 + M_T^2} \left( \dfrac{4 \pi \Omega_T(M_T,Q_{n \alpha,3})}{\kappa} \right)^{2 m}  \right) \nonumber  \\
    &= V_2 \beta \left( \dfrac{ 4 \pi}{\kappa} \right)^2 \dfrac{\Phi_T(M_T)^3}{6}.
\end{align}
In going from the first to the second equality of \eqref{ione}, we have performed a Taylor series expansion of the cosine in the integrand on the RHS. In going from the second to the third equality of \eqref{ione}, we have used the identities,
\begin{align} \label{eq:omega_identity_4_int}
    \dfrac{1}{\beta} \int \ddm{q} \sum_{\substack{\alpha \\ m \in \mathbb{Z} + \frac{1}{2}}} \dfrac{Q_{m \alpha,3}^n}{Q_{m \alpha}^2 + M_T^2} \Omega_T(M_T, Q_{m \alpha,3})^j = \dfrac{1}{n+1}\begin{cases}
        \Phi_T(M_T)^{n+1}, & j=n \\
        0, & j>n
    \end{cases}\;,
\end{align}
and,
\begin{equation}\label{eq:omega_identity_3_int}
    \dfrac{1}{\beta} \int\ddm{q} \sum_{\substack{\alpha \\ m \in \mathbb{Z} + \frac{1}{2}}} \dfrac{1}{Q_{m \alpha}^2+M_T^2} \Omega_T(M_T,Q_{m \alpha,3})^n= 0, \quad \text{for  } n>0,
\end{equation}
which we have derived in appendix \ref{app:omega_iden} (see around \eqref{eq:omega_identity_4} and \eqref{eq:omega_identity_3}, respectively). 

Now lets evaluate $I_2$:
\begin{align} \label{itwo}
    I_2 &= - \dfrac{4 \pi V_2 }{\kappa} \Phi_T(M_T) \int \ddm{q} \sum_{\substack{\alpha \\ n \in \mathbb{Z} + \frac{1}{2}}} \dfrac{Q_{n\alpha,3} \sin({\dfrac{4 \pi}{\kappa}\Omega_T(M_T,Q_{n\alpha,3})})+ M_T\cos({\dfrac{4 \pi}{\kappa}\Omega_T(M_T,Q_{n\alpha,3})})}{(Q_{n\alpha})^2+M_T^2} \nonumber \\
    &= - \dfrac{4 \pi V_2 \beta }{\kappa} \Phi_T(M_T) \left( \dfrac{4 \pi \Phi_{T}(M_T)^2 }{2 \kappa} + M_T \Phi_T(M_T) \right),
\end{align}
where in going from the first to the second equality, we have once again Taylor-expanded the trigonometric functions in the integrand on the RHS of the first equality in \eqref{itwo}, and have used identities \eqref{eq:omega_identity_3_int} and \eqref{eq:omega_identity_4_int} together with the definition \eqref{eq:integral_phi_FT} of the function $\Phi_T(M_T).$

Note that the identities \eqref{eq:omega_identity_3_int} and \eqref{eq:omega_identity_4_int} have allowed us to find an explicit expression 
for the free energy, even though we do not have a similarly explicit expression for the function $\Omega_T(M_T,Q_{m \alpha,3})$ itself. 

Combining \eqref{detexp} with \eqref{eq:S_T_2nd_term_intermed} (through \eqref{ione} and \eqref{itwo}) in \eqref{eq:action_CS_m_FT_2}, we find,
\begin{align}
    S_T = -\dfrac{4\pi\beta V_2}{\kappa}\left(\Phi_T(M_T)\right)^2 \left(\dfrac{4\pi}{3\kappa}\Phi_T(M_T)+M_T\right) - V_2 \sum_{\substack{\alpha \\ n \in \mathbb{Z} + \frac{1}{2}}} \int \ddm{q} \ln{\left((Q_{n\alpha})^2+M_T^2\right)}.
\end{align}
Similarly, $S_0$ evaluates to,
\begin{align}
    S_0 = -\dfrac{4\pi\beta V_2}{\kappa}\left(\Phi_0(M_0)\right)^2 \left(\dfrac{4\pi}{3\kappa}\Phi_0(M_0)+M_0\right) - NV_2\beta \int \dm{q} \ln{\left(q^2+M_0^2\right)}.
\end{align}
Thus,
\begin{align}\label{eq:Free_E}
    S_T - S_0 =& -\dfrac{4\pi\beta V_2}{\kappa}\left(\Phi_T(M_T)\right)^2 \left(\dfrac{4\pi}{3\kappa}\Phi_T(M_T)+M_T\right) + \dfrac{4\pi\beta V_2}{\kappa}\left(\Phi_0(M_0)\right)^2 \left(\dfrac{4\pi}{3\kappa}\Phi_0(M_0)+M_0\right) \nonumber \\
    &\resizebox{1\hsize}{!}{$\displaystyle -V_2 \sum_{\substack{\alpha \\ n \in \mathbb{Z} + \frac{1}{2}}} \int \ddm{q} \ln{\left((Q_{n\alpha})^2+M_T^2\right)} + V_2 \beta N\int \dm{q} \ln{\left(q^2+M_T^2\right)} - NV_2\beta \int \dm{q} \ln{\left(\dfrac{q^2+M_T^2}{q^2+M_0^2}\right)}.$}
\end{align}
Recall that $\Phi_T(M_T)$ is a simple explicit function listed in \eqref{phit}, and so the first line of \eqref{eq:Free_E} is completely explicit. It remains to evaluate the second line of this expression.

Though each of the first two terms in the second line of (\ref{eq:Free_E}) suffers from a UV divergence, the sum of these terms converges (this follows because the summation in the first term is excellently approximated by an integral at large $N$). We demonstrate this in appendix \ref{app:contour}, namely,\footnote{\eqref{eq:Free_E_term34} essentially asserts the that the path integral expression for the free energy of a harmonic oscillator equals the Hamiltonian expression for the same quantity, and so is physically obvious (see below \eqref{hamilot}). In appendix \ref{app:contour}, we present a purely mathematical derivation (involving contour manipulations) of this identity.}
\begin{align}\label{eq:Free_E_term34}
    S_{\rm free} =&-V_2 \sum_{\substack{\alpha \\ n \in \mathbb{Z} + \frac{1}{2}}} \int \ddm{q} \ln{\left((Q_{n\alpha})^2+M_T^2\right)} + V_2 \beta N\int \dm{q} \ln{\left(q^2+M_T^2\right)} \nonumber \\
    =& -V_2 \sum_{\mu} \int \ddm{q} \ln\left\lvert \left( 1 + e^{ -\beta\sqrt{q^2 +M_T^2} + i  \lambda_{\mu}} \right) \left( 1 + e^{-\beta\sqrt{q^2 +M_T^2} - i \lambda_{\mu}} \right) \right\rvert \nonumber \\
    =& - \dfrac{V_2}{2 \pi \beta^2} \sum_{\mu} \int\limits_{\beta M_T}^{\infty} \dd{y} \, y \, \ln \left\lvert \left( 1 + e^{-y + i \lambda_{\mu}} \right) \left( 1 + e^{-y - i \lambda_{\mu}} \right) \right\rvert.
\end{align}
The last term in (\ref{eq:Free_E}) has a linear divergence, which disappears in the dimensional regularization scheme. In this scheme, the integral is easily evaluated, and we find,
\begin{align}\label{eq:Free_E_term5}
    - NV_2\beta \int \dfrac{d^{3-\epsilon} q}{(2 \pi)^{3 - \epsilon}} \ln{\left(\dfrac{q^2+M_T^2}{q^2+M_0^2}\right)} &= - NV_2 \beta M_T^{3} \int  \dfrac{d^{3-\epsilon} y}{(2 \pi)^{3 - \epsilon}} \ln{\left(\dfrac{y^2+1}{y^2+ \dfrac{M_0^2}{M_T^2}}\right)} \nonumber \\
    &= - N V_2 \beta M_T^3 \times \dfrac{1}{6 \pi} \left( \left( \dfrac{M_0}{M_T} \right)^3 - 1 \right) \nonumber \\
    &= \dfrac{N V_2}{6 \pi \beta^2} \left( (\beta M_T)^3 - (\beta M_0)^3 \right).
\end{align}
Substituting (\ref{eq:Free_E_term34}), (\ref{eq:Free_E_term5}), and the explicit functional forms of $\Phi_T(M_T)$ and $\Phi_0(M_0)$ (see \eqref{phit} and \eqref{phi_0}) into  (\ref{eq:Free_E}), we find,
\begin{align}\label{eq:Free_E2}
    S_T - S_0 &= - \dfrac{V_2}{2\pi \beta^2} \sum_{\mu} \int\limits_{\beta M_T}^{\infty} \dd{y} \, y \, \ln \left\lvert\left( 1 + e^{-y + i \lambda_{\mu} )} \right) \left( 1 + e^{-y - i \lambda_{\mu} )} \right) \right\rvert - \dfrac{V_2 N}{6 \pi \beta^2} \left( (\beta M_0)^3 - (\beta M_T)^3 \right) \nonumber \\
    &+ \dfrac{V_2 N\lambda}{4 \pi \beta^2} (\beta M_0)^3 \left(1 - \dfrac{\lambda}{3} \right)  - \dfrac{V_2 \lambda}{4\pi \beta^2 N} \left(\sum_{\mu} \ln\left|{e^{\beta M_T}(1 + e^{-\beta M_T + i  \lambda_{\mu}})(1 + e^{- \beta M_T - i \lambda_{\mu}})}\right| \right)^2  \nonumber \\
    & \times\left( \beta M_T - \dfrac{\lambda}{3N} \sum_{\mu} \ln\left|{e^{\beta M_T}(1 + e^{-\beta M_T + i  \lambda_{\mu}})(1 + e^{- \beta M_T - i \lambda_{\mu}})}\right| \right),
\end{align}
where $\lambda = \dfrac{N}{\kappa}$ is the 't Hooft coupling.

In terms of the eigenvalue density function $\rho(\alpha)$ of the holonomy matrix,
\begin{align}\label{eq:holonomy_density}
    \rho(\alpha) = \dfrac{1}{N} \sum_{\mu} \delta(\alpha - \lambda_{\mu}),
\end{align}
(\ref{eq:Free_E2}) can be rewritten as,
\begin{align} \label{fen}
    S_T - S_0 =& - \dfrac{V_2 N}{2\pi \beta^2} \int\dd{\alpha} \rho(\alpha) \int\limits_{\beta M_T}^{\infty} \dd{y} \, y \, \ln \left\lvert\left( 1 + e^{-y + i\alpha)} \right) \left( 1 + e^{-y - i\alpha)} \right) \right\rvert - \dfrac{V_2 N}{6 \pi \beta^2} \left( (\beta M_0)^3 - (\beta M_T)^3 \right) \nonumber \\
    &+ \dfrac{V_2 N\lambda}{4 \pi \beta^2} (\beta M_0)^3 \left(1 - \dfrac{\lambda}{3} \right)  - \dfrac{V_2 N \lambda}{4\pi \beta^2} \left(\int\dd{\alpha} \rho(\alpha) \ln\left|{e^{\beta M_T}(1 + e^{-\beta M_T + i\alpha})(1 + e^{- \beta M_T - i\alpha})}\right| \right)^2  \nonumber \\
    & \times\left( \beta M_T - \dfrac{\lambda}{3}\int\dd{\alpha} \rho(\alpha) \ln\left|{e^{\beta M_T}(1 + e^{-\beta M_T + i\alpha})(1 + e^{- \beta M_T - i\alpha})}\right| \right).
\end{align}
\eqref{fen} is our final result for the `on-shell' free energy, i.e. for the function $S_{\rm eff}(\{\lambda_\alpha\})$.


\subsection{Off-shell free energy density and comparison with results from the light-cone gauge}

It has been noticed in earlier light-cone gauge works that the free energy \eqref{fen} and the mass gap equation \eqref{ge} can be packaged into a single structure. One can define a, so-called, off-shell free energy density functional $F(c, \Tilde{\mathcal{C}})$ that has the following properties:\footnote{$F(c, \Tilde{\mathcal{C}})$ is also a function of the holonomies $\{\lambda_\alpha\}$, but that plays no role in this discussion, so we suppress the holonomy-dependence in our notation.}
\begin{itemize}
    \item The variation of $F(c, \Tilde{\mathcal{C}})$ with respect to $c$ and $\Tilde{\mathcal{C}}$ produces the mass gap equation \eqref{ge}.
    \item $V_2 \,\beta\, F(c, \Tilde{\mathcal{C}})$, evaluated on its extremum, equals the on-shell free energy \eqref{fen}.
\end{itemize}

It is easy to verify that an off-shell free energy density $F(c, \Tilde{\mathcal{C}})$ that satisfies these properties is given by, 
\begin{align}\label{eq:off-shell_free_energy}
    F(c,\Tilde{\mathcal{C}}) = \dfrac{N}{6 \pi \beta^3} & \Bigg( -8 \lambda^2 \Tilde{\mathcal{C}}^3 -3 \Tilde{\mathcal{C}} \left( c^2 - ( 2 \lambda \Tilde{\mathcal{C}} + \beta M )^2 \right) -6 \lambda \beta M \Tilde{\mathcal{C}}^2 + c^3 - (\beta M_0)^3 + \dfrac{3}{2} \lambda \left( 1- \dfrac{\lambda}{3} \right) (\beta M_0)^3 \nonumber \\
    &\qquad -3 \int \limits_{c}^{\infty} dy ~ y \int d\alpha ~ \rho(\alpha) \ln \left\lvert\left( 1 + e^{-y + i\alpha)} \right) \left( 1 + e^{-y - i\alpha)} \right) \right\rvert \Bigg).
\end{align}
Extremizing \eqref{eq:off-shell_free_energy} with respect to $c$ and $\Tilde{\mathcal{C}}$ gives (respectively),
\begin{align}
    \Tilde{\mathcal{C}} = \mathcal{C}(c),
\end{align}
and,
\begin{align}
    c^2 = \left( 2 \lambda \Tilde{\mathcal{C}} + \beta M \right)^2, \label{eq:c_gap}
\end{align}
where,
\begin{align}
     \mathcal{C}(c) = \dfrac{1}{2} \int d\alpha ~ \rho(\alpha)\, \ln \left\lvert e^c \left( 1 + e^{-c + i\alpha)} \right) \left( 1 + e^{-c - i\alpha)} \right) \right\rvert.
\end{align}
From \eqref{ge} and \eqref{eq:c_gap}, one can identify,
\begin{align}
    c= \beta M_T.
\end{align}
Thus, one can easily verify that,
\begin{align}
    F\left( \beta M_T, \mathcal{C}(\beta M_T) \right) = \dfrac{S_T -S_0}{\beta V_2}.
\end{align}
The off-shell free energy $ F(c,\Tilde{\mathcal{C}}) $ \eqref{eq:off-shell_free_energy} is in perfect agreement with the light-cone gauge result (e.g., see equation (3.7) of \cite{Minwalla:2020ysu})). $F(c, \Tilde{\mathcal{C}})$ matches with the previous results, modulo terms that are independent of both temperature and $\{\lambda_\alpha\}$. Such terms are unphysical as they can be absorbed into a renormalization of the vacuum energy of the theory (i.e. they can be absorbed into a `cosmological constant' counter-term). Hence, such terms have no nontrivial impact on the thermodynamics of the theory.

Throughout the main text of this paper, we have focused on the study of the simplest fermion theory, namely the regular fermion theory. The generalization of the results of this paper to the so-called critical fermion theory is straightforward (the interested reader will find details in appendix \ref{critical}).


\section{Discussion}\label{disc}

In this paper, we have set up the framework for the computation of the free energy of fermionic matter-Chern--Simons theories on $\Sigma_g \times S^1$ in the `temporal' gauge. As a check, we have used our formalism to compute the free energy on $\mathbb{R}_2 \times S^1$. The final result for this quantity turns out to be in perfect agreement with the results for the same quantity, previously obtained in the light-cone gauge, yielding a significant check of the framework developed in this paper. 

As mentioned in the introduction, our work is of interest for two reasons. First, it provides a check on the (numerous) previous computations performed in the light-cone gauge; such a check is of particular urgency as the implementation of the light-cone gauge involved a change of integration contour, a step that could involve subtleties. Second (and more importantly), it sets the stage for extremely interesting new computations that have never previously been performed (and would be extremely awkward to perform in light-cone gauge). 

In a work that is currently being written up \cite{minwalla202X:CSMonS2}, we have used the formalism developed in this paper to evaluate the free energy of the fermionic theories on $S^2 \times S^1$ in a regime of parameters in which $R$ (the radius of the $S^2$) and $\beta$ (the circumference of the $S^1$) have arbitrary sizes; such computation has never previously been performed. The main difference between this computation and the $\mathbb{R}^2 \times S^1$ computation presented in the current paper is the following. As explained in \ref{subsec:large_V_2} (see around \eqref{eq:flux-F12}), finite numbers of flux units yield zero local flux in an infinite volume space. In this limit, consequently $S_{\rm eff}$ (the effective action for holonomies and fluxes, obtained by integrating out the fermionic fields) was independent of the flux sector number $n_\alpha$. For this reason, the $n_\alpha$ dependence of the RHS of \eqref{finans} was extremely simple. This fact allowed us to sum over fluxes at the beginning of the computation, a procedure that led to precisely the same quantization of eigenvalues that one sees in pure Chern--Simons theory (\cite{Blau:1993tv, Minwalla:2022sef}) or in the Verlinde formula of the associated Wess–Zumino–Witten theory (see \cite{Minwalla:2022sef}). 

When working with a finite-sized sphere, on the other hand, the dependence on flux sectors turns out to be more involved \cite{minwalla202X:CSMonS2}. In this case, $S_{\rm eff}$ (and thus also the gap equation that determines the mass, sector by sector) is a nontrivial function of the fluxes. One is forced to solve the gap equation separately, sector by sector, and perform both the summation over $n_\alpha$ as well as the integral over holonomies, right at the end of the computation. In particular, the holonomies are no longer effectively quantized, and so the net effect of the combined operation of summing over fluxes and integrating over holonomies is no longer to impose the Wess–Zumino–Witten singlet condition (as was the case in \cite{Minwalla:2022sef}) but instead imposes an effective new constraint. 

Previous verifications of the Bose--Fermi duality of the partition function (on very large $S^2$, i.e. $S^2s$ of with radius of order $\sqrt{N}$) made crucial use of the quantization of the eigenvalues \cite{Aharony:2012ns, Jain:2013py}. The discussion above tells us that the verification of the Bose--Fermi duality of the partition function on spheres of finite (rather than very large) radii must proceed through a different mechanism, one that is likely to have similarities with the verification of the level-rank duality of the superconformal index \cite{Kim:2009wb, Imamura:2011su, Fujitsuka:2013fga, Aharony:2013dha, Dimofte:2011py, Hwang:2012jh} but in a non-supersymmetric context. It would be very interesting to study this point further\footnote{For the same reasons, it appears that the effective description of the Hilbert space structure of matter-Chern--Simons theories (see \cite{Minwalla:2022sef} will be modified away from the `Wess–Zumino–Witten singlet projection structure' found at large volume. It would be fascinating to obtain a structural understanding of the modified structure.} 

As mentioned above, we have already computed the free energy of the fermionic theory on $S^2 \times S^1$ \cite{minwalla202X:CSMonS2}. In order to investigate this quantity in the context of the Bose--Fermi duality, we need an independent computation of the free energy of the bosonic theory. While it seems very likely that this computation can also be carried out in the `temporal' gauge, implementation requires dealing with new technical subtleties related to the $\phi^2 A^2$ coupling. We hope to report on the bosonic gauge computation in `temporal' gauge in the future.

At the technical level, the analysis leading to the solution of the gap equation (see section \ref{sge} of this paper) is heavily inspired by the beautiful zero temperature analysis of \cite{Moshe:2014bja}. Apart from generalizing \cite{Moshe:2014bja} to finite temperature, the analysis presented in this paper also sheds light on a confusing point in \cite{Moshe:2014bja}, namely the treatment of IR singularities in the zero temperature, temporal gauge, gauge boson propagator. As we have discussed in detail in section \ref{subsubsec:zt_exclusion} and appendix \ref{zertemp}, this propagator is ill-defined at zero temporal momentum. In contrast, as explained in section \ref{sup}, the path integral at finite temperature is completely well-defined, without any ambiguities. If we first take the limit $N \to \infty$, and then take $T \to 0$, the unambiguous finite temperature results of this paper reduce to those of \cite{Moshe:2014bja} for the physical quantities that we have computed, namely the thermal mass and the thermal free energy, but not for the (likely unphysical) detailed form of the self-energy itself. In our opinion, it is still unclear whether the results of \cite{Moshe:2014bja} hold if we first take $T$ to zero, and then send $N$ to infinity (see section \ref{subsubsec:zt_exclusion} and appendix \ref{zertemp} for a more detailed discussion). We leave the resolution of this and other confusing questions relating to the zero temperature limit to future work.\footnote{A related confusion goes as follows: The symmetrization identities used in the paper \cite{Moshe:2014bja} are sensitive to the resolution of the singularity in the gauge boson propagator, but do not apply in the principal value prescription of appendix B.2 of \cite{Giombi:2011kc}. See appendix \ref{zertemp} for more details.}

Apart from the study of free energy, it is possible that the `temporal' gauge may prove useful in computing other quantities in the large-$N$ limit. It would, for instance, be interesting to see if this gauge could shed new light on the study of the unusual crossing symmetry properties of S-matrices in these theories \cite{Jain:2013py, Jain:2014nza, Minwalla:2022sef}, or perhaps to compute new correlation functions in these theories (see sections 6 and 8 of \cite{Moshe:2014bja} for useful preliminary results), or study them in the presence of a background flux (global symmetry) of a magnetic field \cite{Dey:2019ihe}. We leave all these questions to future work. 

We find the exact (finite-$N$) expression \eqref{exact} for the free energy of matter-Chern--Simons theories fascinating.\footnote{The RHS of \eqref{exact} is, quite remarkably,  completely local on the base manifold $\Sigma$, even though it is presented after integrating out the gauge field.  As far as we can tell, \eqref{exact} is exact, even at finite $N$.}  It would be interesting to find a world line representation of the integral over fermionic fields in \eqref{exact}. It would also be interesting to obtain the bosonic analogue of \eqref{exact} and perhaps recast the same in terms of world lines. Very optimistically, such manipulations could, conceivably, constitute the starting point for a systematic derivation of the Bose--Fermi duality at finite $N$.


\acknowledgments

We would like to thank S. Bhattacharyya, S. Jain, and V. Umesh for useful discussions (over a period of years) on the rederivation of the light-cone gauge results from the `temporal' gauge. NT and V would like to thank Diksha Jain, G. Mandal, O. Parrikar, and S.P. Trivedi for helpful discussions about our choice of gauge and concerns about our renormalization scheme. SN would like to thank D.P. Lorenzo for discussions regarding difficulties with diagrammatic evaluation of the gap equation, and Akashdeep Roy for providing ideas to compute multiple mathematical identities. We would also like to thank S. Jain, Chintan Patel, N. Prabhakar, S. Prakash, T. Sharma, and S. Wadia for their comments on the manuscript. The work of SM, SN, and NT was supported by the Infosys Endowment for the study of the Quantum Structure of Spacetime and the J C Bose Fellowship JCB/2019/000052. SN would like to thank GGI for their hospitality during the `LACES 2022' winter school. V would like to thank TIFR for their hospitality during this project. We would all also like to acknowledge our debt to the people of India for their steady support of the study of the basic sciences.


\appendix


\section{Notation adopted in this paper} \label{notation}

In this appendix, we describe the notation and conventions adopted in this paper. 

Throughout this paper, the spacetime indices are denoted by dotted Greek indices (like $\dot{\mu},\,\dot{\nu}$) that run from 1 to 3 (with 1 and 2 being the spatial directions and 3 being the temporal direction), unless otherwise specified. At times, we need to deal with purely spatial vectors, which we denote using an overhead arrow (for example, ${\vec p}\,$). 

In this paper, we are interested in computing the finite temperature partition function of a class of the theory defined by \eqref{eq:CSm_action}. For this purpose we evaluate the Euclidean partition function of our theory on $\mathbb{R}^2 \times S^1$. On this manifold, the field configurations we deal with, have continuous spatial momenta and discrete temporal momenta. Throughout this paper, we use the symbol $P_m$ for the 3-tuple of momenta,
\begin{align}\label{eq:mom_matsubara}
    P_{m} = \left(p_1, p_2, \dfrac{2 \pi m}{\beta}\right) = \left(\vec{p}, \dfrac{2 \pi m}{\beta}\right),
\end{align}
where $m$ is half-integral for fermions and integral for the gauge boson. We use Latin indices (like $m$ here) to parameterize the momentum of the Matsubara mode. Correspondingly, the zero-temperature counterpart of $P_m$ is simply denoted by $p$, i.e.,
\begin{align}
    p = (p_1,p_2,p_3).
\end{align}
In this paper, we evaluate the finite temperature partition function in the background of the (spatially constant) holonomy matrix \eqref{eq:holonomyfield}. From gauge-invariance, all our computations involve the eigenvalues of the operator $D_0$, rather than the ordinary derivative $\partial_0$. For this reason, we find it convenient to reserve the symbol $P_{m \alpha}$ to denote the eigenvectors of the covariantized momenta,
\begin{align} \label{gimom} 
    P_{m \alpha} = \left(p_1, p_2, \dfrac{2 \pi m - \lambda_{\alpha}}{\beta} \right) = \left(\vec{p}, \dfrac{2 \pi m  - \lambda_{\alpha}}{\beta}\right),
\end{align}
where $\lambda_\alpha/ \beta$ is the $\alpha^{\rm th}$ eigenvalue of the $A_3$ matrix. We use Greek indices (like $\alpha, \, \sigma, \, \mu$) for subscripts on the covariantized momenta to denote the eigenvalues of the $A_3$ matrix. At times, we have many such momenta, and to distinguish them we use additional (Latin) alphanumeric subscripts on the covariantized momenta enclosed within parenthesis. For example,
\begin{align}
    (Q_{i-2})_{j_{i-2} \sigma_{i-2}} = \left((q_{i-2})_1, (q_{i-2})_2, \dfrac{2 \pi j_{i-2} - \lambda_{\sigma_{i-2}}}{\beta} \right),
\end{align}
where $\left((q_{i-2})_1, (q_{i-2})_2 \right)$ are the components of the 2-vector $\vec{q}_{i-2}$.

In our gauge, the $A_3$ matrix is diagonal, so we use $A_3^{\alpha}$ to denote the diagonal $\alpha \alpha^\text{th}$ component of the $A_3$ field.

We often need to focus on the $3^{\text{rd}}$ component of the 3-vectors above. Hence, we denote these by the symbols $P_{m,3}$, $P_{m \alpha,3}$, and $(Q_{i-2})_{j_{i-2} \sigma_{i-2},3}$, respectively.

Finally, we use the following conventions for the quantities computed. Any quantity with a subscript $T$ is evaluated at temperature $T$, and consequently, any quantity computed at zero temperature is denoted by a subscript $0$. In this paper, $M$ denotes the bare mass of the fermions, $M_0$ the renormalized mass of the fermions at zero temperature, and $M_T$ the thermal mass of the fermions at temperature $T$.


\section{Rederivation of the gap equation as a Schwinger–Dyson equation}\label{schwinger}

In this appendix, we rederive `temporal' gauge gap equation using the Schwinger--Dyson techniques. This appendix closely follows the discussion of section 2.1.2 of \cite{Giombi:2011kc} (also see \cite{Wadia:1980rb}).

Starting with the the path integral \eqref{pfwcnn}, one can derive the Schwinger–Dyson equation for the fermion self-energy using, 
\begin{align}\label{eq:Schwinger-Dy_Eq}
    0 &=\int\dpi{\psi}\dpi{\bar{\psi}}\dfrac{\delta}{\delta \bar{\psi}^{\alpha}(-P_m)}\left(e^{-S_{\rm f}} \bar{\psi}^{\sigma}\left(P^{\prime}_{n}\right)\right) \nonumber\\
    &=\int\dpi{\psi}\dpi{\bar{\psi}}\left(\delta_{\alpha}^{\sigma} \delta^2\left(\vec{p} \, '+\vec{p}\right) \delta_{m,-n} -\dfrac{\delta S_{\rm f}}{\delta \bar{\psi}^{\alpha}(-P_m)} \bar{\psi}^{\sigma}\left(P^{\prime}_{n}\right)\right) e^{-S_{\rm f}},
\end{align}
where $S_{\rm f}$ is the action given in \eqref{exactnn}. Using the detailed form of $S_{\rm f}$ (\eqref{exactnn} written in the momentum basis), it follows that,
\begin{align} \label{vari}
   (2 \pi)^2 \, \beta \, \dfrac{\delta S_{\rm f}}{\delta \bar{\psi}^{\alpha}(-P_m)} =& (i \slashed{P}_{m \alpha} + M) \psi_{\alpha}(P_m) \nonumber \\  
   &\resizebox{0.85\hsize}{!}{$\displaystyle-\dfrac{2 \pi}{\kappa \beta^2} \int \ddm{q} \ddm{r} \sum_{\substack{ \sigma \\ l \in \mathbb{Z} \\ n \in \mathbb{Z} + \frac{1}{2} }}  \dfrac{1}{Q_{l,3} - \dfrac{\lambda_{\alpha}}{ \beta} + \dfrac{\lambda_{\sigma}}{ \beta } } \, H\left(\psi_{\sigma}(P_m-Q_l)\,\bar{\psi}^{\sigma}(-R_n)\right) \,\psi_{\alpha}(Q_l+R_n),$}
\end{align}
where the matrix-valued function $H(A)$ was defined in (\ref{Hdef}). Substituting \eqref{vari} back in (\ref{eq:Schwinger-Dy_Eq}), we obtain the  following relation involving the exact fermion propagator and the four-point functions,
\begin{align}
    &(2 \pi)^2 \, \beta \, \delta^{\sigma}_\alpha \delta^2\left(\vec{p} \, '+\vec{p}\right) \delta_{m,-n} = (i\slashed{P}_{m \alpha}+M)\expval{\psi_\alpha(P_m) \bar{\psi}^\sigma\left(P_n'\right)} \nonumber \\
    &\resizebox{1\hsize}{!}{$\displaystyle -\dfrac{2 \pi}{\kappa \beta^2} \int \ddm{q} \ddm{r} \sum_{\substack{ \rho \\ l \in \mathbb{Z} \\ j \in \mathbb{Z} + \frac{1}{2} }}  \dfrac{1}{Q_{l,3}- \dfrac{\lambda_\alpha}{\beta} + \dfrac{\lambda_\rho}{\beta} } \expval{H\left(\psi_{\rho}(P_m-Q_l)\,\bar{\psi}^{\rho}(-R_j)\right) \,\psi_{\alpha}(Q_l+R_j) \bar{\psi}^\sigma\left(P_n^{\prime}\right)}.$}
\end{align}
In the large-$N$ limit, this factorizes to yield,
\begin{align}
    &(2 \pi)^2 \, \beta \, \delta^{\sigma}_\alpha \delta^2\left(\vec{p} \, '+\vec{p}\right) \, \delta_{m,-n} = (i\slashed{P}_{m \alpha}+M)\expval{\psi_\alpha(P_m) \bar{\psi}^\sigma\left(P_n'\right)} \nonumber \\
    &\resizebox{1\hsize}{!}{$\displaystyle-\dfrac{2 \pi}{\kappa \beta^2} \int \ddm{q} \ddm{r} \sum_{\substack{ \rho \\ l \in \mathbb{Z} \\ j \in \mathbb{Z} + \frac{1}{2} }}  \dfrac{1}{Q_{l,3}- \dfrac{\lambda_\alpha}{\beta} + \dfrac{\lambda_\rho}{\beta} } \expval{H\left(\psi_{\rho}(P_m-Q_l)\,\bar{\psi}^{\rho}(-R_j)\right)} \, \expval{\psi_{\alpha}(Q_l+R_j) \bar{\psi}^\sigma\left(P_n^{\prime}\right)},$}
\end{align}
which upon substituting with the exact fermion propagator (\ref{eq:fermi_prop}), gives the gap equation \eqref{eq:gap_eq_FT},
\begin{equation}
    \Sigma_T(P_{m \alpha})= -\dfrac{2 \pi}{\kappa \beta} \int \ddm{q} \sum_{\substack{\sigma \\ n \in \mathbb{Z} + \frac{1}{2} \\ \sigma \neq \alpha \text{ at } n=m }} \dfrac{1}{P_{m \alpha,3}-Q_{n \sigma,3}} H\left(\dfrac{1}{i\slashed{Q}_{n \sigma}+M+\Sigma_T(Q_{n \sigma})}\right).
\end{equation}


\section{Illustration of integral identities: accounting for exclusions}
\label{careful}

In this appendix, we redo the computations presented in subsection \ref{naive}, this time carefully accounting for the exclusions in the summation ranges. Recall that the summation that appears in \eqref{ntwo} can be written, more explicitly, as,
\begin{align}
    \sum_{\substack{j_1,j_2 \in \mathbb{Z} + \frac{1}{2} \\ \sigma_1 \neq \alpha \text{ at }j_1=m \\\sigma_2 \neq \sigma_1 \text{ at }j_2=j_1} }.
\end{align}
Let $f(j_1,j_2,\sigma_1,\sigma_2)$ be any function of its arguments. Then,
\begin{align}
    \sum_{\substack{j_1,j_2 \in \mathbb{Z} + \frac{1}{2} \\ \sigma_1 \neq \alpha \text{ at }j_1=m \\\sigma_2 \neq \sigma_1 \text{ at }j_2=j_1} } f(j_1,j_2,\sigma_1,\sigma_2) &=  \sum_{\substack{j_1,j_2 \in \mathbb{Z} + \frac{1}{2} \\ \sigma_1, \sigma_2} } f(j_1,j_2,\sigma_1,\sigma_2) \Big( 1 - \delta_{\sigma_1,\alpha} \delta_{j_1,m} \Big) \Big( 1 - \delta_{\sigma_2,\sigma_1} \delta_{j_2,j_1} \Big) \nonumber\\
    &= \resizebox{0.75\hsize}{!}{$\displaystyle\sum_{\substack{j_1,j_2 \in \mathbb{Z} + \frac{1}{2}  \\ \sigma_1, \sigma_2} } f(j_1,j_2,\sigma_1,\sigma_2) \Bigg( \left( 1 -\delta_{\sigma_1,\alpha} \delta_{j_1,m} \right) \left( 1 -\delta_{\sigma_2,\sigma_1} \delta_{j_2,j_1} \right) \left( 1 -\delta_{\sigma_2,\alpha} \delta_{j_2,m} \right) $}\nonumber \\
    &\qquad \qquad \qquad + \delta_{\sigma_2,\alpha} \delta_{j_2,m} \left( 1 -\delta_{\sigma_1,\alpha} \delta_{j_1,m} \right) \left( 1 -\delta_{\sigma_2,\sigma_1} \delta_{j_2,j_1} \right)  \Bigg).
\end{align} 
We thus see that,
\begin{align} 
    \sum_{\substack{j_1,j_2 \in \mathbb{Z} + \frac{1}{2} \\ \sigma_1 \neq \alpha \text{ at }j_1=m \\\sigma_2 \neq \sigma_1 \text{ at }j_2=j_1} } f(j_1,j_2,\sigma_1,\sigma_2) = \sum_{\substack{j_1,j_2 \in \mathbb{Z} + \frac{1}{2} \\ \sigma_1 \neq \alpha \text{ at }j_1=m \\ \sigma_2 \neq \alpha \text{ at }j_2=m \\ \sigma_2 \neq \sigma_1 \text{ at }j_2=j_1 } } f(j_1,j_2,\sigma_1,\sigma_2) + \sum_{\substack{j_1 \in \mathbb{Z} + \frac{1}{2} \\ \sigma_1 \neq \alpha \text{ at }j_1=m } } f(j_1,m,\sigma_1,\alpha).  \label{eq:expanded_sum}
\end{align}
The first term on the RHS of \eqref{eq:expanded_sum} is a sum over $N^2$ terms. On the other hand, the second term on the RHS of \eqref{eq:expanded_sum} is a sum over $N$ terms. It follows that the second term is subleading compared to the first (by order $1/N$) in the large-$N$ limit. In analyzing \eqref{ntwo}, we will study only the first term here (however, we include the contribution from the second term in the final answer presented at the end of this appendix). 
   
Let us start with a more precise computation of $I_1$ (see \eqref{expione}). We have,
\begin{align} \label{expione1}
    \resizebox{1.1\hsize}{!}{$\displaystyle I_1 = \sum_{\substack{j_1,j_2 \in \mathbb{Z} + \frac{1}{2} \\ \sigma_1 \neq \alpha \text{ at }j_1=m \\ \sigma_2 \neq \alpha \text{ at }j_2=m \\ \sigma_2 \neq \sigma_1 \text{ at }j_2=j_1 }} \int\ddm{q_1} \ddm{q_2} \dfrac{1}{\left[ P_{m\alpha,3}-(Q_1)_{j_1\sigma_1,3} \right] \left[ (Q_1)_{j_1\sigma_1}^2 + M_T^2 \right]  \left[  (Q_{1})_{j_{1} \sigma_{1},3} - (Q_{2})_{j_{2} \sigma_{2},3} \right]  \left[ (Q_{2})_{j_{2} \sigma_{2}}^2 + M_T^2  \right] }.$}
\end{align}
In order to evaluate $I_1$ in \eqref{expione1}, we first obtain an alternate expression for $I_1$ by interchanging the dummy variables, and then add it to the original expression of $I_1$ with appropriate factors of one-half for both the expressions. This way we get,
\begin{align}
    I_1 &= 
    \begin{aligned}[t]
        \dfrac{1}{2} \sum_{\substack{j_1,j_2 \in \mathbb{Z} + \frac{1}{2} \\ \sigma_1 \neq \alpha \text{ at }j_1=m \\ \sigma_2 \neq \alpha \text{ at }j_2=m \\ \sigma_2 \neq \sigma_1 \text{ at }j_2=j_1 }}  \int \ddm{q_1} \ddm{q_2} \left( \dfrac{1}{P_{m\alpha,3}-(Q_1)_{j_1\sigma_1,3}} - \dfrac{1}{P_{m\alpha,3}-(Q_2)_{j_2\sigma_2,3}}  \right)&  \\
        \times \dfrac{1}{(Q_{1})_{j_{1} \sigma_{1},3} - (Q_{2})_{j_{2} \sigma_{2},3} } \, \dfrac{1}{(Q_1)_{j_1\sigma_1}^2 + M_T^2} \, \dfrac{1}{(Q_2)_{j_2\sigma_2}^2 + M_T^2} &
    \end{aligned} \nonumber \\
    &= 
    \begin{aligned}[t]
        \dfrac{1}{2} \sum_{\substack{j_1,j_2 \in \mathbb{Z} + \frac{1}{2} \\ \sigma_1 \neq \alpha \text{ at }j_1=m \\ \sigma_2 \neq \alpha \text{ at }j_2=m \\ \sigma_2 \neq \sigma_1 \text{ at }j_2=j_1 }}  \int \ddm{q_1} \ddm{q_2}  \dfrac{(Q_{1})_{j_{1} \sigma_{1},3} - (Q_{2})_{j_{2} \sigma_{2},3} }{ \left[P_{m\alpha,3}-(Q_1)_{j_1\sigma_1,3} \right] \left[ P_{m\alpha,3}-(Q_2)_{j_2\sigma_2,3} \right]} &\\
        \times \dfrac{1}{(Q_{1})_{j_{1} \sigma_{1},3} - (Q_{2})_{j_{2} \sigma_{2},3} } \, \dfrac{1}{(Q_1)_{j_1\sigma_1}^2 + M_T^2} \, \dfrac{1}{(Q_2)_{j_2\sigma_2}^2 + M_T^2}, &
    \end{aligned}
\end{align}
or,
\begin{align}
    I_1 &= 
    \begin{aligned}[t]
        \dfrac{ 1 }{2} \prod_{i=1}^2 \left( \,  \sum_{\substack{ j_i \in \mathbb{Z} + \frac{1}{2} \\ \sigma_{i} \neq \alpha \text{ at } j_i = m }} \int \ddm{q_i} \dfrac{1}{P_{m\alpha,3}-(Q_i)_{j_i\sigma_i,3}} \dfrac{1}{(Q_i)_{j_i\sigma_i}^2 + M_T^2}  \right) &\\
        - \dfrac{1}{2} \sum_{\substack{ j \in \mathbb{Z} + \frac{1}{2} \\ \sigma \neq \alpha \text{ at } j = m }} \left( \int \ddm{q} \dfrac{1}{\left[ P_{m\alpha,3}-Q_{j\sigma,3} \right]  \left[ Q_{j\sigma}^2 + M_T^2 \right]  } \right)^2 &
    \end{aligned}\nonumber \\
    &= 
    \begin{aligned}[t]
        &\dfrac{1}{2} \left( \,  \sum_{\substack{ j \in \mathbb{Z} + \frac{1}{2} \\ \sigma \neq \alpha \text{ at } j=m  }} \int \ddm{q} \dfrac{1}{P_{m\alpha,3}-Q_{j\sigma,3}} \dfrac{1}{Q_{j\sigma}^2 + M_T^2}   \right)^2 \\
        &- \dfrac{1}{2} \sum_{\substack{ j \in \mathbb{Z} + \frac{1}{2} \\ \sigma \neq \alpha \text{ at } j = m }} \left( \int \ddm{q} \dfrac{1}{\left[ P_{m\alpha,3}-Q_{j\sigma,3} \right]  \left[ Q_{j\sigma}^2 + M_T^2 \right]  } \right)^2
    \end{aligned}\nonumber\\
    &= \dfrac{\beta^2 \, \Big( \Omega_T(M_T,P_{m\alpha,3}) \Big)^2}{2} - \dfrac{1}{2} \sum_{\substack{ j \in \mathbb{Z} + \frac{1}{2} \\ \sigma \neq \alpha \text{ at } j = m }} \left( \int \ddm{q} \dfrac{1}{\left[ P_{m\alpha,3}-Q_{j\sigma,3} \right]  \left[ Q_{j\sigma}^2 + M_T^2 \right]  } \right)^2. \label{eq:N_suppressed_M1}
\end{align}
We can see that the first term in the last line of \eqref{eq:N_suppressed_M1} is of the order $N^2$, and the second term is of the order $N$.  

Now turning to the other integral,
\begin{align}
    I_2 &= \resizebox{1\hsize}{!}{$\displaystyle \sum_{ \substack{j_1,j_2 \in \mathbb{Z} + \frac{1}{2} \\ \sigma_1 \neq \alpha \text{ at }j_1=m \\ \sigma_2 \neq \alpha \text{ at }j_2=m \\ \sigma_2 \neq \sigma_1 \text{ at }j_2=j_1 } } \int \ddm{q_1} \ddm{q_2} \dfrac{ \pm i (Q_2)_{j_2 \sigma_2,3}}{ \left[ P_{m\alpha,3}-(Q_1)_{j_1\sigma_1,3} \right] \left[ (Q_1)_{j_1\sigma_1}^2 + M_T^2 \right] \left[ (Q_{1})_{j_{1} \sigma_{1},3} - (Q_{2})_{j_{2} \sigma_{2},3} \right] \left[ (Q_{2})_{j_{2} \sigma_{2}}^2 + M_T^2 \right]  } $} \nonumber\\
    &= \resizebox{1.05\hsize}{!}{$\displaystyle \sum_{ \substack{j_1,j_2 \in \mathbb{Z} + \frac{1}{2} \\ \sigma_1 \neq \alpha \text{ at }j_1=m \\ \sigma_2 \neq \alpha \text{ at }j_2=m \\ \sigma_2 \neq \sigma_1 \text{ at }j_2=j_1 } } \int \ddm{q_1} \ddm{q_2} \dfrac{ \pm i \left[ (Q_{2})_{j_{2} \sigma_{2},3} -(Q_{1})_{j_{1} \sigma_{1},3} + (Q_{1})_{j_{1} \sigma_{1},3} \right] }{ \left[ P_{m\alpha,3}-(Q_1)_{j_1\sigma_1,3} \right] \left[ (Q_1)_{j_1\sigma_1}^2 + M_T^2 \right] \left[ (Q_{1})_{j_{1} \sigma_{1},3} - (Q_{2})_{j_{2} \sigma_{2},3} \right] \left[ (Q_{2})_{j_{2} \sigma_{2}}^2 + M_T^2 \right]  } $} \nonumber\\
    &=
    \begin{aligned}[t]
        &\mp i \sum_{ \substack{j_1,j_2 \in \mathbb{Z} + \frac{1}{2} \\ \sigma_1 \neq \alpha \text{ at }j_1=m \\ \sigma_2 \neq \alpha \text{ at }j_2=m \\ \sigma_2 \neq \sigma_1 \text{ at }j_2=j_1 } } \int \ddm{q_1} \ddm{q_2} \dfrac{ 1 }{ \left[ P_{m\alpha,3}-(Q_1)_{j_1\sigma_1,3} \right] \left[ (Q_1)_{j_1\sigma_1}^2 + M_T^2 \right]  \left[ (Q_{2})_{j_{2} \sigma_{2}}^2 + M_T^2 \right]  }\\
        & \resizebox{1.05\hsize}{!}{$\displaystyle\pm i  \sum_{ \substack{j_1,j_2 \in \mathbb{Z} + \frac{1}{2} \\ \sigma_1 \neq \alpha \text{ at }j_1=m \\ \sigma_2 \neq \alpha \text{ at }j_2=m \\ \sigma_2 \neq \sigma_1 \text{ at }j_2=j_1 } } \int \ddm{q_1} \ddm{q_2} \dfrac{(Q_1)_{j_1 \sigma_1,3}}{ \left[ P_{m\alpha,3}-(Q_1)_{j_1\sigma_1,3} \right] \left[ (Q_1)_{j_1\sigma_1}^2 + M_T^2 \right] \left[ (Q_{1})_{j_{1} \sigma_{1},3} - (Q_{2})_{j_{2} \sigma_{2},3} \right] \left[ (Q_{2})_{j_{2} \sigma_{2}}^2 + M_T^2 \right]  },$}
    \end{aligned}
\end{align}
or,
\begin{align}\label{eq:N_suppressed_Q1}
    I_2 &= 
    \begin{aligned}[t]
        & \mp i  \sum_{ \substack{j_1 \in \mathbb{Z} + \frac{1}{2} \\ \sigma_1 \neq \alpha \text{ at }j_1=m } } \int\ddm{q_1} \dfrac{1}{P_{m\alpha,3}-(Q_1)_{j_1\sigma_1,3}} \dfrac{1}{(Q_1)_{j_1\sigma_1}^2 + M_T^2}    \int \ddm{q_{2}} \sum_{\substack{ j_{2} \in \mathbb{Z} + \frac{1}{2} \\ \sigma_2 } } \dfrac{1}{(Q_{2})_{j_{2} \sigma_{2}}^2 + M_T^2}\\
        & \pm i   \sum_{ \substack{j_1 \in \mathbb{Z} + \frac{1}{2} \\ \sigma_1 \neq \alpha \text{ at }j_1=m } } \int\ddm{q_1} \dfrac{1}{P_{m\alpha,3}-(Q_1)_{j_1\sigma_1,3}} \dfrac{1}{(Q_1)_{j_1\sigma_1}^2 + M_T^2}    \int \ddm{q_{2}} \dfrac{1}{(Q_{2})_{j_{1} \sigma_{1}}^2 + M_T^2}\\
        & \pm i  \sum_{ \substack{j \in \mathbb{Z} + \frac{1}{2} \\ \sigma \neq \alpha \text{ at }j=m } } \int\ddm{q_1} \dfrac{1}{P_{m\alpha,3}-(Q_1)_{j\sigma,3}} \dfrac{1}{(Q_1)_{j\sigma}^2 + M_T^2}    \int \ddm{q_{2}} \dfrac{1}{(Q_{2})_{m \alpha}^2 + M_T^2} \\
        & \resizebox{1.05\hsize}{!}{$\displaystyle\pm i   \sum_{ \substack{j_1,j_2 \in \mathbb{Z} + \frac{1}{2} \\ \sigma_1 \neq \alpha \text{ at }j_1=m \\ \sigma_2 \neq \alpha \text{ at }j_2=m \\ \sigma_2 \neq \sigma_1 \text{ at }j_2=j_1 } } \int \ddm{q_1} \ddm{q_2} \dfrac{(Q_1)_{j_1 \sigma_1,3}}{ \left[ P_{m\alpha,3}-(Q_1)_{j_1\sigma_1,3} \right] \left[ (Q_1)_{j_1\sigma_1}^2 + M_T^2 \right] \left[ (Q_{1})_{j_{1} \sigma_{1},3} - (Q_{2})_{j_{2} \sigma_{2},3} \right] \left[ (Q_{2})_{j_{2} \sigma_{2}}^2 + M_T^2 \right]  }  $}
    \end{aligned}\nonumber\\
    &=
    \begin{aligned}[t]
        &\beta^2 \, \Phi_T(M_T) \Big(\mp i \Omega_T(M_T,P_{m \alpha,3}) \Big) + \tilde{I}_2\\
        & \pm i  \sum_{ \substack{j_1 \in \mathbb{Z} + \frac{1}{2} \\ \sigma_1 \neq \alpha \text{ at }j_1=m } } \int\ddm{q_1} \dfrac{1}{P_{m\alpha,3}-(Q_1)_{j_1\sigma_1,3}} \dfrac{1}{(Q_1)_{j_1\sigma_1}^2 + M_T^2}    \int \ddm{q_{2}} \dfrac{1}{(Q_{2})_{j_{1} \sigma_{1}}^2 + M_T^2}\\
        & \pm i \sum_{ \substack{j \in \mathbb{Z} + \frac{1}{2} \\ \sigma \neq \alpha \text{ at }j=m } } \int\ddm{q_1} \dfrac{1}{P_{m\alpha,3}-(Q_1)_{j\sigma,3}} \dfrac{1}{(Q_1)_{j\sigma}^2 + M_T^2}    \int \ddm{q_{2}} \dfrac{1}{(Q_{2})_{m \alpha}^2 + M_T^2},
    \end{aligned}
\end{align}
where we have used the definitions of $\Phi_T(M_T)$ \eqref{eq:integral_phi_FT} and $\Omega_T(M_T, P_{m \alpha,3})$ \eqref{eq:integral_omega_FT} to obtain the first integral of \eqref{eq:N_suppressed_Q1}. To solve $\tilde{I}_2$, we will symmetrize the integrand of $\tilde{I}_2$ in $q_1$ and $q_2$, and the sum in $(\sigma_1, \, j_1)$ and $(\sigma_2, \, j_2)$. This way we get,
\begin{align}
    \tilde{I}_2 &=
    \begin{aligned}[t]
        \dfrac{\pm i}{2}  \sum_{ \substack{j_1,j_2 \in \mathbb{Z} + \frac{1}{2} \\ \sigma_1 \neq \alpha \text{ at }j_1=m \\ \sigma_2 \neq \alpha \text{ at }j_2=m \\ \sigma_2 \neq \sigma_1 \text{ at }j_2=j_1 } }  \int \ddm{q_1} \ddm{q_2} \left( \dfrac{(Q_{1})_{j_{1} \sigma_{1},3} }{P_{m\alpha,3}-(Q_1)_{j_1\sigma_1,3}} - \dfrac{(Q_{2})_{j_{2} \sigma_{2},3} }{P_{m\alpha,3}-(Q_2)_{j_2\sigma_2,3}}  \right)& \\
        \times \dfrac{1}{(Q_{1})_{j_{1} \sigma_{1},3} - (Q_{2})_{j_{2} \sigma_{2},3} } \, \dfrac{1}{(Q_1)_{j_1\sigma_1}^2 + M_T^2} \, \dfrac{1}{(Q_2)_{j_2\sigma_2}^2 + M_T^2} &
    \end{aligned}\nonumber\\
    &=
    \begin{aligned}[t]
        \dfrac{\pm i}{2}  \sum_{ \substack{j_1,j_2 \in \mathbb{Z} + \frac{1}{2} \\ \sigma_1 \neq \alpha \text{ at }j_1=m \\ \sigma_2 \neq \alpha \text{ at }j_2=m \\ \sigma_2 \neq \sigma_1 \text{ at }j_2=j_1 } }  \int \ddm{q_1} \ddm{q_2} \dfrac{ P_{m\alpha,3} \Big((Q_1)_{j_1\sigma_1,3} -(Q_{2})_{j_{2} \sigma_{2},3} \Big) }{\Big[ P_{m\alpha,3}-(Q_1)_{j_1\sigma_1,3} \Big] \Big[ P_{m\alpha,3}-(Q_2)_{j_2\sigma_2,3} \Big]} &\\
        \times \dfrac{1}{(Q_{1})_{j_{1} \sigma_{1},3} - (Q_{2})_{j_{2} \sigma_{2},3} } \, \dfrac{1}{(Q_1)_{j_1\sigma_1}^2 + M_T^2} \, \dfrac{1}{(Q_2)_{j_2\sigma_2}^2 + M_T^2} ,&
    \end{aligned}
\end{align}
or,
\begin{align} \label{eq:N_suppressed_Q2}
    \tilde{I}_2 &=
    \begin{aligned}[t]
        &\dfrac{\pm i P_{m\alpha,3}}{2} \prod_{i=1}^2 \left( \, \sum_{\substack{ j_i \in \mathbb{Z} + \frac{1}{2} \\ \sigma_{i} \neq \alpha \text{ at } j_{i} = m }}  \int \ddm{q_i}  \dfrac{1}{P_{m\alpha,3}-(Q_i)_{j_i\sigma_i,3}}  \dfrac{1}{(Q_i)_{j_i\sigma_i}^2 + M_T^2} \right)\\
        & +  \left(\dfrac{\mp i P_{m\alpha,3}}{2}\right) \sum_{\substack{ j \in \mathbb{Z} + \frac{1}{2} \\ \sigma \neq \alpha \text{ at } j = m }} \left(  \int \ddm{q}  \dfrac{1}{P_{m\alpha,3}-Q_{j\sigma,3}}  \dfrac{1}{(Q_{j\sigma})^2 + M_T^2} \right)^2
    \end{aligned}\nonumber\\
    &=
    \begin{aligned}[t]
        & \dfrac{\pm i P_{m\alpha,3}}{2}  \left( \, \sum_{\substack{\sigma \neq \alpha \\ j \in \mathbb{Z} + \frac{1}{2}}}  \int \ddm{q}  \dfrac{1}{P_{m\alpha,3}-Q_{j\sigma,3}}  \dfrac{1}{(Q_{j\sigma})^2 + M_T^2} \right)^2\\
        & +  \left(\dfrac{\mp i P_{m\alpha,3}}{2}\right) \sum_{\substack{ j \in \mathbb{Z} + \frac{1}{2} \\ \sigma \neq \alpha \text{ at } j = m }} \left(  \int \ddm{q}  \dfrac{1}{P_{m\alpha,3}-Q_{j\sigma,3}}  \dfrac{1}{(Q_{j\sigma})^2 + M_T^2} \right)^2 
    \end{aligned}\nonumber\\
    &=\dfrac{\pm i P_{m\alpha,3}}{2} \Big( \beta \, \Omega_T(M_T,P_{m \alpha,3}) \Big)^2 +  \left(\dfrac{\mp i P_{m\alpha,3}}{2}\right) \sum_{\substack{ j \in \mathbb{Z} + \frac{1}{2} \\ \sigma \neq \alpha \text{ at } j = m }} \left( \int \ddm{q}  \dfrac{1}{P_{m\alpha,3}-Q_{j\sigma,3}}  \dfrac{1}{(Q_{j\sigma})^2 + M_T^2} \right)^2.    
\end{align}
Substituting \eqref{eq:N_suppressed_Q2} in \eqref{eq:N_suppressed_Q1} we get,
\begin{align}\label{eq:N_suppressed_Q_final}
    I_2 &= \beta^2 \, \Phi_T(M_T) \Big( \mp i \Omega_T(M_T,P_{m \alpha,3}) \Big) \pm \dfrac{ i P_{m\alpha,3}}{2} \Big( \beta \, \Omega_T(M_T,P_{m \alpha,3}) \Big)^2 \nonumber \\
    & \;\;\;\; \pm i  \sum_{ \substack{j_1 \in \mathbb{Z} + \frac{1}{2} \\ \sigma_1 \neq \alpha \text{ at }j_1=m } } \int\ddm{q_1} \dfrac{1}{P_{m\alpha,3}-(Q_1)_{j_1\sigma_1,3}} \dfrac{1}{(Q_1)_{j_1\sigma_1}^2 + M_T^2}    \int \ddm{q_{2}} \dfrac{1}{(Q_{2})_{j_{1} \sigma_{1}}^2 + M_T^2} \nonumber \\
    & \;\;\;\; \pm i   \sum_{ \substack{j \in \mathbb{Z} + \frac{1}{2} \\ \sigma \neq \alpha \text{ at }j=m } } \int\ddm{q_1} \dfrac{1}{P_{m\alpha,3}-(Q_1)_{j\sigma,3}} \dfrac{1}{(Q_1)_{j\sigma}^2 + M_T^2}    \int \ddm{q_{2}} \dfrac{1}{(Q_{2})_{m \alpha}^2 + M_T^2} \nonumber \\
    & \;\;\;\; +  \left(\dfrac{\mp i P_{m\alpha,3}}{2}\right) \sum_{\substack{ j \in \mathbb{Z} + \frac{1}{2} \\ \sigma \neq \alpha \text{ at } j = m }} \left( \int \ddm{q}  \dfrac{1}{P_{m\alpha,3}-Q_{j\sigma,3}}  \dfrac{1}{(Q_{j\sigma})^2 + M_T^2} \right)^2.
\end{align}
Collecting both the integrals \eqref{eq:N_suppressed_M1} and \eqref{eq:N_suppressed_Q_final} in \eqref{signt}, we get,
\begin{align}\label{eq:exact_ntwo}
    \Sigma^{(2)}_{T,\pm}(P_{m\alpha,3}) &= 
    \begin{aligned}[t]
        &\resizebox{0.9\hsize}{!}{$\displaystyle\dfrac{M}{2} \big( \mp i \Omega_T(M_T,P_{m\alpha,3}) \big)^2 - \Phi_T(M_T) \big(\mp i \Omega_T(M_T,P_{m \alpha,3}) \big) \pm \dfrac{ i P_{m\alpha,3}}{2} \big( \mp i \Omega_T(M_T,P_{m \alpha,3}) \big)^2$} \\
        & - \dfrac{M}{2} \sum_{\substack{ j \in \mathbb{Z} + \frac{1}{2} \\ \sigma \neq \alpha \text{ at } j = m }} \left( \dfrac{\mp i}{\beta} \int \ddm{q} \dfrac{1}{\left[ P_{m\alpha,3}-Q_{j\sigma,3} \right]  \left[ Q_{j\sigma}^2 + M_T^2 \right]  } \right)^2 \\
        & \resizebox{0.9\hsize}{!}{$\displaystyle\pm i \left( \mp \dfrac{i}{\beta} \right)^2  \sum_{ \substack{j_1 \in \mathbb{Z} + \frac{1}{2} \\ \sigma_1 \neq \alpha \text{ at }j_1=m } } \int\ddm{q_1} \dfrac{1}{P_{m\alpha,3}-(Q_1)_{j_1\sigma_1,3}} \dfrac{1}{(Q_1)_{j_1\sigma_1}^2 + M_T^2} \int \ddm{q_{2}} \dfrac{1}{(Q_{2})_{j_{1} \sigma_{1}}^2 + M_T^2}$} \\
        & \pm i \left( \mp \dfrac{i}{\beta} \right)^2  \sum_{ \substack{j \in \mathbb{Z} + \frac{1}{2} \\ \sigma \neq \alpha \text{ at }j=m } } \int\ddm{q_1} \dfrac{1}{P_{m\alpha,3}-(Q_1)_{j\sigma,3}} \dfrac{1}{(Q_1)_{j\sigma}^2 + M_T^2}    \int \ddm{q_{2}} \dfrac{1}{(Q_{2})_{m \alpha}^2 + M_T^2} \\
        & \mp  \dfrac{ i P_{m\alpha,3}}{2} \sum_{\substack{ j \in \mathbb{Z} + \frac{1}{2} \\ \sigma \neq \alpha \text{ at } j = m }} \left( \dfrac{\mp i}{\beta}  \int \ddm{q}  \dfrac{1}{P_{m\alpha,3}-Q_{j\sigma,3}}  \dfrac{1}{(Q_{j\sigma})^2 + M_T^2} \right)^2 \\
        & \resizebox{0.95\hsize}{!}{$\displaystyle\dfrac{\mp i}{\beta^2}  \sum_{ \substack{j_1 \in \mathbb{Z} + \frac{1}{2} \\ \sigma_1 \neq \alpha \text{ at }j_1=m } }  \int\ddm{q_1}\ddm{q_2} \dfrac{(Q_2)_{m \alpha,3} \mp i M }{\left[P_{m\alpha,3}-(Q_1)_{j_1\sigma_1,3}\right]  \left[(Q_1)^2_{j_1\sigma_1} + M_T^2 \right]  \left[(Q_{1})_{j_{1} \sigma_{1},3} - (Q_{2})_{m \alpha,3} \right] \left[ (Q_{2})_{m \alpha }^2 + M_T^2\right]}.$}
    \end{aligned}
\end{align}
In \eqref{eq:exact_ntwo}, only the terms in the first line are of the order $N^2$ while all the other terms are of the order $N$. So we see that the leading order result of $\Sigma^{(2)}_{T,\pm}(P_{m\alpha,3})$ matches with the naive calculation \eqref{eq:sigma_pm_naive_result} performed in section \ref{naive}. This feature is shared by all the other $\Sigma^{(n)}_{T,\pm}(P_{m\alpha,3})$, and each of the identities that we will derive in appendix \ref{app:omega_iden}. So in this paper, we get away with doing just the naive calculations everywhere without worrying about this subtlety appearing in the sum.


\section{Relevant algebraic identities}\label{app:omega_iden}

In this appendix, we will derive the integral identities used in sections \ref{sge} and \ref{sec:free_energy_FT}.

Consider the algebraic identity,
\begin{align} \label{eq:identitygen}
    \dfrac{x^m}{(x-a_1)(x-a_2)\cdots(x-a_n)} &= \sum_{i=1}^n \dfrac{a_i^m}{x-a_i} \prod_{j\neq i} \dfrac{1}{a_i-a_j}, \quad\text{for   }\,\, m = 0,1,2,\ldots, n-1.
\end{align}
\eqref{eq:identitygen} may be understood as follows. The expression on the LHS of this equation has a simple pole at $x=a_i$, for $i=1, \ldots, n$. The residue of the pole at $x=a_i$ is given by $\displaystyle a_i^m \prod_{j\neq i} \dfrac{1}{a_i-a_j}$.  It follows that the LHS$-$RHS is a function that is analytic everywhere. For $m <n$, this analytic function also vanishes in the limit $x \to \infty$. However, the only function with both these properties is the function that is zero everywhere. It follows that the LHS must equal the RHS.

In a similar manner, it is easy to convince oneself that,
\begin{align} \label{eq:identitygen_m=n}
    \dfrac{x^n}{(x-a_1)(x-a_2)\cdots(x-a_n)} &= 1 + \sum_{i=1}^n \dfrac{a_i^n}{x-a_i} \prod_{j\neq i} \dfrac{1}{a_i-a_j}.
\end{align}
Similar derivation as above continues to work, but this time the LHS$-$RHS is an everywhere-analytic function that reduces to unity at infinity, and the only function with both these properties is the function $f(z)=1$.

We will make use of the following special cases of \eqref{eq:identitygen} and \eqref{eq:identitygen_m=n},
\begin{align}
    \dfrac{1}{(x-a_1)(x-a_2)\cdots(x-a_n)} &= \sum_{i=1}^n \dfrac{1}{x-a_i} \prod_{j\neq i} \dfrac{1}{a_i-a_j},\label{eq:identity1}\\
    \dfrac{x}{(x-a_1)(x-a_2)\cdots(x-a_n)} &= \sum_{i=1}^n \dfrac{a_i}{x-a_i} \prod_{j\neq i} \dfrac{1}{a_i-a_j},\label{eq:identity2}\\
    \sum\limits_{i=1}^{n} a_i^m \prod\limits_{\substack{j=1\\ j\neq i}}^{n}\dfrac{1}{a_i-a_j} &= \begin{cases} 1, & n=m+1\\ 0, & n>m+1
    \end{cases}\,.\label{eq:identity3}
\end{align}
The first and second of these follow from \eqref{eq:identitygen} upon setting $m=0$ and $m=1$, respectively. The last follows upon setting $x=0$ in \eqref{eq:identitygen} and \eqref{eq:identitygen_m=n}. 


\subsection{Integral formulae for \texorpdfstring{$\Omega_T(M_T,P_{m\alpha,3})$}{Ω\_T(M\_T,P\_{mα,3})}}\label{subsec:omega_identity}

We look to evaluate,
\begin{align}\label{eq:omega_proof_1}
    &\dfrac{1}{\beta} \int\ddm{q}  \sum_{\substack{\sigma \\ l \in \mathbb{Z} + \frac{1}{2}}} \dfrac{1}{Q_{l \sigma}^2+M_T^2} \dfrac{1}{P_{m \alpha,3}-Q_{l \sigma,3}} \Omega_T(M_T,Q_{l \sigma,3})^{(n-1)} \nonumber \\
    &\resizebox{1.05\hsize}{!}{$\displaystyle = \dfrac{1}{\beta} \int\ddm{r_1} \sum_{\substack{\sigma_1 \\ l_1 \in \mathbb{Z} + \frac{1}{2}}} \dfrac{1}{(R_1)_{l_1 \sigma_1}^2+M_T^2} \dfrac{1}{P_{m \alpha,3}-(R_1)_{l_1 \sigma_1,3}} \prod_{i=2}^n \left( \dfrac{1}{\beta} \ddm{r_i} \sum_{\substack{\sigma_i \\ l_i \in \mathbb{Z} + \frac{1}{2}}} \dfrac{1}{(R_i)_{l_i \sigma_i}^2+M_T^2} \, \dfrac{1}{(R_1)_{l_1 \sigma_1,3}-(R_i)_{l_i \sigma_i,3}}\right).$}
\end{align}
The integrand in \eqref{eq:omega_proof_1} can be written as a product of two terms. The first of these terms, 
\begin{align}
    \prod_{i=1}^n \left( \dfrac{1}{\beta}  \dfrac{1}{(R_i)_{l_i \sigma_i}^2+M_T^2} \right),
\end{align}
is completely symmetric in the $n$ integration (and summation) variables. However the second of these factors, 
\begin{align}
    \dfrac{1}{P_{m \alpha,3}-(R_1)_{l_1 \sigma_1,3}} \prod_{i=2}^n \left(\dfrac{1}{(R_1)_{l_1 \sigma_1,3}-(R_i)_{l_i \sigma_i,3}}\right),
\end{align}
is invariant only under permutations of the variables with labels $2,\ldots,n$ (the dummy variables with the label 1 are special). This lack of manifest symmetry can be cured as follows. Of course, our integral remains unchanged if we interchange the indices $1 \leftrightarrow i$ (where $i= 2, \ldots, n$) in the integrand above. Adding the $n$ integrals that we obtain in this manner (the one original integral and the $n-1$ new integrals that we get from the dummy variable interchanges above) gives us $n$ times the original integral. The advantage of this manipulation is that our new expression for the integral \ref{eq:omega_proof_1} is manifestly symmetric in all dummy variables. Dividing this expression (obtained after summing those $n$ integrals) by $n$, the integral in \eqref{eq:omega_proof_1} may be rewritten as,

\begin{align}\label{eq:omega_proof_1_int}
 \dfrac{1}{\beta} &\int\ddm{q}  \sum_{\substack{\sigma \\ l \in \mathbb{Z} + \frac{1}{2}}} \dfrac{1}{Q_{l \sigma}^2+M_T^2} \dfrac{1}{P_{m \alpha,3}-Q_{l \sigma,3}} \Omega_T(M_T,Q_{l \sigma,3})^{(n-1)} \nonumber \\
 &=\dfrac{1}{n}\int\prod_{i=1}^n \left( \dfrac{1}{\beta} \ddm{r_i} \sum_{\substack{\sigma_i \\ l_i \in \mathbb{Z} + \frac{1}{2}}} \dfrac{1}{(R_i)_{l_i \sigma_i}^2+M_T^2} \right) f\left(P_{m \alpha,3},(R_1)_{l_1 \sigma_1,3},\dots,(R_n)_{l_n \sigma_n,3}\right),
\end{align}
where the function $f$ is given by,
\begin{align}
    f\left(x,a_1,\dots,a_n\right) =& \dfrac{1}{(x-a_1)(a_1-a_2)\cdots(a_1-a_n)} + \dfrac{1}{(x-a_2)(a_2-a_1)\cdots(a_2-a_n)} +\cdots \nonumber\\
    &+ \dfrac{1}{(x-a_n)(a_n-a_1)\cdots(a_n-a_{n-1})}. \label{eq:f_partial_frac_1} 
\end{align}
Using the identity in (\ref{eq:identity1}), (\ref{eq:f_partial_frac_1}) simply becomes,
\begin{align}
    f\left(x,a_1,\dots,a_n\right) = \dfrac{1}{(x-a_1)\cdots(x-a_n)}. \label{eq:f_partial_frac_2}
\end{align}
Substituting (\ref{eq:f_partial_frac_2}) in (\ref{eq:omega_proof_1_int}), we obtain,
\begin{align}\label{eq:omega_identity_1}
    &\dfrac{1}{\beta} \int\ddm{q}  \sum_{\substack{\sigma \\ l \in \mathbb{Z} + \frac{1}{2}}} \dfrac{1}{Q_{l \sigma}^2+M_T^2} \dfrac{1}{P_{m \alpha,3}-Q_{l \sigma,3}} \Omega_T(M_T,Q_{l \sigma,3})^{(n-1)} \nonumber\\
    &= \dfrac{1}{n} \int\prod_{i=1}^n \left( \dfrac{1}{\beta} \ddm{r_i} \sum_{\substack{\sigma_i \\ l_i \in \mathbb{Z} + \frac{1}{2}}} \dfrac{1}{(R_i)_{l_i \sigma_i}^2+M_T^2} \dfrac{1}{P_{m \alpha,3}-(R_i)_{l_i \sigma_i,3}} \right) \nonumber\\
    &= \dfrac{1}{n}\Bigg( \dfrac{1}{\beta} \int\ddm{r} \sum_{\substack{ \sigma \\ l \in \mathbb{Z} + \frac{1}{2}}} \dfrac{1}{R_{l \sigma}^2+M_T^2}\dfrac{1}{P_{m \alpha,3}-R_{l \sigma,3}}\Bigg)^n  \nonumber\\
    &= \dfrac{1}{n}\Omega_T(M_T,P_{m \alpha,3})^{n}.
\end{align}
We have thus established the identity \eqref{ident-maintext12}.

Similarly,
\begin{align}\label{eq:omega_proof_2}
    &\dfrac{1}{\beta} \int\ddm{q}  \sum_{\substack{\sigma \\ l \in \mathbb{Z} + \frac{1}{2}}} \dfrac{Q_{l\sigma,3}}{Q_{l \sigma}^2+M_T^2} \dfrac{1}{P_{m \alpha,3}-Q_{l \sigma,3}} \Omega_T(M_T,Q_{l \sigma,3})^{(n-1)} \nonumber \\
    &= \text{\small $\dfrac{1}{\beta} \int\ddm{r_1} \sum_{\substack{\sigma_1 \\ l_1 \in \mathbb{Z} + \frac{1}{2}}} \dfrac{(R_1)_{l_1 \sigma_1,3}}{(R_1)_{l_1 \sigma_1}^2+M_T^2} \dfrac{1}{P_{m \alpha,3}-(R_1)_{l_1 \sigma_1,3}} \prod_{i=2}^n \left( \dfrac{1}{\beta} \ddm{r_i} \sum_{\substack{\sigma_i \\ l_i \in \mathbb{Z} + \frac{1}{2}}} \dfrac{1}{(R_i)_{l_i \sigma_i}^2+M_T^2} \dfrac{1}{((R_1)_{l_1 \sigma_1,3}-(R_i)_{l_i \sigma_i,3})}\right)$ } \nonumber\\
    &= \dfrac{1}{n}\int\prod_{i=1}^n \left( \dfrac{1}{\beta} \ddm{r_i} \sum_{\substack{\sigma_i \\ l_i \in \mathbb{Z} + \frac{1}{2}}} \dfrac{1}{(R_i)_{l_i \sigma_i}^2+M_T^2} \right) g\left(P_{m \alpha,3},(R_1)_{l_1 \sigma_1,3},\dots,(R_n)_{l_n \sigma_n,3}\right),
\end{align}
where we symmetrize the integral with respect to the integration variables $r_1,\dots,r_n$ and the sum with respect to the summation variables $l_1, \dots, l_n$ and $\sigma_1, \dots, \sigma_n$, in going from the second line in (\ref{eq:omega_proof_2}) to the third line, and the function $g$ is given by,
\begin{align}\label{eq:g_partial_frac_1}
    g\left(x,a_1,\dots,a_n\right) =& \dfrac{a_1}{(x-a_1)(a_1-a_2)\cdots(a_1-a_n)} + \dfrac{a_2}{(x-a_2)(a_2-a_1)\cdots(a_2-a_n)} +\cdots \nonumber\\
    &+ \dfrac{a_n}{(x-a_n)(a_n-a_1)\cdots(a_n-a_{n-1})}.
\end{align}
Using the identity in (\ref{eq:identity2}), (\ref{eq:g_partial_frac_1}) simply becomes,
\begin{align}
    g\left(x,a_1,\dots,a_n\right) = \dfrac{x}{(x-a_1)\cdots(x-a_n)}. \label{eq:g_partial_frac_2}
\end{align}
Substituting (\ref{eq:g_partial_frac_2}) in (\ref{eq:omega_proof_2}), we obtain,
\begin{align}\label{eq:omega_identity_2}
    &\dfrac{1}{\beta} \int\ddm{q}  \sum_{\substack{\sigma \\ l \in \mathbb{Z} + \frac{1}{2}}} \dfrac{Q_{l\sigma,3}}{Q_{l \sigma}^2+M_T^2} \dfrac{1}{P_{m \alpha,3}-Q_{l \sigma,3}} \Omega_T(M_T,Q_{l \sigma,3})^{(n-1)} \nonumber \\
    &= \dfrac{P_{m \alpha,3}}{n}\int\prod_{i=1}^n \left( \dfrac{1}{\beta} \ddm{r_i} \sum_{\substack{\sigma_i \\ l_i \in \mathbb{Z} + \frac{1}{2}}} \dfrac{1}{(R_i)_{l_i \sigma_i}^2+M_T^2} \dfrac{1}{P_{m \alpha,3}-(R_i)_{l_i \sigma_i,3}} \right) \nonumber\\
    &= \dfrac{P_{m \alpha,3}}{n}\Bigg( \dfrac{1}{\beta} \int\ddm{r} \sum_{\substack{ \sigma \\ l \in \mathbb{Z} + \frac{1}{2}}} \dfrac{1}{R_{l \sigma}^2+M_T^2}\dfrac{1}{P_{m \alpha,3}-R_{l \sigma,3}}\Bigg)^n \nonumber\\
    &= \dfrac{P_{m \alpha,3}}{n}\Omega_T(M_T,P_{m \alpha,3})^{n}.
\end{align}
Using (\ref{eq:omega_identity_1}) and (\ref{eq:omega_identity_2}), we obtain for $n>0$,
\begin{equation}\label{eq:omega_identity}
    \dfrac{1}{\beta} \int\ddm{q} \sum_{\substack{\sigma \\ l \in \mathbb{Z} + \frac{1}{2}}} \dfrac{M_T\pm iQ_{l \sigma,3}}{Q_{l \sigma}^2+M_T^2} \dfrac{1}{P_{m \alpha,3}-Q_{l \sigma,3}} \Omega_T(M_T,Q_{l \sigma,3})^n= \dfrac{1}{n+1}(M_T\pm iP_{m \alpha,3})\Omega_T(M_T,P_{m \alpha,3})^{n+1}.
\end{equation}
By adding and subtracting $P_{m \alpha,3}$ in the numerator on the LHS of \eqref{eq:omega_identity_2}, dividing the resulting integral into two terms, and using \eqref{eq:omega_identity_1}, it is easy to verify that,
\begin{equation}\label{eq:omega_identity_3}
    \dfrac{1}{\beta} \int\ddm{q} \sum_{\substack{\alpha \\ m \in \mathbb{Z} + \frac{1}{2}}} \dfrac{1}{Q_{m \alpha}^2+M_T^2} \Omega_T(M_T,Q_{m \alpha,3})^n= 0, \qquad \text{for} \quad n>0.
\end{equation}
We will require one final identity to evaluate the free energy. In order to obtain it, let us evaluate for $n \leq j$,
\begin{align}\label{eq:omega_proof_4}
    &\dfrac{1}{\beta} \int \ddm{q} \sum_{\substack{\alpha \\ m \in \mathbb{Z} + \frac{1}{2}}} \dfrac{Q_{m \alpha,3}^n}{Q_{m \alpha}^2 + M_T^2} \Omega_T(M_T, Q_{m \alpha,3})^j \nonumber \\
    &= \dfrac{1}{\beta} \int \ddm{q_1} \sum_{\substack{\alpha_1 \\ m_1 \in \mathbb{Z} + \frac{1}{2}}} \dfrac{(Q_1)_{m_1 \alpha_1,3}^n}{(Q_1)_{m_1 \alpha_1}^2 + M_T^2} \prod_{i=2}^{j+1} \left(  \dfrac{1}{\beta} \ddm{q_i} \sum_{\substack{\alpha_i \\ m_i \in \mathbb{Z} + \frac{1}{2}}} \dfrac{1}{(Q_i)_{m_i \alpha_i}^2+M_T^2} \dfrac{1}{(Q_1)_{m_1 \alpha_1,3}-(Q_i)_{m_i \alpha_i,3})} \right) \nonumber \\
    &= \dfrac{1}{j+1} \int \prod_{i=1}^{j+1}\left( \dfrac{1}{\beta} \ddm{q_i} \sum_{\substack{\alpha_i \\ m_i \in \mathbb{Z} + \frac{1}{2}}} \dfrac{1}{(Q_i)_{m_i \alpha_i}^2 + M_T^2}  \right) h\left((Q_1)_{m_1 \alpha_1,3}, \dots , (Q_{j+1})_{m_{j+1} \alpha_{j+1},3}\right),
\end{align}
where we have symmetrized the integral with respect to the integration variables $q_1,\dots,q_{j+1}$ and sum with respect to the summation variables $m_1, \dots, m_{j+1}$ and $\alpha_1, \dots, \alpha_{j+1}$ in going from the second line in (\ref{eq:omega_proof_4}) to the third line, and the function $h$ is given by,
\begin{align}
    h(a_1, \dots, a_{j+1}) &= \dfrac{a_1^n}{(a_1 - a_2)(a_1-a_3) \cdots (a_1 - a_{j+1})} + \dfrac{a_2^n}{(a_2 - a_1)(a_2-a_3) \cdots (a_2 - a_{j+1})} + \dots \nonumber \\
    &+ \dfrac{a_{j+1}^n}{(a_{j+1} - a_1)(a_{j+1}-a_2) \cdots (a_{j+1} - a_{j})} \label{eq:h_partial_frac_1}
\end{align}
Using the identity in (\ref{eq:identity3}), (\ref{eq:h_partial_frac_1}) becomes,
\begin{align}
    h(a_1, \dots, a_{j+1}) = \begin{cases} \label{eq:h_partial_frac_2}
        1, & j=n \\
        0, & j>n
    \end{cases}\;.
\end{align}
We thus conclude that the integral in \eqref{eq:omega_proof_4} vanishes for  
$j>n$.

On the other hand, when $j=n$, by substituting (\ref{eq:h_partial_frac_2}) in (\ref{eq:omega_proof_4}), it follows that,
\begin{align}
    \dfrac{1}{\beta} \int \ddm{q} \sum_{\substack{\alpha \\ m \in \mathbb{Z} + \frac{1}{2}}} \dfrac{Q_{m \alpha,3}^n}{Q_{m \alpha}^2 + M_T^2} \Omega_T(M_T, Q_{m \alpha,3})^n  &= \dfrac{1}{n+1} \int \prod_{i=1}^{n+1}\left( \dfrac{1}{\beta} \ddm{q_i} \sum_{\substack{\alpha_i \\ m_i \in \mathbb{Z} + \frac{1}{2}}} \dfrac{1}{(Q_i)_{m_i \alpha_i}^2 + M_T^2}  \right) \nonumber \\
    &= \dfrac{1}{n+1} \left( \dfrac{1}{\beta} \int  \ddm{q} \sum_{\substack{\alpha \\ m \in \mathbb{Z} + \frac{1}{2}}} \dfrac{1}{Q_{ m \alpha}^2 + M_T^2}  \right)^{n+1} \nonumber \\
    &= \dfrac{1}{n+1} \Phi_T(M_T)^{n+1}.
\end{align}
In summary, we have established that,
\begin{align} \label{eq:omega_identity_4}
    \dfrac{1}{\beta} \int \ddm{q} \sum_{\substack{\alpha \\ m \in \mathbb{Z} + \frac{1}{2}}} \dfrac{Q_{m \alpha,3}^n}{Q_{m \alpha}^2 + M_T^2} \Omega_T(M_T, Q_{m \alpha,3})^j = \dfrac{1}{n+1}\begin{cases}
        \Phi_T(M_T)^{n+1}, & j=n \\
        0, & j>n
    \end{cases}\;.
\end{align}
In the $T \to 0$ limit,
\begin{align}
    \lim_{\beta \to \infty} \dfrac{1}{\beta} \int \ddm{q} \sum_{\substack{\alpha \\ m \in \mathbb{Z} + \frac{1}{2}}} \dfrac{Q_{m \alpha,3}^n}{Q_{m \alpha}^2 + M_T^2} \Omega_T(M_T, Q_{m \alpha,3})^j &= \dfrac{1}{n+1}\begin{cases}
        \lim \limits_{\beta \to \infty} \Phi_T(M_T)^{n+1}, & j=n \\
        0, & j>n
    \end{cases}\;,
\end{align}
or,
\begin{align}
    \int \dm{q} \sum_{\substack{\alpha}} \dfrac{q_{3}^n}{q^2 + M_0^2} \Omega_0(M_0, q_{3})^j &= \dfrac{1}{n+1}\begin{cases}
        \Phi_0(M_0)^{n+1}, & j=n \\
        0, & j>n
    \end{cases}\;.
\end{align}


\section{Solving the gap equation}\label{app:sol_gap_eq}

In section \ref{subsec:sol_gpe}, we briefly described how to solve the gap equation. In this appendix, we will provide more details about the method outlined in the section \ref{subsec:sol_gpe}.

We need to evaluate,
\begin{align}
    \Sigma^{(n)}_{T,\pm}(P_{m\alpha,3}) &= \left( \mp \dfrac{i}{\beta} \right)^{n-1} \dfrac{1}{\beta}  \sum_{\substack{\sigma_1\\ j_1 \in \mathbb{Z} + \frac{1}{2}}} \int\ddm{q_1} \dfrac{1}{P_{m\alpha,3}-(Q_1)_{j_1\sigma_1,3}} \dfrac{1}{((Q_1)_{j_1\sigma_1})^2 + M_T^2}   \nonumber \\
    & \qquad\times \prod_{i=2}^{n} \left( \int \ddm{q_{i}} \sum_{\substack{\sigma_{i} \\ j_{i} \in \mathbb{Z} + \frac{1}{2}} } \dfrac{1}{(Q_{i-1})_{j_{i-1} \sigma_{i-1},3} - (Q_{i})_{j_{i} \sigma_{i},3} } \dfrac{1}{(Q_{i})_{j_{i} \sigma_{i}}^2 + M_T^2} \right) \nonumber \\
    &\qquad\times \Big( (Q_n)_{j_n \sigma_n,3} \mp i M  \Big), \quad \text{for   }\, n\geq 1. \label{nthorder_app}
\end{align}
We divide \eqref{nthorder_app} into a sum of two integrals,
\begin{align}\label{nthorder_app_decom}
    \Sigma^{(n)}_{T,\pm}(P_{m\alpha,3}) =  M \left( \mp \dfrac{i}{\beta} \right)^n I_{1,n} + \left( \mp \dfrac{i}{\beta} \right)^n I_{2,n},
\end{align}
where,
\begin{flalign}\label{eq:I1,n_def}
    I_{1,n} =& \sum_{\substack{\sigma_1\\ j_1 \in \mathbb{Z} + \frac{1}{2}}} \int\ddm{q_1} \dfrac{1}{P_{m\alpha,3}-(Q_1)_{j_1\sigma_1,3}} \dfrac{1}{((Q_1)_{j_1\sigma_1})^2 + M_T^2}  &&  \nonumber \\
    & \quad \times \prod_{i=2}^{n} \left( \int \ddm{q_{i}} \sum_{\substack{\sigma_{i} \\ j_{i} \in \mathbb{Z} + \frac{1}{2}} } \dfrac{1}{(Q_{i-1})_{j_{i-1} \sigma_{i-1},3} - (Q_{i})_{j_{i} \sigma_{i},3} } \dfrac{1}{(Q_{i})_{j_{i} \sigma_{i}}^2 + M_T^2} \right), && 
\end{flalign}
and,
\begin{flalign}\label{eq:I2,n_def}
    I_{2,n} =& \sum_{\substack{\sigma_1\\ j_1 \in \mathbb{Z} + \frac{1}{2}}} \int\ddm{q_1} \dfrac{1}{P_{m\alpha,3}-(Q_1)_{j_1\sigma_1,3}} \dfrac{1}{((Q_1)_{j_1\sigma_1})^2 + M_T^2} &&   \nonumber \\
    & \quad \times \prod_{i=2}^{n} \left( \int \ddm{q_{i}} \sum_{\substack{\sigma_{i} \\ j_{i} \in \mathbb{Z} + \frac{1}{2}} } \dfrac{1}{(Q_{i-1})_{j_{i-1} \sigma_{i-1},3} - (Q_{i})_{j_{i} \sigma_{i},3} } \dfrac{1}{(Q_{i})_{j_{i} \sigma_{i}}^2 + M_T^2} \right) \Big(  \pm i (Q_n)_{j_n \sigma_n,3}  \Big). &&
\end{flalign}
Let us start with $I_{1,n}$ \eqref{eq:I1,n_def}. We will first evaluate the integral/summation over the dummy variables $(Q_{n})_{j_{n} \sigma_{n}}$, then the integral/summation over $(Q_{n-1})_{j_{n-1} \sigma_{n-1}}$, and so on. The integral/summation over $(Q_{n})_{j_{n} \sigma_{n}}$ is easily evaluated as,
\begin{align}
    I_{1,n} =& \sum_{\substack{\sigma_1\\ j_1 \in \mathbb{Z} + \frac{1}{2}}} \int\ddm{q_1} \dfrac{1}{P_{m\alpha,3}-(Q_1)_{j_1\sigma_1,3}} \dfrac{1}{((Q_1)_{j_1\sigma_1})^2 + M_T^2}   \nonumber \\
    & \quad \times \prod_{i=2}^{n} \left( \int \ddm{q_{i}} \sum_{\substack{\sigma_{i} \\ j_{i} \in \mathbb{Z} + \frac{1}{2}} } \dfrac{1}{(Q_{i-1})_{j_{i-1} \sigma_{i-1},3} - (Q_{i})_{j_{i} \sigma_{i},3} } \dfrac{1}{(Q_{i})_{j_{i} \sigma_{i}}^2 + M_T^2} \right)\nonumber\\
    =& \sum_{\substack{\sigma_1\\ j_1 \in \mathbb{Z} + \frac{1}{2}}} \int\ddm{q_1} \dfrac{1}{P_{m\alpha,3}-(Q_1)_{j_1\sigma_1,3}} \dfrac{1}{((Q_1)_{j_1\sigma_1})^2 + M_T^2}   \nonumber \\
    & \quad \resizebox{1\hsize}{!}{$\displaystyle\times \prod_{i=2}^{n-1} \left( \int \ddm{q_{i}} \sum_{\substack{\sigma_{i} \\ j_{i} \in \mathbb{Z} + \frac{1}{2}} } \dfrac{1}{(Q_{i-1})_{j_{i-1} \sigma_{i-1},3} - (Q_{i})_{j_{i} \sigma_{i},3} } \dfrac{1}{(Q_{i})_{j_{i} \sigma_{i}}^2 + M_T^2} \right) \Bigg( \beta \Omega_T(M_T,(Q_{n-1})_{j_{n-1}\sigma_{n-1},3}) \Bigg),$}
\end{align}
where the summation/integral over $(Q_{n})_{j_{n} \sigma_{n}}$ has been evaluated using the definition of $\Omega_T$ \eqref{eq:integral_omega_FT}. Now to evaluate the summation/integral over $(Q_{n-1})_{j_{n-1}\sigma_{n-1}}$, $(Q_{n-2})_{j_{n-2}\sigma_{n-2}}$, $\dots$, we will use the identity \eqref{eq:omega_identity_1} repeatedly to get,
\begin{align}
    I_{1,n} =& \sum_{\substack{\sigma_1\\ j_1 \in \mathbb{Z} + \frac{1}{2}}} \int\ddm{q_1} \dfrac{1}{P_{m\alpha,3}-(Q_1)_{j_1\sigma_1,3}} \dfrac{1}{((Q_1)_{j_1\sigma_1})^2 + M_T^2} \nonumber \\
    & \quad \resizebox{1\hsize}{!}{$\displaystyle\times \prod_{i=2}^{n-2} \left( \int \ddm{q_{i}} \sum_{\substack{\sigma_{i} \\ j_{i} \in \mathbb{Z} + \frac{1}{2}} } \dfrac{1}{(Q_{i-1})_{j_{i-1} \sigma_{i-1},3} - (Q_{i})_{j_{i} \sigma_{i},3} } \dfrac{1}{(Q_{i})_{j_{i} \sigma_{i}}^2 + M_T^2} \right) \Bigg( \dfrac{\beta^2}{2!} \Omega_T(M_T,(Q_{n-2})_{j_{n-2}\sigma_{n-2},3})^2 \Bigg)$}\nonumber \\
    =& \sum_{\substack{\sigma_1\\ j_1 \in \mathbb{Z} + \frac{1}{2}}} \int\ddm{q_1} \dfrac{1}{P_{m\alpha,3}-(Q_1)_{j_1\sigma_1,3}} \dfrac{1}{((Q_1)_{j_1\sigma_1})^2 + M_T^2} \nonumber \\
    & \quad \resizebox{1\hsize}{!}{$\displaystyle\times \prod_{i=2}^{n-3} \left( \int \ddm{q_{i}} \sum_{\substack{\sigma_{i} \\ j_{i} \in \mathbb{Z} + \frac{1}{2}} } \dfrac{1}{(Q_{i-1})_{j_{i-1} \sigma_{i-1},3} - (Q_{i})_{j_{i} \sigma_{i},3} } \dfrac{1}{(Q_{i})_{j_{i} \sigma_{i}}^2 + M_T^2} \right) \Bigg( \dfrac{\beta^3}{3!} \Omega_T(M_T,(Q_{n-3})_{j_{n-3}\sigma_{n-3},3})^3 \Bigg)$} \nonumber \\
    =& \quad\vdots \nonumber \\
    =& \sum_{\substack{\sigma_1\\ j_1 \in \mathbb{Z} + \frac{1}{2}}} \int\ddm{q_1} \dfrac{1}{P_{m\alpha,3}-(Q_1)_{j_1\sigma_1,3}} \dfrac{1}{((Q_1)_{j_1\sigma_1})^2 + M_T^2}   \nonumber \\
    & \quad \times  \int \ddm{q_{2}} \sum_{\substack{\sigma_{2} \\ j_{2} \in \mathbb{Z} + \frac{1}{2}} } \dfrac{1}{(Q_{1})_{j_{1} \sigma_{1},3} - (Q_{2})_{j_{2} \sigma_{2},3} } \dfrac{1}{(Q_{2})_{j_{2} \sigma_{2}}^2 + M_T^2} \Bigg( \dfrac{\beta^{n-2}}{(n-2)!} \Omega_T(M_T,(Q_{2})_{j_{2}\sigma_{2},3})^{n-2} \Bigg),
\end{align}
or,
\begin{align}
    I_{1,n} =& \sum_{\substack{\sigma_1\\ j_1 \in \mathbb{Z} + \frac{1}{2}}} \int\ddm{q_1} \dfrac{1}{P_{m\alpha,3}-(Q_1)_{j_1\sigma_1,3}} \dfrac{1}{((Q_1)_{j_1\sigma_1})^2 + M_T^2} \Bigg( \dfrac{\beta^{n-1}}{(n-1)!} \Omega_T(M_T,(Q_{1})_{j_{1}\sigma_{1},3})^{n-1} \Bigg)  \nonumber \\
   =& \dfrac{\beta^{n}}{n!} \Omega_T(M_T,P_{m \alpha,3})^{n}.\label{eq:I_1n}
\end{align}
We now turn to $I_{2,n}$ \eqref{eq:I2,n_def}. It is convenient to first rearrange the expression for $I_{2,n}$ as follows,
\begin{flalign}
    I_{2,n} =& \sum_{\substack{\sigma_1\\ j_1 \in \mathbb{Z} + \frac{1}{2}}} \int\ddm{q_1} \dfrac{1}{P_{m\alpha,3}-(Q_1)_{j_1\sigma_1,3}} \dfrac{1}{((Q_1)_{j_1\sigma_1})^2 + M_T^2} &&   \nonumber \\
    & \quad \times \prod_{i=2}^{n} \left( \int \ddm{q_{i}} \sum_{\substack{\sigma_{i} \\ j_{i} \in \mathbb{Z} + \frac{1}{2}} } \dfrac{1}{(Q_{i-1})_{j_{i-1} \sigma_{i-1},3} - (Q_{i})_{j_{i} \sigma_{i},3} } \dfrac{1}{(Q_{i})_{j_{i} \sigma_{i}}^2 + M_T^2} \right) \Big(  \pm i (Q_n)_{j_n \sigma_n,3}  \Big). && \label{eq:I_2n_intm_0}
\end{flalign}
Adding and subtracting $(Q_{n-1})_{j_{n-1} \sigma_{n-1},3}$ in the numerator in \eqref{eq:I_2n_intm_0}, we can write,
\begin{flalign}
     I_{2,n} =& \sum_{\substack{\sigma_1\\ j_1 \in \mathbb{Z} + \frac{1}{2}}} \int\ddm{q_1} \dfrac{1}{P_{m\alpha,3}-(Q_1)_{j_1\sigma_1,3}} \dfrac{1}{((Q_1)_{j_1\sigma_1})^2 + M_T^2} &&   \nonumber \\
    & \quad \times \prod_{i=2}^{n} \left( \int \ddm{q_{i}} \sum_{\substack{\sigma_{i} \\ j_{i} \in \mathbb{Z} + \frac{1}{2}} } \dfrac{1}{(Q_{i-1})_{j_{i-1} \sigma_{i-1},3} - (Q_{i})_{j_{i} \sigma_{i},3} } \dfrac{1}{(Q_{i})_{j_{i} \sigma_{i}}^2 + M_T^2} \right) && \nonumber \\
    & \qquad \times \pm i \Bigg(  (Q_n)_{j_n \sigma_n,3} - (Q_{n-1})_{j_{n-1} \sigma_{n-1},3} + (Q_{n-1})_{j_{n-1} \sigma_{n-1},3}  \Bigg). && \label{eq:I_2n_intm_1}
\end{flalign}
Breaking up the numerator in \eqref{eq:I_2n_intm_1} into two terms, $\displaystyle\left((Q_n)_{j_n \sigma_n,3} - (Q_{n-1})_{j_{n-1} \sigma_{n-1},3}\right)$ and $(Q_{n-1})_{j_{n-1} \sigma_{n-1},3}$,
\begin{flalign}
     I_{2,n}=& \mp i \sum_{\substack{\sigma_1\\ j_1 \in \mathbb{Z} + \frac{1}{2}}} \int\ddm{q_1} \dfrac{1}{P_{m\alpha,3}-(Q_1)_{j_1\sigma_1,3}} \dfrac{1}{((Q_1)_{j_1\sigma_1})^2 + M_T^2} &&   \nonumber \\
    & \quad \times \prod_{i=2}^{n-1} \left( \int \ddm{q_{i}} \sum_{\substack{\sigma_{i} \\ j_{i} \in \mathbb{Z} + \frac{1}{2}} } \dfrac{1}{(Q_{i-1})_{j_{i-1} \sigma_{i-1},3} - (Q_{i})_{j_{i} \sigma_{i},3} } \dfrac{1}{(Q_{i})_{j_{i} \sigma_{i}}^2 + M_T^2} \right) \beta \Phi_T(M_T) && \nonumber \\
    & + \sum_{\substack{\sigma_1\\ j_1 \in \mathbb{Z} + \frac{1}{2}}} \int\ddm{q_1} \dfrac{1}{P_{m\alpha,3}-(Q_1)_{j_1\sigma_1,3}} \dfrac{1}{((Q_1)_{j_1\sigma_1})^2 + M_T^2} &&   \nonumber \\
    & \quad \times \prod_{i=2}^{n-1} \left( \int \ddm{q_{i}} \sum_{\substack{\sigma_{i} \\ j_{i} \in \mathbb{Z} + \frac{1}{2}} } \dfrac{1}{(Q_{i-1})_{j_{i-1} \sigma_{i-1},3} - (Q_{i})_{j_{i} \sigma_{i},3} } \dfrac{1}{(Q_{i})_{j_{i} \sigma_{i}}^2 + M_T^2} \right) && \nonumber \\
    & \qquad \times \Bigg( \pm i (Q_{n-1})_{j_{n-1} \sigma_{n-1},3} ~ \beta \Omega_T(M_T,(Q_{n-1})_{j_{n-1} \sigma_{n-1},3})  \Bigg). && \label{eq:I_2n_intm_2}
\end{flalign}
In the first term of \eqref{eq:I_2n_intm_2}, we have cancelled the factor of $(Q_n)_{j_n \sigma_n,3} - (Q_{n-1})_{j_{n-1} \sigma_{n-1},3}$ between the numerator and denominator, and have used the definition of $\Phi_T(M_T)$ \eqref{eq:integral_phi_FT} to evaluate the integral/summation over $(Q_{n})_{j_{n} \sigma_{n}}$. In the second term of \eqref{eq:I_2n_intm_2}, we have used the definition of $\Omega_T(M_T,P_{m \alpha,3})$ \eqref{eq:integral_omega_FT} to evaluate the integral/summation over $(Q_{n})_{j_{n} \sigma_{n}}$.

Using the definition of $I_{1,n}$ \eqref{eq:I1,n_def} in the first term of \eqref{eq:I_2n_intm_2}, we obtain,
\begin{align}\label{eq:I2,n_split}
    I_{2,n} = \mp i \beta \Phi_T(M_T) I_{1,n-1} + \Tilde{I}_{2,n},
\end{align}
where,
\begin{flalign}\label{eq:I2tilde,n_def}
    \Tilde{I}_{2,n} =& \sum_{\substack{\sigma_1\\ j_1 \in \mathbb{Z} + \frac{1}{2}}} \int\ddm{q_1} \dfrac{1}{P_{m\alpha,3}-(Q_1)_{j_1\sigma_1,3}} \dfrac{1}{((Q_1)_{j_1\sigma_1})^2 + M_T^2} &&   \nonumber \\
    & \quad \times \prod_{i=2}^{n-1} \left( \int \ddm{q_{i}} \sum_{\substack{\sigma_{i} \\ j_{i} \in \mathbb{Z} + \frac{1}{2}} } \dfrac{1}{(Q_{i-1})_{j_{i-1} \sigma_{i-1},3} - (Q_{i})_{j_{i} \sigma_{i},3} } \dfrac{1}{(Q_{i})_{j_{i} \sigma_{i}}^2 + M_T^2} \right) && \nonumber \\
    & \qquad \times \Bigg( \pm i (Q_{n-1})_{j_{n-1} \sigma_{n-1},3} ~ \beta \Omega_T(M_T,(Q_{n-1})_{j_{n-1} \sigma_{n-1},3})  \Bigg). &&
\end{flalign}
To evaluate $\Tilde{I}_{2,n}$ \eqref{eq:I2tilde,n_def}, we will first evaluate the integral/summation over  $(Q_{n-1})_{j_{n-1} \sigma_{n-1}}$, then over  $(Q_{n-2})_{j_{n-2} \sigma_{n-2}}$, and so on (using the identity \eqref{eq:omega_identity_2}),
\begin{flalign}
    \Tilde{I}_{2,n} =& \sum_{\substack{\sigma_1\\ j_1 \in \mathbb{Z} + \frac{1}{2}}} \int\ddm{q_1} \dfrac{1}{P_{m\alpha,3}-(Q_1)_{j_1\sigma_1,3}} \dfrac{1}{((Q_1)_{j_1\sigma_1})^2 + M_T^2} &&   \nonumber \\
    & \quad \times \prod_{i=2}^{n-2} \left( \int \ddm{q_{i}} \sum_{\substack{\sigma_{i} \\ j_{i} \in \mathbb{Z} + \frac{1}{2}} } \dfrac{1}{(Q_{i-1})_{j_{i-1} \sigma_{i-1},3} - (Q_{i})_{j_{i} \sigma_{i},3} } \dfrac{1}{(Q_{i})_{j_{i} \sigma_{i}}^2 + M_T^2} \right) && \nonumber \\
    & \qquad \times \Bigg( \pm i (Q_{n-2})_{j_{n-2} \sigma_{n-2},3} ~ \dfrac{\beta^{2}}{2!} \Omega_T(M_T,(Q_{n-2})_{j_{n-2} \sigma_{n-2},3})^{2}  \Bigg) && \nonumber\\
    =& \sum_{\substack{\sigma_1\\ j_1 \in \mathbb{Z} + \frac{1}{2}}} \int\ddm{q_1} \dfrac{1}{P_{m\alpha,3}-(Q_1)_{j_1\sigma_1,3}} \dfrac{1}{((Q_1)_{j_1\sigma_1})^2 + M_T^2} &&   \nonumber \\
    & \quad \times \prod_{i=2}^{n-3} \left( \int \ddm{q_{i}} \sum_{\substack{\sigma_{i} \\ j_{i} \in \mathbb{Z} + \frac{1}{2}} } \dfrac{1}{(Q_{i-1})_{j_{i-1} \sigma_{i-1},3} - (Q_{i})_{j_{i} \sigma_{i},3} } \dfrac{1}{(Q_{i})_{j_{i} \sigma_{i}}^2 + M_T^2} \right) && \nonumber \\
    & \qquad \times \Bigg( \pm i (Q_{n-3})_{j_{n-3} \sigma_{n-3},3} ~ \dfrac{\beta^{3}}{3!} \Omega_T(M_T,(Q_{n-3})_{j_{n-3} \sigma_{n-3},3})^{3}  \Bigg) && \nonumber \\
    =& \quad\vdots && \nonumber \\
    =& \sum_{\substack{\sigma_1\\ j_1 \in \mathbb{Z} + \frac{1}{2}}} \int\ddm{q_1} \dfrac{1}{P_{m\alpha,3}-(Q_1)_{j_1\sigma_1,3}} \dfrac{1}{((Q_1)_{j_1\sigma_1})^2 + M_T^2} &&   \nonumber \\
    &  \resizebox{1\hsize}{!}{$\displaystyle\times \int \ddm{q_{i}} \sum_{\substack{\sigma_{i} \\ j_{i} \in \mathbb{Z} + \frac{1}{2}} } \dfrac{1}{(Q_{1})_{j_{1} \sigma_{1},3} - (Q_{2})_{j_{2} \sigma_{2},3} } \dfrac{1}{(Q_{2})_{j_{2} \sigma_{2}}^2 + M_T^2} \Bigg( \pm i (Q_{2})_{j_{2} \sigma_{2},3} ~ \dfrac{\beta^{n-2}}{(n-2)!} \Omega_T(M_T,(Q_{2})_{j_{2} \sigma_{2},3})^{n-2}  \Bigg)$}, &&
\end{flalign}
or,
\begin{flalign}\label{eq:I2tilde,n_final}
    \Tilde{I}_{2,n} =& \resizebox{1\hsize}{!}{$\displaystyle \sum_{\substack{\sigma_1\\ j_1 \in \mathbb{Z} + \frac{1}{2}}} \int\ddm{q_1} \dfrac{1}{P_{m\alpha,3}-(Q_1)_{j_1\sigma_1,3}} \dfrac{1}{((Q_1)_{j_1\sigma_1})^2 + M_T^2}  \Bigg( \pm i (Q_{1})_{j_{1} \sigma_{1},3} ~ \dfrac{\beta^{n-1}}{(n-1)!} \Omega_T(M_T,(Q_{1})_{j_{1} \sigma_{1},3})^{n-1}  \Bigg)$} &&\nonumber \\
    =&\pm i P_{m \alpha,3} ~ \dfrac{\beta^{n}}{n!} \Omega_T(M_T,P_{m \alpha,3})^{n}. &&
\end{flalign}
Substituting \eqref{eq:I_1n} and \eqref{eq:I2tilde,n_final} in \eqref{eq:I2,n_split}, we get, 
\begin{align}
    I_{2,n} = \mp i \dfrac{\beta^{n}}{(n-1)!} \Phi_T(M_T)  \Omega_T(M_T,P_{m \alpha,3})^{n-1} \pm i ~ \dfrac{\beta^{n}}{n!} P_{m \alpha,3} ~ \Omega_T(M_T,P_{m \alpha,3})^{n}. \label{eq:I_2n}
\end{align}
Substituting \eqref{eq:I_1n} and \eqref{eq:I_2n} in \eqref{nthorder_app_decom}, we get,
\begin{align}
    \resizebox{1.1\hsize}{!}{$\displaystyle\Sigma^{(n)}_{T,\pm}(P_{m\alpha,3}) =  \dfrac{M}{n!} \big( \mp i \Omega_T(M_T,P_{m \alpha,3}) \big)^n - \dfrac{\Phi_T(M_T)}{(n-1)!} \big( \mp i \Omega_T(M_T,P_{m \alpha,3}) \big)^{n-1}  \pm \dfrac{i P_{m \alpha,3}}{n!} \big( \mp i \Omega_T(M_T,P_{m \alpha,3}) \big)^n .$}
\end{align}


\section{Some explicit expressions}


\subsection{Evaluation of \texorpdfstring{$\Phi_T(M_T)$}{Φ\_T(M\_T)}}\label{app:phi_eval}
The integral $\Phi_T(M_T)$ is defined as,
\begin{align}\label{eq:finite_temp_phi}
    \Phi_T(M_T) &= \dfrac{1}{\beta}\sum_{\substack{\mu \\ j \in \mathbb{Z} + \frac{1}{2}}}\int\ddm{q} \dfrac{1}{(Q_{j\mu})^2+M_T^2}\nonumber\\
    &= \dfrac{1}{\beta}\sum_{\substack{\sigma \\ j \in \mathbb{Z}}}\int\ddm{q} \dfrac{1}{\left(\dfrac{2\pi}{\beta}\left(j+\dfrac{1}{2}\right)- \dfrac{\lambda_\sigma}{\beta}\right)^2+{\vec{q}}^{\,\,2}+M_T^2}\,.
\end{align}
Using the following summation identity,
\begin{equation}
    \sum\limits_{n=-\infty}^\infty \dfrac{1}{(n+a)^2+b^2} = \dfrac{\pi}{b} \dfrac{\sinh{(2\pi b)}}{\cosh{(2\pi b)}-\cos{(2\pi a)}}\,,
\end{equation}
the summand in (\ref{eq:finite_temp_phi}) can be written as,
\begin{align}\label{eq:finite_temp_phi_eval1}
    \Phi_T(M_T) &= \dfrac{\beta}{(2\pi)^2}\sum_{\substack{\sigma}}\int\ddm{q}\dfrac{\pi}{\left(\dfrac{\beta}{2\pi}\right)\sqrt{\vec{q}^{\,\,2}+M_T^2}} \, \dfrac{\sinh{(\beta \sqrt{\vec{q}^{\,\,2}+M_T^2})}}{\cosh{(\beta \sqrt{\vec{q}^{\,\,2}+M_T^2})}+\cos{(\lambda_\sigma)}} \nonumber\\
    &= \dfrac{1}{2} \sum_{\substack{\sigma}} \int \ddm{q} \dfrac{1}{\sqrt{\vec{q}^{\,\,2}+M_T^2}} \, \dfrac{\sinh{(\beta \sqrt{\vec{q}^{\,\,2}+M_T^2})}}{\cosh{(\beta \sqrt{\vec{q}^{\,\,2}+M_T^2})}+\cos{(\lambda_\sigma)}} \nonumber\\
    &= \resizebox{0.95\hsize}{!}{$\displaystyle\dfrac{1}{2} \sum_{\substack{\sigma}} \int \ddm{q} \dfrac{1}{\sqrt{\vec{q}^{\,\,2}+M_T^2}} \, \left(\dfrac{\sinh{(\beta \sqrt{\vec{q}^{\,\,2}+M_T^2})}}{\cosh{(\beta \sqrt{\vec{q}^{\,\,2}+M_T^2})}+\cos{(\lambda_\sigma)}} -1 \right) + \dfrac{N}{2} \int \dfrac{d^{2- \epsilon} q}{(2 \pi)^{2 - \epsilon}} \dfrac{1}{\sqrt{\vec{q}^{\,\,2} + M_T^2}} .$}\nonumber\\
\end{align}
The first integral in (\ref{eq:finite_temp_phi_eval1}) is convergent. The second integral is divergent and will be evaluated in the dimensional regularization scheme.  

The first integral may be evaluated as follows:
\begin{align}\label{eq:phi_proof_I1}
    I_1 &= \dfrac{1}{2} \sum_{\substack{\sigma}} \int \ddm{q} \dfrac{1}{\sqrt{\vec{q}^{\,\,2}+M_T^2}} \, \left(\dfrac{\sinh{(\beta \sqrt{\vec{q}^{\,\,2}+M_T^2})}}{\cosh{(\beta \sqrt{\vec{q}^{\,\,2}+M_T^2})}+\cos{(\lambda_\sigma)}} -1 \right) \nonumber \\
    &= \dfrac{1}{4 \pi} \sum_{\substack{\sigma}} \int_{0}^{\infty} dq \, \dfrac{q}{\sqrt{q^2 + M_T^2}} \left(\dfrac{\sinh{(\beta \sqrt{q^{2}+M_T^2})}}{\cosh{(\beta \sqrt{q^{2}+M_T^2})}+\cos{(\lambda_\sigma)}} -1 \right) \nonumber \\
    &= \dfrac{1}{ 4 \pi \beta} \sum_{\substack{\sigma}} \int \limits_{\beta M_T}^{\infty} dy \left( \dfrac{\sinh{y}}{\cosh{y}+\cos{(\lambda_\sigma)}} -1  \right) \nonumber \\
    &= \dfrac{1}{4 \pi \beta} \sum_{\substack{\sigma}} \left( \lim_{\Lambda \to \infty} \left( \ln| \cosh(\Lambda) + \cos( \lambda_{\sigma}) | - \Lambda \right) - \ln|\cosh(\beta M_T) + \cos( \lambda_{\sigma}) | + \beta M_T \right) \nonumber \\
    &= -\dfrac{1}{4 \pi \beta} \sum_{\sigma} \ln \left| 2 (\cosh(\beta M_T) + \cos( \lambda_{\sigma})) \right| + \dfrac{N M_T}{4 \pi}.
\end{align}

We now turn to the evaluation of the second (divergent) integral using dimensional regularization. We will use the following general result for dimensional regularization,
\begin{equation}\label{eq:dim_reg_identity}
    \int\dfrac{\dd[d]{k}}{(2\pi)^d} \dfrac{k^a}{(k^2+\Delta)^b} = \dfrac{1}{(4\pi)^{d/2}} \Delta^{(a-2b+d)/2} \dfrac{\Gamma\left(b-\dfrac{a+d}{2}\right)\, \Gamma\left(\dfrac{a+d}{2}\right)}{\Gamma\left(\dfrac{d}{2}\right)\,\Gamma\left(b\right)}.
\end{equation}
It follows that,
\begin{align}\label{eq:phi_proof_I2}
    I_2 &= \dfrac{N}{2} \int \dfrac{d^{2- \epsilon} q}{(2 \pi)^{2 - \epsilon}} \dfrac{1}{\sqrt{\vec{q}^{\,\,2} + M_T^2}} \nonumber\\
    &= \dfrac{N}{2} \dfrac{1}{(4 \pi)^{1 - \epsilon/2}} (M_T)^{1 - \epsilon} \dfrac{\Gamma\left( \dfrac{\epsilon -1}{2} \right)}{\Gamma(\dfrac{1}{2})} \nonumber \\
    &= \dfrac{N M_T}{8 \pi} \dfrac{(-2 \sqrt{\pi})}{\sqrt{\pi}} \nonumber \\
    &= - \dfrac{N M_T}{4 \pi},
\end{align}
where we have used \eqref{eq:dim_reg_identity} to obtain the second equality.

Thus, combining \eqref{eq:phi_proof_I1} and \eqref{eq:phi_proof_I2}, we get (from \eqref{eq:finite_temp_phi_eval1}),
\begin{align}
    \Phi_T(M_T) &= -\dfrac{1}{4 \pi \beta} \sum_{\sigma} \ln \left| 2 (\cosh(\beta M_T) + \cos( \lambda_{\sigma})) \right| \nonumber\\
    &= -\dfrac{1}{4 \pi \beta} \sum_{\sigma} \, \ln\left|{e^{\beta M_T}(1 + e^{-\beta M_T + i  \lambda_{\sigma}})(1 + e^{- \beta M_T - i \lambda_{\sigma}})}\right|.
\end{align}
Taking the limit $T \to  0$, we obtain,
\begin{align}
    \Phi_{0}(M_0) &= \lim_{\beta \to \infty} -\dfrac{1}{4 \pi \beta} \sum_{\sigma} \, \ln\left|{e^{\beta M_T}(1 + e^{-\beta M_T +  i\lambda_{\sigma}})(1 + e^{- \beta M_T - i \lambda_{\sigma}})}\right| \nonumber \\
    &= - \dfrac{M_0}{4 \pi} \sum_{\sigma} 1 \nonumber \\
    &= - \dfrac{N M_0}{4 \pi}. \label{phi_0}
\end{align}
$\Phi_0(M_0)$ can also be computed directly at zero temperature. Indeed we obtain the zero temperature integral expression for $\Phi_0(M_0)$ by converting the sum over Matsubara frequencies into an integral. The integral is evaluated easily in the dimensional regularization scheme. We obtain,
\begin{align}\label{eq:integral_phi_ZT}
    \Phi_0(M_0) &= \lim_{\beta \to \infty} \dfrac{1}{\beta}\sum_{\substack{\sigma \\ j \in \mathbb{Z} + \frac{1}{2}}}\int\ddm{q} \dfrac{1}{(Q_{j\sigma})^2+M_T^2}\nonumber\\ 
    &= \sum_{\sigma} \int\dm{q} \dfrac{1}{q^2+M_0^2}\nonumber\\
    &= -\frac{N}{4\pi}M_0,
\end{align}
in agreement with \eqref{phi_0}.


\subsection{Evaluation of free thermal determinants}\label{app:contour}

In this subsection, we will prove the result \eqref{eq:Free_E_term34}, i.e., we will evaluate,
\begin{align}\label{eq:S_int_eval}
    S_{\rm free} =&-V_2 \sum_{\substack{\alpha \\ n \in \mathbb{Z} + \frac{1}{2}}} \int \ddm{q} \ln{\left((Q_{n\alpha})^2+M_T^2\right)} + V_2 \beta N\int \dm{q} \ln{\left(q^2+M_T^2\right)} \nonumber\\
    =&-V_2 \int \ddm{q} \left( \sum_{\substack{\alpha \\ n \in \mathbb{Z} + \frac{1}{2}}} \ln{\left((Q_{n\alpha})^2+M_T^2\right)} - \beta \sum_\alpha \int \dsm{q_3} \ln{\left(q^2+M_T^2\right)} \right) \nonumber\\
    =& -V_2 \sum_\alpha\int \ddm{q} \left( S_1 - S_2 \right),
\end{align}
where,
\begin{align}
    S_1 &= \sum_{\substack{n \in \mathbb{Z} + \frac{1}{2}}} \ln{\left((Q_{n\alpha})^2+M_T^2\right)} ,\\
    S_2 &= \beta \int \dsm{q_3} \ln{\left(q^2+M_T^2\right)}.\label{eq:contour_S2_def}
\end{align}
We demonstrate that the expression in \eqref{eq:S_int_eval} evaluates to, 
\begin{equation}\label{hamilot}
    S_{\rm free} = -V_2 \sum_{\alpha} \int \ddm{q} \ln\left\lvert \left( 1 + e^{ -\beta\sqrt{q^2 +M_T^2} + i \lambda_{\alpha}} \right) \left( 1 + e^{-\beta\sqrt{q^2 +M_T^2} - i \lambda_{\alpha}} \right) \right\rvert.
\end{equation} 
As mentioned in the main text, our final result is obvious on physical grounds. Consider any given value of the two-momentum ${\vec q}$, and any particular value of the flavour index $\alpha$. At those values, the summation over $n$ in \eqref{eq:S_int_eval} is the path integral expression for the logarithm of the partition function of a complex fermionic harmonic oscillator of frequency $\omega=\sqrt{M^2+{\vec q}^{\,\,2}}$, twisted with an imaginary chemical potential $i \lambda_\alpha/\beta$. In a renormalization scheme that sets the zero point energy of the oscillator to zero, it follows immediately from the Hamiltonian formalism, that this partition function is given by,
\begin{align}
    \ln\left\lvert \left( 1 + e^{ -\beta\sqrt{q^2 +M_T^2} + i \lambda_{\alpha}} \right) \right\rvert+ \ln \left\lvert \left( 1 + e^{-\beta\sqrt{q^2 +M_T^2} - i \lambda_{\alpha}} \right) \right\rvert.
\end{align} 
The two terms above are, respectively, the contributions of the $\psi$ and ${\bar \psi}$ oscillators, which have opposite signs. Applying this result to each value of ${\vec q}$ and $\alpha$, the result \eqref{hamilot} follows. 

In this appendix, we present a mathematical proof of the (physically obvious) equivalence of \eqref{eq:S_int_eval} and \eqref{hamilot}.

Expressing $S_1$ as a contour integral,
\begin{align}\label{eq:S1_contour}
    S_1 &= \sum_{\substack{n \in \mathbb{Z} + \frac{1}{2}}} \ln{\left(\left(\dfrac{2\pi n}{\beta} - \dfrac{\lambda_\alpha}{\beta}\right)^2+\vec{q}^{\;2}+M_T^2\right)} \nonumber\\
    &= \dfrac{-\pi}{2\pi i}\oint_\mathcal{C}\dd{z} \tan{(\pi z)} \ln{\left(\left(\dfrac{2\pi z - \lambda_\alpha}{\beta}  \right)^2+\vec{q}^{\;2}+M_T^2\right)}.
\end{align}
The second equality in \eqref{eq:S1_contour} follows from the fact that the $\tan{(\pi z)}$ function has poles at all half-integers, and the residue at each pole is $-\dfrac{1}{\pi}$. Upon a change of variable,
\begin{align}
    S_1 = -\dfrac{\beta}{4\pi i}\oint_\mathcal{C}\dd{z} \tan{\left(\dfrac{\beta z}{2} + \dfrac{\lambda_\alpha}{2}\right)} \ln{\left(z^2+\Tilde{c}^2\right)},
\end{align}
where,
\begin{align}\label{eq:c_tilde}
    \Tilde{c}^2 = \vec{q}^{\;2}+M_T^2,
\end{align}
and the contour $\mathcal{C}$ is chosen (as depicted in \eqref{Contour_1}) to enclose all the poles of the tangent function, but to exclude the branch cuts of the logarithmic function.
\begin{align}\label{Contour_1}
    \begin{tikzpicture}[scale = 0.45]
        \draw[thick] (-5,0) -- (5,0) coordinate (xaxis);
        \node[cross out, draw = black,scale = 0.6] at (-4.5,0) {};
        \node[cross out, scale = 0.6, draw = black] at (-3.5,0) {};
        \node[cross out, scale = 0.6,draw = black] at (-2.5,0) {};
        \node[cross out, scale = 0.6, draw = black] at (-1.5,0) {};
        \node[cross out, scale = 0.6, draw = black] at (-0.5,0) {};
        \node[cross out, scale = 0.6, draw = black] at (0.5,0) {};
        \node[cross out, scale = 0.6, draw = black] at (1.5,0) {};
        \node[cross out, scale = 0.6, draw = black] at (2.5,0) {};
        \node[cross out, scale = 0.6, draw = black] at (3.5,0) {};
        \node[cross out, scale = 0.6, draw = black] at (4.5,0) {};
        \node[draw = white] at (6,0) {=};
        \begin{scope}[thick,black, decoration={
            markings,
            mark=at position 0.5 with {\arrow[line width = 0.6 pt]{>}},
            }]
        \draw[postaction={decorate}] (0.3,4)--(0.3,1);
        \draw[postaction={decorate}] (-0.3,1)--(-0.3,4);
        \draw[postaction={decorate}] (-0.3,-4)--(-0.3,-1);
        \draw[postaction={decorate}] (0.3,-1)--(0.3,-4);
        \end{scope}
        \begin{scope}[thick,black, decoration={
            markings,
            mark=at position 0.25 with {\arrow[line width = 0.6 pt]{>}},
            mark=at position 0.75 with {\arrow[line width = 0.6 pt]{>}}
            }]
        \draw[postaction={decorate}] (0.3,-4) arc (-90:90:4);
        \draw[postaction={decorate}] (-0.3,4) arc (90:270:4);
        \end{scope}
        \begin{scope}[thick,black, decoration={
            markings,
            }]
        \draw[postaction={decorate}] (0.3,1) arc (0:-180:0.3);
        \draw[postaction={decorate}] (-0.3,-1) arc (180:0:0.3);
        \end{scope}
        \filldraw[red] (0,1) circle (2pt);
        \filldraw[red] (0,-1) circle (2pt);
        \draw[snake = snake, segment amplitude = 0.5mm, segment length = 3.5mm, thick] (0,1) -- (0,4);
        \draw[snake = snake,segment amplitude = 0.5mm, segment length = 3.5mm, thick] (0,-1) -- (0,-4);
        \node[right] at (0.3,1) {$\color{red} \tilde{c}$};
        \node[left] at (-0.3,-1) {$\color{red} -\tilde{c}$};
        \node[above left] at (2.5,-3.5) {$\mathcal{C}$};
    \end{tikzpicture}
    \hbox{
    \begin{tikzpicture}[scale = 0.45]
        \draw[thick] (-5,0) -- (5,0) coordinate (xaxis);
        \node[cross out, draw = black,scale = 0.6] at (-4.5,0) {};
        \node[cross out, scale = 0.6, draw = black] at (-3.5,0) {};
        \node[cross out, scale = 0.6,draw = black] at (-2.5,0) {};
        \node[cross out, scale = 0.6, draw = black] at (-1.5,0) {};
        \node[cross out, scale = 0.6, draw = black] at (-0.5,0) {};
        \node[cross out, scale = 0.6, draw = black] at (0.5,0) {};
        \node[cross out, scale = 0.6, draw = black] at (1.5,0) {};
        \node[cross out, scale = 0.6, draw = black] at (2.5,0) {};
        \node[cross out, scale = 0.6, draw = black] at (3.5,0) {};
        \node[cross out, scale = 0.6, draw = black] at (4.5,0) {};
        \node[draw = white] at (6,0) {+};
        \begin{scope}[thick,black, decoration={
            markings,
            mark=at position 0.5 with {\arrow[line width = 0.6 pt]{>}},
            }]
        \draw[postaction={decorate}] (0.3,4)--(0.3,1);
        \draw[postaction={decorate}] (-0.3,1)--(-0.3,4);
        \draw[postaction={decorate}] (-0.3,-4)--(-0.3,-1);
        \draw[postaction={decorate}] (0.3,-1)--(0.3,-4);
        \end{scope}
        \begin{scope}[thick,black, decoration={
            markings,
            }]
        \draw[postaction={decorate}] (0.3,1) arc (0:-180:0.3);
        \draw[postaction={decorate}] (-0.3,-1) arc (180:0:0.3);
        \end{scope}
        \filldraw[red] (0,1) circle (2pt);
        \filldraw[red] (0,-1) circle (2pt);
        \draw[snake = snake,segment amplitude = 0.5mm, segment length = 3.5mm, thick] (0,1) -- (0,4);
        \draw[snake = snake,segment amplitude = 0.5mm, segment length = 3.5mm, thick] (0,-1) -- (0,-4);
        \node[right] at (0.3,1) {$\color{red} \tilde{c}$};
        \node[left] at (-0.3,-1) {$\color{red} -\tilde{c}$};
        \node[right] at (0.3,2.1) {I};
        \node[left] at (-0.3,2.1) {II};
        \node[right] at (0.3,-2.1) {IV};
        \node[left] at (-0.3,-2.1) {III};
    \end{tikzpicture}
    }
    \hbox{
    \begin{tikzpicture}[scale=0.45]
        \draw[thick] (-5,0) -- (5,0) coordinate (xaxis);
        \node[cross out, draw = black,scale = 0.6] at (-4.5,0) {};
        \node[cross out, scale = 0.6, draw = black] at (-3.5,0) {};
        \node[cross out, scale = 0.6,draw = black] at (-2.5,0) {};
        \node[cross out, scale = 0.6, draw = black] at (-1.5,0) {};
        \node[cross out, scale = 0.6, draw = black] at (-0.5,0) {};
        \node[cross out, scale = 0.6, draw = black] at (0.5,0) {};
        \node[cross out, scale = 0.6, draw = black] at (1.5,0) {};
        \node[cross out, scale = 0.6, draw = black] at (2.5,0) {};
        \node[cross out, scale = 0.6, draw = black] at (3.5,0) {};
        \node[cross out, scale = 0.6, draw = black] at (4.5,0) {};
        \node[draw = white] at (6,0) {.};
        \begin{scope}[thick,black, decoration={
            markings,
            mark=at position 0.25 with {\arrow[line width = 0.6 pt]{>}},
            mark=at position 0.75 with {\arrow[line width = 0.6 pt]{>}}
            }]
        \draw[postaction={decorate}] (0.3,-4) arc (-90:90:4);
        \draw[postaction={decorate}] (-0.3,4) arc (90:270:4);
        \end{scope}
        \filldraw[red] (0,1) circle (2pt);
        \filldraw[red] (0,-1) circle (2pt);
        \draw[snake = snake,segment amplitude = 0.5mm, segment length = 3.5mm, thick] (0,1) -- (0,4);
        \draw[snake = snake, segment amplitude = 0.5mm, segment length = 3.5mm, thick] (0,-1) -- (0,-4);
        \node[right] at (0.3,1) {$\color{red} \tilde{c}$};
        \node[left] at (-0.3,-1) {$\color{red} -\tilde{c}$};
        \node[above left] at (2.5,-3.5) {$\mathcal{C}'$};
    \end{tikzpicture}
    }
\end{align}
The contour integral over $\mathcal{C}'$ can be simplified as follows. Recall that,
\begin{equation}\label{tan}
    \tan w=\dfrac{1}{i} \left(   \dfrac{ e^{i w}- e^{-i w}}{e^{i w}+ e^{-iw}} \right).
\end{equation}
Whenever $|w|$ is large, and $0 < Arg(w) < \pi$ (i.e. when $w$ is in the upper half plane), \eqref{tan} is well approximated by,
\begin{equation}\label{tanapo}
    \tan w \approx \dfrac{1}{i}  \left( \dfrac{- e^{-i w}}{ e^{-iw}} \right)  = i.
\end{equation}
On the other hand, whenever $|w|$ is large and $-\pi < Arg(w)<0$ (i.e. when $w$ is in the lower half plane), \eqref{tan} is well approximated by.
\begin{equation}\label{tanapt}
    \tan w \approx \dfrac{1}{i}  \dfrac{ e^{i w}} {e^{i w}}=- i.
\end{equation}
It follows that the integrand in the integral over $\mathcal{C}'$ simplifies to, 
\begin{align}
    - \dfrac{\beta}{4\pi}\oint_\mathcal{C'}\dd{z} \Theta(z) \ln{\left(z^2+\Tilde{c}^2\right)},
\end{align}
where $\Theta(z)$ is $1$ in the upper half-plane and $-1$ in the lower half-plane.

Let us now turn to $S_2$. We first write $S_2$ as the sum over two identical terms (see \eqref{eq:contour_S_2_split}), each of which is half of the RHS of \eqref{eq:contour_S2_def} that defines $S_2$. We then deform the contour in the first of these terms to a large semicircular arc around infinity in the upper half-plane (upper half of the $\mathcal{C}_2'$ contour in \eqref{Contour_2}), plus a contour that encloses the branch cut of the logarithmic function in the upper half-plane (I$'$ and II$'$ in \eqref{Contour_2}). We perform a similar manipulation for the second term, but this time we move all contours to the lower half plane. All these manipulations are displayed in \eqref{Contour_2}. In equations,
\begin{align}\label{eq:contour_S_2_split}
    S_2 &= \beta \int \dsm{q_3} \ln{\left(q_3^2+ \Tilde{c}^2\right)} \nonumber\\
    &= \dfrac{\beta}{4 \pi}\int_{\mathcal{C}_1}\dd{z} \ln{\left(z^2+ \Tilde{c}^2\right)} + \dfrac{\beta}{4 \pi}\int_{\mathcal{C}_2}\dd{z} \ln{\left(z^2+ \Tilde{c}^2\right)},
\end{align}
where all contours are displayed in \eqref{Contour_2}:
\begin{align}\label{Contour_2}
    \hbox{
    \begin{tikzpicture}[scale = 0.45]
        \draw[thick] (-5,0) -- (5,0) coordinate (xaxis);
        \node[draw = white] at (6,0) {=};
        \begin{scope}[thick,black, decoration={
            markings,
            mark=at position 0.25 with {\arrow[line width = 0.1 pt]{>}},
            mark=at position 0.75 with {\arrow[line width = 0.1 pt]{>}}
            }]
        \draw[postaction={decorate}] (-4.5,0.3)--(4.5,0.3);
        \draw[postaction={decorate}] (-4.5,-0.3)--(4.5,-0.3);
        \end{scope}
        \filldraw[red] (0,1) circle (2pt);
        \filldraw[red] (0,-1) circle (2pt);
        \draw[snake = snake,segment amplitude = 0.5mm, segment length = 3.5mm, thick] (0,1) -- (0,4);
        \draw[snake = snake, segment amplitude = 0.5mm, segment length = 3.5mm, thick] (0,-1) -- (0,-4);
        \node[right] at (0.3,1) {$\color{red} \tilde{c}$};
        \node[left] at (-0.3,-1) {$\color{red} -\tilde{c}$};
        \node[above] at (3.5,0.5) {$\mathcal{C}_1$};
        \node[below] at (3.5,-0.5) {$\mathcal{C}_2$};
    \end{tikzpicture}
    }
    \hbox{
    \begin{tikzpicture}[scale = 0.45]
        \draw[thick] (-5,0) -- (5,0) coordinate (xaxis);
        \node[draw = white] at (6,0) {+};
        \begin{scope}[thick,black, decoration={
            markings,
            mark=at position 0.5 with {\arrow[line width = 0.6 pt]{>}},
            }]
        \draw[postaction={decorate}] (-0.3,4)--(-0.3,1);
        \draw[postaction={decorate}] (0.3,1)--(0.3,4);
        \draw[postaction={decorate}] (-0.3,-4)--(-0.3,-1);
        \draw[postaction={decorate}] (0.3,-1)--(0.3,-4);
        \end{scope}
        \begin{scope}[thick,black, decoration={
            markings,
            }]
        \draw[postaction={decorate}] (0.3,1) arc (0:-180:0.3);
        \draw[postaction={decorate}] (-0.3,-1) arc (180:0:0.3);
        \end{scope}
        \filldraw[red] (0,1) circle (2pt);
        \filldraw[red] (0,-1) circle (2pt);
        \draw[snake = snake,segment amplitude = 0.5mm, segment length = 3.5mm, thick] (0,1) -- (0,4);
        \draw[snake = snake,segment amplitude = 0.5mm, segment length = 3.5mm, thick] (0,-1) -- (0,-4);
        \node[right] at (0.3,1) {$\color{red} \tilde{c}$};
        \node[left] at (-0.3,-1) {$\color{red} -\tilde{c}$};
        \node[right] at (0.3,2.1) {I$'$};
        \node[left] at (-0.3,2.1) {II$'$};
        \node[right] at (0.3,-2.1) {IV$'$};
        \node[left] at (-0.3,-2.1) {III$'$};
    \end{tikzpicture}
    }
    \hbox{
    \begin{tikzpicture}[scale=0.45]
        \draw[thick] (-5,0) -- (5,0) coordinate (xaxis);
        \node[draw = white] at (6,0) {.};
        \begin{scope}[thick,black, decoration={
            markings,
            mark=at position 0.5 with {\arrow[line width = 0.6 pt]{>}},
            }]
        \draw[postaction={decorate}] (0.3,4) arc (85:5:4);
        \draw[postaction={decorate}] (0.3,-4) arc (-85:-5:4);
        \draw[postaction={decorate}] (-4,0.3) arc (175:95:4);
        \draw[postaction={decorate}] (-4,-0.3) arc (185:265:4);
        \end{scope}
        \filldraw[red] (0,1) circle (2pt);
        \filldraw[red] (0,-1) circle (2pt);
        \draw[snake = snake,segment amplitude = 0.5mm, segment length = 3.5mm, thick] (0,1) -- (0,4);
        \draw[snake = snake, segment amplitude = 0.5mm, segment length = 3.5mm, thick] (0,-1) -- (0,-4);
        \node[right] at (0.3,1) {$\color{red} \tilde{c}$};
        \node[left] at (-0.3,-1) {$\color{red} -\tilde{c}$};
        \node[above left] at (2.5,-3.5) {$\mathcal{C}_2'$};
    \end{tikzpicture}
    }
\end{align}
Now the contour integrals along $\mathcal{C}_2'$ (in \eqref{Contour_2}) and $\mathcal{C}'$ (in \eqref{Contour_1}) are equal, and so their difference is zero.

It follows that $S_1-S_2$ in \eqref{eq:S_int_eval} is given by the integrals I$+$II$+$III$+$IV$-$I$'-$II$'-$III$'-$IV$'$, where,
\begin{itemize}
    \item[I:]
    \begin{align}
        -\dfrac{\beta}{4\pi}\int\limits_\infty^{\Tilde{c}}\dd{y}\, \tan{\left(\dfrac{ i \beta y}{2} +\dfrac{\lambda_\alpha}{2} \right)} \ln{\left(-y^2+\Tilde{c}^2\right)},
    \end{align}
    \item[II:]
    \begin{align}
        -\dfrac{\beta}{4\pi}\int\limits^\infty_{\Tilde{c}}\dd{y} \,\tan{\left(\dfrac{ i \beta y}{2} +\dfrac{\lambda_\alpha}{2} \right)} \ln{\left(-y^2+\Tilde{c}^2\right)} +\dfrac{\beta}{4\pi}\int\limits^\infty_{\Tilde{c}}\dd{y}\, \tan{\left( \dfrac{ i \beta y}{2} +\dfrac{\lambda_\alpha}{2} \right)} 2\pi i ,
    \end{align}
    \item[III:]
    \begin{align}
        \dfrac{\beta}{4\pi}\int\limits_\infty^{\Tilde{c}}\dd{y}\, \tan{\left( -\dfrac{ i \beta y}{2} +\dfrac{\lambda_\alpha}{2} \right)} \ln{\left(-y^2+\Tilde{c}^2\right)},
    \end{align}
    \item[IV:]
    \begin{align}
        \dfrac{\beta}{4\pi}\int\limits^\infty_{\Tilde{c}}\dd{y}\, \tan{\left(-\dfrac{ i \beta y}{2} +\dfrac{\lambda_\alpha}{2} \right)} \ln{\left(-y^2+\Tilde{c}^2\right)} - \dfrac{\beta}{4\pi}\int\limits^\infty_{\Tilde{c}}\dd{y} \,\tan{\left(-\dfrac{ i \beta y}{2} +\dfrac{\lambda_\alpha}{2} \right)} 2\pi i,
    \end{align}
    \item[I$'$:]
    \begin{align}
        -\dfrac{i\beta}{4\pi}\int\limits_\infty^{\Tilde{c}}\dd{y}\, \ln{\left(-y^2+\Tilde{c}^2\right)},
    \end{align}
    \item[II$'$:]
    \begin{align}
        -\dfrac{i\beta}{4\pi}\int\limits^\infty_{\Tilde{c}}\dd{y}\, \ln{\left(-y^2+\Tilde{c}^2\right)} +\dfrac{i\beta}{4\pi}\int\limits^\infty_{\Tilde{c}}\dd{y}\, 2\pi i ,
    \end{align}
    \item[III$'$:]
    \begin{align}
        -\dfrac{i\beta}{4\pi}\int\limits_\infty^{\Tilde{c}}\dd{y}\, \ln{\left(-y^2+\Tilde{c}^2\right)},
    \end{align}
    \item[IV$'$:]
    \begin{align}
        -\dfrac{i\beta}{4\pi}\int\limits^\infty_{\Tilde{c}}\dd{y}\, \ln{\left(-y^2+\Tilde{c}^2\right)} + \dfrac{i\beta}{4\pi}\int\limits^\infty_{\Tilde{c}}\dd{y}\, 2\pi i.
    \end{align}
\end{itemize}
Simplifying, we find,
\begin{align}\label{eq:S1_lambda}
    \text{I+II+II+IV} &= \dfrac{i\beta}{2}\int\limits^\infty_{\Tilde{c}}\dd{y}\, \left(\tan{\left( \dfrac{ i \beta y}{2} +\dfrac{\lambda_\alpha}{2} \right)} - \tan{\left( -\dfrac{ i \beta y}{2} +\dfrac{\lambda_\alpha}{2} \right)}\right) \nonumber\\
    &= \dfrac{i\beta}{2}\int\limits^\infty_{\Tilde{c}}\dd{y} \,\left(\dfrac{2i\sinh{(\beta y)}}{\cosh{(\beta y)}+\cos{(\lambda_\alpha)}}\right) \nonumber\\
    &= -\beta \int\limits^\infty_{\Tilde{c}}\dd{y}\, \left(\dfrac{\sinh{(\beta y)}}{\cosh{(\beta y)}+\cos{(\lambda_\alpha)}} -1\right) - \beta \int\limits^\Lambda_{\Tilde{c}}\dd{y}\,,
\end{align}
and,
\begin{align}\label{eq:S2_lambda}
   \text{I$'$+II$'$+III$'$+IV$'$}  &= -\beta\int\limits^\Lambda_{\Tilde{c}}\dd{y}\,.
\end{align}
Therefore, from \eqref{eq:S1_lambda} and \eqref{eq:S2_lambda},
\begin{align}\label{eq:S_1-S_2}
    S_1 - S_2 &= -\beta \int\limits^\infty_{\Tilde{c}}\dd{y} \,\left(\dfrac{\sinh{(\beta y)}}{\cosh{(\beta y)}+\cos{(\lambda_\alpha)}} -1\right) \nonumber\\
    &= -\left.\bigg( \ln{\left\lvert \cosh{(\beta y)}+\cos{(\lambda_\alpha)} \right\rvert} - \beta y \bigg)\right\rvert^\infty_{\Tilde{c}} \nonumber\\
    &= \ln{\left\lvert \cosh{(\beta\Tilde{c})}+\cos{(\lambda_\alpha)} \right\rvert} + \ln{2} - \beta\Tilde{c}\nonumber\\
    &= \ln{\left\lvert \left( 1 + e^{-\beta\Tilde{c} + i\lambda_\alpha} \right) \left( 1 + e^{-\beta \Tilde{c} - i\lambda_\alpha} \right) \right\rvert}.
\end{align}
Substituting \eqref{eq:S_1-S_2} and \eqref{eq:c_tilde} in \eqref{eq:S_int_eval},
\begin{align}
    S_{\rm free} &= -V_2\sum_\alpha\int\ddm{q} \ln{\left\lvert \left( 1 + e^{-\beta \Tilde{c} + i \lambda_\alpha} \right) \left( 1 + e^{-\beta \Tilde{c} - i\lambda_\alpha} \right) \right\rvert}\nonumber \\
    &= -V_2 \sum_{\alpha} \int \ddm{q} \ln\left\lvert \left( 1 + e^{ -\beta\sqrt{q^2 +M_T^2} + i \lambda_{\alpha}} \right) \left( 1 + e^{-\beta\sqrt{q^2 +M_T^2} - i \lambda_{\alpha}} \right) \right\rvert.
\end{align}


\section{Evaluation of \texorpdfstring{$\Omega_T(M_T, P_{m\alpha,3})$}{Ω\_T(M\_T,P\_{mα,3})}}\label{app:Omegadiv}
The quantity $\Omega_T(M_T, P_{m\alpha,3})$ is defined as,
\begin{equation}\label{eq:omega_eval}
\begin{split}
\Omega_T(M_T,P_{m\alpha,3}) =& \dfrac{1}{\beta}\sum_{\substack{ j \in \mathbb{Z} + \frac{1}{2} \\ \sigma \\ \sigma \neq \alpha \text{ at } j=m}}\int\ddm{q} \dfrac{1}{(Q_{j \sigma})^2+M_T^2} \dfrac{1}{(P_{m\alpha,3}-Q_{j \sigma,3})}\\
=& \dfrac{1}{\beta}\sum_{\substack{\sigma \\ \sigma \neq \alpha \\ j \in \mathbb{Z}}}\int\ddm{q} \dfrac{1}{\dfrac{2\pi}{\beta}\left(m-j + \dfrac{\lambda_\sigma - \lambda_\alpha}{2\pi}\right)}\dfrac{1}{\left(\dfrac{2\pi}{\beta}\left(j+\dfrac{1}{2}- \dfrac{\lambda_\sigma}{2\pi}\right)\right)^2+{\vec{q}}^{\,\,2}+M_T^2}\\
&+ \dfrac{1}{\beta}\sum_{\substack{j \in \mathbb{Z}\\ j \neq m}}\int\ddm{q} \dfrac{1}{\dfrac{2\pi}{\beta}\left(m-j \right)}\dfrac{1}{\left(\dfrac{2\pi}{\beta}\left(j+\dfrac{1}{2}- \dfrac{\lambda_\alpha}{2\pi}\right)\right)^2+{\vec{q}}^{\,\,2}+M_T^2}\,.
\end{split}
\end{equation}
In going from the first to the second expression in \eqref{eq:omega_eval} we have separated the sum into the terms with $\sigma \neq \alpha$ (second line of \eqref{eq:omega_eval}) and terms with $\sigma=\alpha$ (third line of \eqref{eq:omega_eval}): note the $j\neq m$ exclusion for the terms on the third line. Collecting factors of $\dfrac{2\pi}{\beta}$ from the denominator, we rewrite \eqref{eq:omega_eval} as,
\begin{equation}\label{eq:omega_eval_1}
\begin{split}
\Omega_T(M_T,P_{m\alpha,3}) =& \dfrac{\beta^2}{(2\pi)^3}\sum_{\substack{\sigma \\ \sigma \neq \alpha \\ j \in \mathbb{Z}}}\int\ddm{q} \dfrac{1}{m-j + \dfrac{\lambda_\sigma - \lambda_\alpha}{2\pi}}\dfrac{1}{\left(j+\dfrac{1}{2}- \dfrac{\lambda_\sigma}{2\pi}\right)^2+\left(\dfrac{\beta}{2\pi}\sqrt{{\vec{q}}^{\,\,2}+M_T^2}\right)^2}\\
&+ \dfrac{\beta^2}{(2\pi)^3}\sum_{\substack{j \in \mathbb{Z}\\ j \neq m}}\int\ddm{q}~ \dfrac{1}{m-j}~\dfrac{1}{\left(j+\dfrac{1}{2}- \dfrac{\lambda_\alpha}{2\pi}\right)^2+\left(\dfrac{\beta}{2\pi}\sqrt{{\vec{q}}^{\,\,2}+M_T^2}\right)^2}\,.
\end{split}
\end{equation}
The expression on the second line of \eqref{eq:omega_eval_1} has no summation over gauge indices, and so is of the order $N^0$. Thus, it can be ignored compared to the expression on the first line of \eqref{eq:omega_eval_1} (which is a sum over $N$ gauge indices, and so is of order $N$). For this reason, we will ignore the second line of \eqref{eq:omega_eval_1}. The summation on the first line can be evaluated using the identity,
\begin{equation}\label{noiinent}
    \sum\limits_{n=-\infty}^\infty \dfrac{1}{a-n}\dfrac{1}{(n+b)^2+c^2} = \dfrac{\pi}{(a+b)^2+c^2} \left(\dfrac{\dfrac{(a+b)}{c}\sinh{(2\pi c)}+\sin{2\pi b}}{\cosh{(2\pi c)}-\cos{(2\pi b)}}\right)+ \dfrac{\pi\cot{\pi a}}{(a+b)^2 + c^2}\,,
\end{equation}
with $a= m+\dfrac{\lambda_\sigma-\lambda_\alpha}{2\pi}$, $b= \dfrac{1}{2}-\dfrac{\lambda_\sigma}{2\pi}$, and $c=\dfrac{\beta}{2\pi}\sqrt{{\vec{q}}^{\,\,2}+M_T^2}$. 
Note that both the LHS and RHS of \eqref{noiinent} diverge when $a$
is an integer (on the RHS the divergence comes form the term proportional to $\cot \pi a$). This is not an issue for the evaluation of 
\eqref{eq:omega_eval_1} as $m+\dfrac{\lambda_\sigma-\lambda_\alpha}{2\pi}$ is never an integer. 

Using \eqref{noiinent}, we find that the first line of \eqref{eq:omega_eval_1} equals,
\begin{equation}\label{eq:omega_eval_2}
    \begin{split}
    \dfrac{1}{2\pi}\sum_{\sigma \neq \alpha} \int\ddm{q} \dfrac{\pi}{(P_{m\alpha,3})^2+\vec{q}^{\,\,2}+M_T^2}\left(\dfrac{P_{m\alpha,3}\frac{\sinh(\beta \sqrt{\vec{q}^{\,\,2}+M_T^2})}{\sqrt{\vec{q}^{\,\,2}+M_T^2}}+\sin(\lambda_\sigma)}{\cosh(\beta \sqrt{\vec{q}^{\,\,2}+M_T^2})+\cos(\lambda_\sigma)}\right)  \\
    + \dfrac{1}{2\pi}\sum_{\sigma \neq \alpha} \int\ddm{q} \dfrac{\pi\cot(m\pi +\dfrac{\lambda_\sigma-\lambda_\alpha}{2})}{(P_{m\alpha,3})^2+\vec{q}^{\,\,2}+M_T^2}.
    \end{split}
\end{equation}
The first term of \eqref{eq:omega_eval_2} can be simplified by changing the integration variable to $x=\sqrt{\vec{q}^{\,\,2}+M_T^2}$; we obtain,
\begin{equation} \label{fl}
\begin{split}
    \dfrac{1}{4\pi}\sum_{\sigma \neq \alpha} \int_{M_T}^{\infty} {\mathrm{d}}x \,\dfrac{1}{(P_{m\alpha,3})^2+x^2}\left(\dfrac{P_{m\alpha,3}\sinh(\beta x) + x\sin(\lambda_\sigma)}{\cosh(\beta x)+\cos( \lambda_\sigma)}\right).
\end{split}
\end{equation}
The integral in the second line of \eqref{eq:omega_eval_2} may be evaluated explicitly: it has a logarithmic divergence in the $\Lambda$-cutoff scheme; in the dimensional regularization scheme, it evaluates to,
\begin{equation}\label{sl}
    \begin{split}
        & \dfrac{1}{8\pi}\sum_{\sigma \neq \alpha} \cot(\dfrac{\lambda_\sigma-\lambda_\alpha}{2})\left(\dfrac{2}{\epsilon}+\ln(\dfrac{4\pi e^{-\gamma_E}}{(P_{m\alpha,3})^2+M_T^2})+\mathcal{O}(\epsilon)\right),
    \end{split}
\end{equation}
where $\gamma_E$ is the Euler--Mascheroni constant.


\section{The zero temperature limit}\label{zertemp}

Throughout this paper, we have worked at a finite, nonzero temperature. This fact allowed us to completely fix all the gauge freedom and formulate the gauge-fixed path integral of our theory in a clean and unambiguous manner, even at finite $N$ (see e.g. \eqref{exact}). A key element in our gauge-fixing procedure (following \cite{Blau:1993tv}) was the use of the gauge holonomies to abelianize the $U(N)$ gauge symmetry that was left unfixed by the condition $\partial_3 A_3=0$. 

The natural zero temperature analogue of our gauge is the condition $A_3=0$. This condition, however, leaves a two-dimensional $U(N)$ gauge-invariance unfixed, and we lack a holonomy matrix that we could use to abelianize the unfixed gauge-invariance in a natural manner. This residual gauge-invariance shows up in the fact that the gauge field propagator, at zero temperature, is singular when the $3^\text{rd}$ component of the momentum of the gauge boson vanishes. In order to give meaning to Feynman diagrams in this gauge at zero temperature, the previous literature \cite{Moshe:2014bja, Giombi:2011kc} resorts to adopting arbitrarily chosen prescriptions (like the principal value prescription). 

One natural way to give meaning to the $A_3=0$ gauge at zero temperature would be to start with the finite-$N$, finite temperature formulation \eqref{exact}, and first take the limit $T \to 0$, followed by the limit $N \to \infty$. A second prescription would be to first take $N \to \infty$ and then take $T \to 0$. In the previous paragraphs, we have also discussed other prescriptions, namely to simply shut one's eyes and naively work with the gauge $A_3=0$, and to cure all singularities in the resultant expressions with a principal value (or some other) prescription. It is natural to wonder about the relationship between these various prescriptions. In this appendix, we make some observations relevant to this question.

\subsection{The relationship between \texorpdfstring{$\Omega_T(M_T, P_{m\alpha,3})$}{Ω\_T(M\_T,P\_{mα,3})} computed in the `second prescription' and principal value prescription}

In this subsection we compare the $T \to 0$ limit of the large-$N$ value of $\Omega_T(M_T, P_{m\alpha,3})$ (the sum of \eqref{fl}
and \eqref{sl}) with $\Omega_0(M_0,p_3)$, computed directly at zero temperature using the principal value prescription. 

Let us set $m= \dfrac{\beta}{2\pi} p_3$, and study $\eqref{fl}$ and $\eqref{sl}$ in the zero temperature limit, $\beta \to \infty$, with $p_3$, the now effectively continuous $3^\text{rd}$ component of momentum, held fixed. In this limit, \eqref{sl} reduces to,
\begin{equation}\label{sln}
    \begin{split}
        & \dfrac{1}{8\pi}\sum_{\sigma \neq \alpha} \cot(\dfrac{\lambda_\sigma-\lambda_\alpha}{2})\left(\dfrac{2}{\epsilon}+\ln(\dfrac{4\pi e^{-\gamma_E}}{p_3^2+M_0^2})+\mathcal{O}(\epsilon)\right),
    \end{split}
\end{equation}
while \eqref{fl} simplifies to,
\begin{equation} \label{fln}
    \begin{split}
        \dfrac{N}{4\pi} \int_{M_0}^{\infty} {\mathrm{d}}x \,\dfrac{p_3}{p_3^2+x^2} = \dfrac{N}{4\pi}\tan^{-1}\left(\dfrac{p_3}{M_0}\right).
    \end{split}
\end{equation}
It is interesting to compare \eqref{fln} and \eqref{sln} with the direct (naive) evaluation of the expression $\Omega_T(M_T,P_{m\alpha,3})$ at zero temperature, that is, 
\begin{align}\label{eq:integral_omega_ZT}
    \Omega_0(M_0,p_3)
    &=\sum_{\sigma}\int\dm{q} \dfrac{1}{q^2+M_0^2} \,\Pv\left( \dfrac{1}{p_{3}-q_{3}} \right)  \nonumber\\
    &= \frac{N}{4\pi} \tan^{-1}\left(\frac{p_3}{M_0}\right),
\end{align}
where $\Pv$ stands for principal value. Notice that \eqref{eq:integral_omega_ZT} is in perfect agreement with \eqref{fln}, but (therefore) does not agree with the sum of \eqref{fln} and \eqref{sln}.

Aspects of the disagreement between the zero temperature limit of $\Omega_T(M_T,P_{m\alpha,3})$ and the (naive) direct computation of $\Omega_0(M_0,p_3)$ may seem puzzling from both the technical and the physical points of view. 

At the technical level, the zero temperature limit of the finite case corresponds, effectively, to studying a summation of the form,
\begin{equation}\label{zerotempsum}
    \beta^2 \sum\limits_{n=-\infty}^\infty \dfrac{1}{a\beta-n}\dfrac{1}{(n+b)^2+\beta^2 c^2},
\end{equation}
in the limit $\beta \to \infty$, with $a$, $b$, and $c$ held fixed. This summation receives contributions of the order unity from two classes of integers $n$:
\begin{itemize}
    \item[(1)] First, we have integers such that $n-a \beta$ is of the order $\beta$. Such terms contribute at the order unity because the summand is of the order $\beta^2/\beta^3$ (the $\beta^2$ in the numerator of this estimate is the overall factor outside the summation in \eqref{zerotempsum}), and the total number of such terms is of the order $\beta$. As the fractional variation between successive terms in the summation here is small, this sum can be well-approximated by an integral. The contribution of these terms yields \eqref{fln}, or, equivalently, \eqref{eq:integral_omega_ZT}.
    \item[(2)] Next, we have integers such that $n-a \beta$ is of the order unity. The summand for such terms is of the order $\beta^2/\beta^2$. As there are order one such terms, their net contribution is also of the order unity. As the fractional variation of successive terms in this sum is not small, the contribution of such terms cannot be approximated by an integral. The contribution of these terms yields \eqref{sln}.
\end{itemize} 
From the physical viewpoint, it is confusing that the zero temperature limit of $\Omega_T(M_T,P_{m\alpha,3})$ retains a reference to the holonomies $\{\lambda_\alpha\}$ (see \eqref{sln}). At zero temperature, time is effectively non-compact, so one should expect the holonomies to be irrelevant. This confusion would have been a sharp one had the form of the function $\Omega_T(M_T,P_{m\alpha,3})$ impacted any physical observable (like the pole mass or the free energy). We have seen in the main text, however, that these physical observables can be evaluated without making any reference to the explicit form of $\Omega_T(M_T,P_{m\alpha,3})$, and yield a physically sensible (in particular, holonomy-independent) zero temperature limit. We conjecture that the holonomy-dependence of $\Omega_T(M_T,P_{m\alpha,3})$ (like its logarithmic-dependence on the cutoff) are gauge-artefacts that will enter no physical, gauge-invariant observables (like the correlators of gauge-invariant operators, or S-matrices; see the discussion in the last paragraph of subsection \ref{subsec:gap_solve_outline}). We leave the important task of  verifying this conjecture to future work.

\subsection{\texorpdfstring{$\Sigma_T$}{Σ\_T} at the second order: the `second prescription' versus computations performed directly at zero temperature }

The fermion self energy has been studied directly at zero temperature on at least two occasions in the previous literature. The zero temperature fermion propagator in the temporal gauge was computed to the second order in the 't Hooft coupling $\lambda$ in appendix B.2 of \cite{Giombi:2011kc}. In the temporal gauge at zero temperature, the gauge boson propagator is proportional to $\dfrac{1}{p_3}$ (here $p_3$ is the component of the momentum in the temporal direction of the gauge boson). In order to perform their computation, the authors of \cite{Giombi:2011kc} needed to give meaning to this propagator at $p_3=0$ (note this is the zero temperature analogue of the region in our parameter space that had finite temperature exclusions, see appendix \ref{careful}). They chose to define this propagator by a principal value prescription,
\begin{equation}\label{pvp}
    \dfrac{1}{p_3} \rightarrow \dfrac{p_3}{p_3^2 + \epsilon^2}\,.
\end{equation}
Using this definition, the authors of \cite{Giombi:2011kc} found that the fermion propagator has a strange (and seemingly physically unacceptable) divergence at the second order in $\lambda$ (see the second term on the RHS of the last line of B.4 in \cite{Giombi:2011kc}). This divergence has its origin in a finite (in $\epsilon$) contribution to the integrals, coming from the neighbourhood of $p_3=0$.

In a separate work, the authors of \cite{Moshe:2014bja} also worked in the temporal gauge. They proceeded to use the zero temperature versions of the finite temperature identities described in subsection \ref{naive} (illustrated more carefully in appendix \ref{careful} and derived generally in \ref{app:omega_iden}) to evaluate the fermion propagator at all the orders in $\lambda$. The beautiful results of \cite{Moshe:2014bja} are in perfect agreement with those obtained within the light-cone gauge, and also agree with the results of this paper (if we take the limit $T \to 0$ after taking the limit $N \to \infty$). For this reason, it seems likely that the final results of \cite{Moshe:2014bja} are correct in some appropriate sense. However, we do not completely understand the derivation of these results. The `symmetrization identities' of \cite{Moshe:2014bja} appear to have been obtained via some naive manipulations (analogous to the manipulations performed in subsection \ref{naive}, i.e. ignoring the effect of exclusions illustrated in \ref{careful}). In particular, the authors of that paper repeatedly use identities like,
\begin{equation}\label{identlike}
    \frac{1}{p_3-r_3} \frac{1} {r_3-s_3} + \frac{1}{p_3-s_3} \frac{1}{s_3-r_3}= \frac{1}{(p_3-r_3)(p_3-s_3)}.
\end{equation} 
While \eqref{identlike} is indeed a true algebraic identity, it is not directly useful for perturbative computations. In order to make sense of the integrals over temporal momenta, we need to supply a definition of the propagator at $p_3=0$. If we adopt the definition \eqref{pvp}, the relevant identity is, 
\begin{equation}\label{identnew}
    \begin{split}
        &\frac{p_3-r_3}{(p_3-r_3)^2 + \epsilon^2} \frac{r_3-s_3}{(r_3-s_3)^2 + \delta^2} + \frac{p_3-s_3}{(p_3-s_3)^2+\epsilon^2} \frac{s_3-r_3}{(s_3-r_3)^2+\delta^2}\\
        &= \frac{(r_3-s_3)^2}{(r_3-s_3)^2 + \delta^2}\left(\frac{(p_3-r_3)(p_3-s_3) - \epsilon^2}{\left((p_3-r_3)^2+\epsilon^2\right)~\left((p_3-s_3)^2+\epsilon^2\right)}\right).
    \end{split}
\end{equation}
The RHS of \eqref{identnew} certainly reduces to the RHS of \eqref{identlike}, when $\epsilon$ and $\delta$ are taken to zero. However, it is easily verified that the integral over $s_3$ and $r_3$ of the term proportional to $\epsilon^2$ in the numerator on the RHS of \eqref{identnew} is actually finite (the $\epsilon^2$ cancels against the part of the integral that turns out to be proportional to $\frac{1}{\epsilon^2}$). Consequently, the $\epsilon^2$ `exclusion' to the symmetrization identity (compare \eqref{identnew} and \eqref{identlike}) results in the leading order corrections to the final result of the integration. Indeed, it is possible to verify that the strange (and unphysical) contribution to the $\lambda^2$ correction to the gauge boson propagator (second term on the RHS of the last line of B.4 in \cite{Giombi:2011kc}) has its origin precisely in the $\epsilon^2$ `exclusion' term described above.

It is possible that some other prescription for regularizing the gauge propagator at zero momentum (see e.g. the discussion in section 3.6 and appendix A1.1\footnote{Note that the analysis of that appendix applies only to the study of the fermion propagator at the order $\lambda$, and not at the order $\lambda^2$, the order at which the confusing ambiguities (as discussed above) arise.} of \cite{Moshe:2014bja}) yields the identities \eqref{identlike} rather than \eqref{identnew}. If this is the case, it would be interesting, both to find such a prescription, and to explain the physical reason for its applicability (e.g. to derive it by taking first the $T\to 0$ and then the $N\to \infty$ limit of the unambiguous finite temperature results of this paper).

\subsection{Summary of the conclusions and confusions}

Let us summarize the discussion of this appendix. The self energy computed by first taking the limit $N \to \infty$, and then taking the limit $T \to 0$ does not agree with the self energy computed directly at zero temperature, using any obvious prescription for smoothing out the singularity in the gauge boson propagator. This is the case both for the principal-value smoothing (the disagreement here is at the first order, i.e. in the computation of $\Omega_0(M_0,p_3)$) and the naive symmetrization procedure of \cite{Moshe:2014bja} (the disagreement here is both at the first and the second orders).\footnote{However, it is possible that these disagreements are all in the `pure gauge' sector, and never show up in any physical observable.}

It is possible, on the other hand, that the `first prescription' of working with the finite temperature formulation developed in this paper (i.e. first taking $T \to 0$ and then taking $N \to \infty$) will actually agree with the beautiful results obtained directly at zero temperature (using naive symmetrization) in \cite{Moshe:2014bja}.\footnote{Even though we have not been able to come up with a clear definition of the zero temperature gauge boson propagator (replacing the principal value prescription) for which these naive manipulations hold.} The fact that \cite{Moshe:2014bja} obtains mass renormalization formula,
\begin{align}\label{gezt}
    M_0 &= M + \lambda M_0,
\end{align}
which is in perfect agreement with zero temperature results within the light-cone gauge, suggests that this might be the case. In other words, it is possible that the self energy reported in \cite{Moshe:2014bja}, i.e.,
\begin{align}
    \Sigma_{0,I}(p_3) &= p_{3} \sin(\lambda\tan^{-1}\left(\frac{p_3}{M_0}\right))+ M_0\cos(\lambda\tan^{-1}\left(\frac{p_3}{M_0}\right)),\label{eq:gap_eq_gen_sol_ZT_I}\\
    \Sigma_{0,3}(p_3) &= p_3 \left(\cos(\lambda\tan^{-1}\left(\frac{p_3}{M_0}\right))-1\right) - M_0\sin(\lambda\tan^{-1}\left(\frac{p_3}{M_0}\right)),\label{eq:gap_eq_gen_sol_ZT_3}
\end{align}
emerges from a careful analysis of the finite temperature path integral, on first taking $T \to 0$, and then taking $N \to \infty$. It would be interesting to investigate this possibility in future work. 


\section{Critical fermion theory}\label{critical} 

In the main text, we have computed the free energy of the so-called `regular fermion' theory, defined by the Euclidean action \eqref{eq:CSm_action} in the `temporal' gauge, and have demonstrated that our final result is in perfect agreement with the previously obtained light-cone gauge results. 

In this appendix, we study the so-called `critical fermion theory' defined by the Euclidean action,
\begin{align}\label{eq:CSm_action_critical_fermion}
    S_{E,{\rm cri}} =& \dfrac{i\kappa}{4\pi}\int_{\mathbb{R}^2 \times S^1} \,\Tr{A\wedge \dd{A}-\dfrac{2i}{3}A\wedge A\wedge A}+\int_{\mathbb{R}^2 \times S^1} \, \bar{\psi}(\slashed{D})\psi\nonumber\\
    &+\int_{\mathbb{R}^2 \times S^1} \left(-\dfrac{4\pi}{\kappa}\zeta\left(\bar{\psi}\psi-\dfrac{\kappa y_2^2}{4\pi}\right)-\dfrac{4\pi y_4}{\kappa}\zeta^2 + \left(\dfrac{2\pi}{\kappa}\right)^2x_6\zeta^3\right),
\end{align}
where $y_2, y_4$ and $x_6$ are coupling constants, and,
\begin{align}
    D_{\dot{\mu}}= \partial_{\dot{\mu}} - i A_{\dot{\mu}}.
\end{align}
Here $\zeta$ is an auxiliary field (we have normalized $\zeta$ so that its expectation value is of the order $N$ on the saddle point). The free energy of the theory \eqref{eq:CSm_action_critical_fermion} can be obtained from a small modification of the analysis of the main text, and once again yields results in perfect agreement with the previous computations performed in the light-cone gauge. In this appendix, we briefly outline the relevant computations, and present our final results. 

The finite temperature gap equations for the theory defined by \eqref{eq:CSm_action_critical_fermion} are,
\begin{align}\label{eq:Gap_eq_sigma_cr}
    \Sigma_{T,{\rm cri}}(P_{m \alpha}) &= -\dfrac{2 \pi}{\kappa \beta} \int \ddm{q} \sum_{\substack{\sigma \\ n \in \mathbb{Z} + \frac{1}{2}}} \dfrac{1}{P_{m \alpha,3}-Q_{n \sigma,3}} H\left(\dfrac{1}{i\slashed{Q}_{n \sigma}-\dfrac{4\pi}{\kappa}\zeta_T+\Sigma_{T,{\rm cri}}(Q_{n \sigma})}\right),
\end{align}
and,
\begin{align}\label{eq:Gap_eq_zeta_cr_intm}
    \dfrac{1}{\beta}\int\ddm{q}\sum_{\substack{\alpha \\ m \in \mathbb{Z} + \frac{1}{2}}} \Tr{\dfrac{1}{i\slashed{Q}_{m\alpha}-\dfrac{4\pi}{\kappa}\zeta_T+\Sigma_{T,{\rm cri}}(Q_{m\alpha,3})}} +\dfrac{\kappa}{4\pi}y_2^2-2y_4\zeta_T+\dfrac{3\pi}{\kappa}x_6\zeta_T^2=0.
\end{align}
\eqref{eq:Gap_eq_sigma_cr} is the analogue of \eqref{gap2} in the main text. \eqref{eq:Gap_eq_zeta_cr_intm} has no analogue in the analysis of the main text: it is the equation of motion with respect to the auxiliary singlet variable $\zeta_T$, which had no counterpart in the regular fermion theory. 

Proceeding exactly along the lines of the analysis of the regular fermion theory, we find that the solution of \eqref{eq:Gap_eq_sigma_cr} is given by,
\begin{align}
    \Sigma_{T,I,{\rm cri}}(P_{m\alpha,3}) &= P_{m\alpha,3} \sin({\dfrac{4 \pi}{\kappa}\Omega_T(M_{T,{\rm cri}},P_{m\alpha,3})})+ M_{T,{\rm cri}}\cos({\dfrac{4 \pi}{\kappa}\Omega_T(M_{T,{\rm cri}},P_{m\alpha,3})}),\label{eq:gap_eq_gen_sol_FT_cr_I}\\
    \Sigma_{T,3,{\rm cri}}(P_{m\alpha,3}) &= P_{m\alpha,3} \left(\cos({\dfrac{4 \pi}{\kappa}\Omega_T(M_{T,{\rm cri}},P_{m\alpha,3})})-1\right) - M_{T,{\rm cri}} \sin({\dfrac{4 \pi}{\kappa}\Omega_T(M_{T,{\rm cri}},P_{m\alpha,3})}).\label{eq:gap_eq_gen_sol_FT_cr_3}
\end{align}
The analogue of the mass gap equation (for the regular fermion theory) \eqref{ge} is the equation,
\begin{align}\label{fge}
    M_{T,{\rm cri}} &= - \dfrac{4 \pi \zeta_T}{\kappa}- \dfrac{4 \pi}{\kappa} \Phi_{T}(M_{T,{\rm cri}}).
\end{align}
Substituting \eqref{eq:gap_eq_gen_sol_FT_cr_I} and \eqref{eq:gap_eq_gen_sol_FT_cr_3} back into the second gap equation \eqref{eq:Gap_eq_zeta_cr_intm}, we find,
\begin{align}
    \dfrac{2}{\beta}\int\ddm{q}\sum_{\substack{\alpha \\ m \in \mathbb{Z} + \frac{1}{2}}} \dfrac{\Sigma_{T,I,{\rm cri}}(Q_{m\alpha,3})}{Q_{m \alpha}^2 + M_{T,{\rm cri}}^2} + \dfrac{\kappa}{4\pi}y_2^2 - 2y_4\zeta_T + \dfrac{3\pi}{\kappa}x_6\zeta_T^2 &=0,
\end{align}
which can be rewritten as,
\begin{align}\label{zeta_ge}
    \Phi_{T}(M_{T,{\rm cri}}) \left( 2 M_{T,{\rm cri}} + \dfrac{4 \pi}{\kappa} \Phi_{T}(M_{T,{\rm cri}}) \right) + \dfrac{\kappa}{4\pi}y_2^2 - 2y_4\zeta_T + \dfrac{3\pi}{\kappa}x_6\zeta_T^2 &=0.
\end{align}
$M_{T,{\rm cri}}$ and the saddle point value of $\zeta_T$ are obtained by simultaneously solving \eqref{fge} and \eqref{zeta_ge}. Note that \eqref{fge} and \eqref{zeta_ge} are in perfect agreement with the light-cone gauge gap equations (3.13) in \cite{Minwalla:2020ysu}.

With $M_{T,{\rm cri}}$ and $\zeta_T$ in hand, the evaluation of the finite temperature free energy is a straightforward exercise. The path integral of our system on $\mathbb{R}^2\times S^1$, with the circumference of $S^1$ given by $\beta=T^{-1}$, is given as an integral of $e^{-S_{T,{\rm cri}}}$ over the holonomy, where,
\begin{align}
    S_{T,{\rm cri}}=& \resizebox{1.05\hsize}{!}{ $\displaystyle - V_2 \sum_{\substack{\alpha \\ n \in \mathbb{Z} + \frac{1}{2}}} \int \ddm{q} \Tr{\ln \left(  i \slashed{Q}_{n\alpha}  - \dfrac{4 \pi}{\kappa} \zeta_T + \Sigma_{T,{\rm cri}}(Q_{n\alpha,3}) \right) -\dfrac{1}{2} \Sigma_{T,{\rm cri}}(Q_{n\alpha,3}) \dfrac{1}{i \slashed{Q}_{n\alpha}  - \dfrac{4 \pi}{\kappa} \zeta_T + \Sigma_{T,{\rm cri}}(Q_{n\alpha,3})}}$ }\nonumber \\
    &+ \beta V_2 \left( y_2^2 \zeta_T - \dfrac{4 \pi y_4}{\kappa} \zeta_T^2 + \left( \dfrac{2 \pi}{\kappa} \right)^2 x_6 \zeta_T^3 \right).
\end{align}
Again, $S_{0,{\rm cri}}$ is defined the same way as for regular fermion theory,
\begin{align}
    S_{0,{\rm cri}} =&- V_2 \beta \lim_{\beta' \to \infty} \Bigg( \dfrac{1}{\beta'} \sum_{\substack{\alpha \\ n \in \mathbb{Z} + \frac{1}{2}}} \int \ddm{q} \Tr\Bigg\{\ln \left(  i \slashed{Q}_{n\alpha}  - \dfrac{4 \pi}{\kappa} \zeta_T + \Sigma_{T,{\rm cri}}(Q_{n\alpha,3}) \right)\nonumber \\
    &-\dfrac{1}{2} \Sigma_{T,{\rm cri}}(Q_{n\alpha,3}) \dfrac{1}{i \slashed{Q}_{n\alpha}  - \dfrac{4 \pi}{\kappa} \zeta_T + \Sigma_{T,{\rm cri}}(Q_{n\alpha,3})} \Bigg\}
     - y_2^2 \zeta_T + \dfrac{4 \pi y_4}{\kappa} \zeta_T^2 - \left( \dfrac{2 \pi}{\kappa} \right)^2 x_6 \zeta_T^3 \Bigg) \nonumber \\
    =& -V_2 \beta \sum_{\alpha} \int \dm{q} \Tr{  \ln \left(  i \slashed{q}  - \dfrac{4 \pi}{\kappa} \zeta_0 + \Sigma_{0,{\rm cri}}(q_{3}) \right) -\dfrac{1}{2} \Sigma_{0,{\rm cri}}(q_{3}) \dfrac{1}{i \slashed{q}  - \dfrac{4 \pi}{\kappa} \zeta_0 + \Sigma_{0,{\rm cri}}(q_{3})}  } \nonumber \\
    & + V_2 \beta \left( y_2^2 \zeta_0 - \dfrac{4 \pi y_4}{\kappa} \zeta_0^2 + \left( \dfrac{2 \pi}{\kappa} \right)^2 x_6 \zeta_0^3 \right) \nonumber \\
    =& -V_2 \beta N \int \dm{q} \Tr{  \ln \left(  i \slashed{q}  - \dfrac{4 \pi}{\kappa} \zeta_0 + \Sigma_{0,{\rm cri}}(q_3) \right) -\dfrac{1}{2} \Sigma_{0,{\rm cri}}(q_3) \dfrac{1}{i \slashed{q}  - \dfrac{4 \pi}{\kappa} \zeta_0 + \Sigma_{0,{\rm cri}}(q_3)}  } \nonumber \\
    & + V_2 \beta \left( y_2^2 \zeta_0 - \dfrac{4 \pi y_4}{\kappa} \zeta_0^2 + \left( \dfrac{2 \pi}{\kappa} \right)^2 x_6 \zeta_0^3 \right),
\end{align}
with zero temperature gap equations given by,
\begin{align}
    M_{0,{\rm cri}} + \dfrac{4 \pi \zeta_0}{\kappa} + \dfrac{4 \pi}{\kappa}  \Phi_{0}(M_{0,{\rm cri}}) &= 0, \\
    \Phi_{0}(M_{0,{\rm cri}}) \left( 2 M_{0,{\rm cri}} + \dfrac{4 \pi}{\kappa} \Phi_{0}(M_{0,{\rm cri}}) \right) + \dfrac{\kappa}{4\pi}y_2^2 - 2y_4\zeta_0 + \dfrac{3\pi}{\kappa}x_6\zeta_0^2 &= 0.
\end{align}
Again, the calculations involved in $S_{T,{\rm cri}} -S_{0,{\rm cri}}$ are the same as that for the regular fermion theory, leading to,
\begin{align} \label{rffinfree}
    S_{T,{\rm cri}} - S_{0,{\rm cri}} =& - \dfrac{V_2}{2\pi \beta^2} \sum_{\mu} \int\limits_{\beta M_{T,{\rm cri}}}^{\infty} \dd{y} \, y \, \ln \left\{\left( 1 + e^{-y + i \lambda_{\mu} )} \right) \left( 1 + e^{-y - i  \lambda_{\mu} )} \right) \right\} \nonumber \\
    &- \dfrac{V_2 N}{6 \pi \beta^2} \left( (\beta M_{0,{\rm cri}})^3 - (\beta M_{T,{\rm cri}})^3 \right) + \dfrac{V_2 N\lambda}{4 \pi \beta^2} (\beta M_{0,{\rm cri}})^3 \left(1 - \dfrac{\lambda}{3} \right) \nonumber \\
    &- \dfrac{V_2 \lambda}{4\pi \beta^2 N} \left(\sum_{\mu} \ln\left|{e^{\beta M_{T,{\rm cri}}}(1 + e^{-\beta M_{T,{\rm cri}} + i \lambda_{\mu}})(1 + e^{- \beta M_{T,{\rm cri}} - i \lambda_{\mu}})}\right| \right)^2  \nonumber \\
    & \times\left( \beta M_{T,{\rm cri}} - \dfrac{\lambda}{3N} \sum_{\mu} \ln\left|{e^{\beta M_{T,{\rm cri}}}(1 + e^{-\beta M_{T,{\rm cri}} + i  \lambda_{\mu}})(1 + e^{- \beta M_{T,{\rm cri}} - i  \lambda_{\mu}})}\right| \right) \nonumber \\
    &+ V_2 \beta \left( y_2^2 \left( \zeta_T - \zeta_0 \right) - \dfrac{4 \pi y_4}{\kappa} \left( \zeta_T^2 - \zeta_0^2 \right) + \left( \dfrac{2 \pi}{\kappa} \right)^2 x_6 \left( \zeta_T^3 - \zeta_0^3 \right) \right),
\end{align}
where $\lambda = \dfrac{N}{\kappa}$ is the 't Hooft coupling.

In terms of the eigenvalue density functional $\rho(\alpha)$ of the holonomy matrix,
\begin{align}\label{eq:holonomy_density_c}
    \rho(\alpha) = \dfrac{1}{N} \sum_{\mu} \delta(\alpha - \lambda_{\mu}),
\end{align}
(\ref{rffinfree}) can be rewritten as,
\begin{align}\label{eq:S_T-S_0_cri}
    S_{T,{\rm cri}} - S_{T,{\rm cri}} &= - \dfrac{V_2 N}{2\pi \beta^2} \int d\alpha ~ \rho(\alpha) \int\limits_{\beta M_{T,{\rm cri}}}^{\infty} \dd{y} \, y \, \ln \left\{\left( 1 + e^{-y + i \alpha )} \right) \left( 1 + e^{-y - i  \alpha )} \right) \right\} \nonumber \\ 
    &- \dfrac{V_2 N}{6 \pi \beta^2} \left( (\beta M_{0,{\rm cri}})^3 - (\beta M_{T,{\rm cri}})^3 \right) + \dfrac{V_2 N\lambda}{4 \pi \beta^2} (\beta M_{0,{\rm cri}})^3 \left(1 - \dfrac{\lambda}{3} \right) \nonumber \\
    &- \dfrac{V_2 \lambda N}{4\pi \beta^2} \left(\int d\alpha ~ \rho(\alpha) \ln\left|{e^{\beta M_{T,{\rm cri}}}(1 + e^{-\beta M_{T,{\rm cri}} + i \alpha})(1 + e^{- \beta M_{T,{\rm cri}} - i \alpha)}}\right| \right)^2  \nonumber \\
    & \times\left( \beta M_{T,{\rm cri}} - \dfrac{\lambda}{3} \int d\alpha ~ \rho(\alpha) \ln\left|{e^{\beta M_{T,{\rm cri}}}(1 + e^{-\beta M_{T,{\rm cri}} + i  \alpha})(1 + e^{- \beta M_{T,{\rm cri}} - i  \alpha})}\right| \right) \nonumber \\
    &+ V_2 \beta \left( y_2^2 \left( \zeta_T - \zeta_0 \right) - \dfrac{4 \pi y_4}{\kappa} \left( \zeta_T^2 - \zeta_0^2 \right) + \left( \dfrac{2 \pi}{\kappa} \right)^2 x_6 \left( \zeta_T^3 - \zeta_0^3 \right) \right).
\end{align}
As in the main text, it is convenient to define an off-shell free energy density $F_{{\rm cri}}(c, \zeta, \Tilde{\mathcal{C}} )$, presented as a function of three auxiliary variables $\Tilde{\mathcal{C}}$, $c$, and $\zeta$, namely,
\begin{align}\label{eq:off-shell_free_energy_cri}
    F_{{\rm cri}}(c,\zeta,\Tilde{\mathcal{C}}) = \dfrac{N}{6 \pi \beta^3} & \Bigg( -8 \lambda^2 \Tilde{\mathcal{C}}^3 -3 \Tilde{\mathcal{C}} \left( c^2 - \left( 2 \lambda \Tilde{\mathcal{C}} - \dfrac{4 \pi \zeta}{\kappa}  \right)^2 \right) +6 \lambda \Tilde{\mathcal{C}}^2 \left( \dfrac{4 \pi \zeta}{\kappa} \right)  + c^3 \nonumber \\
    &- (\beta M_{0,{\rm cri}})^3 + \dfrac{3}{2} \lambda \left( 1- \dfrac{\lambda}{3} \right) (\beta M_{0,{\rm cri}})^3 + \dfrac{6 \pi \zeta }{N} \left( \hat{y}_2^2 - \hat{y}_4 \dfrac{4 \pi \zeta}{\kappa} + x_6 \left(\dfrac{2 \pi \zeta}{\kappa}\right)^2  \right) \nonumber \\
    &- \dfrac{6 \pi \beta^3 \zeta_0 }{N} \left( y_2^2 - y_4 \dfrac{4 \pi \zeta_0}{\kappa} + x_6 \left(\dfrac{2 \pi \zeta_0}{\kappa}\right)^2  \right) \nonumber \\
    &-3 \int \limits_{c}^{\infty} dy ~ y \int d\alpha ~ \rho(\alpha) \ln \left\lvert\left( 1 + e^{-y + i\alpha)} \right) \left( 1 + e^{-y - i\alpha)} \right) \right\rvert \Bigg),
\end{align}
where $\hat{y}_2 = \beta y_2$, and $\hat{y}_4 = \beta y_4$.

Extremizing \eqref{eq:off-shell_free_energy_cri} with respect to $c$, $\Tilde{\mathcal{C}}$, and $\zeta$ gives (respectively),
\begin{align}
    \Tilde{\mathcal{C}} &= \mathcal{C}(c), \\
    c^2 &= \left( 2 \lambda \Tilde{\mathcal{C}} - \dfrac{4 \pi \zeta}{\kappa} \right)^2, \label{eq:c_gap_cri} \\
    0 &= \hat{y}_2^2 + \dfrac{8 \pi \zeta}{\kappa} (2 \lambda \Tilde{\mathcal{C}} - \hat{y}_4) + \dfrac{3 x_6}{4} \left( \dfrac{4 \pi \zeta}{\kappa} \right)^2 - 4 \lambda^2  \Tilde{\mathcal{C}}^2, \label{eq:zeta_gap_cri}
\end{align}
where,
\begin{align}
     \mathcal{C}(c) = \dfrac{1}{2} \int d\alpha ~ \rho(\alpha) \ln \left\lvert e^c \left( 1 + e^{-c + i\alpha)} \right) \left( 1 + e^{-c - i\alpha)} \right) \right\rvert.
\end{align}
From \eqref{fge} and \eqref{eq:c_gap_cri}, one can identify,
\begin{align}
    c&= \beta M_{T,{\rm cri}}, &\text{and},& &\zeta &= \beta \zeta_T.
\end{align}
One can easily verify that,
\begin{align}
    F_{{\rm cri}}\left( \beta M_{T,{\rm cri}}, \beta \zeta_T, \mathcal{C}(\beta M_{T,{\rm cri}}) \right) = \dfrac{S_{T,{\rm cri}} -S_{0,{\rm cri}}}{\beta V_2}.
\end{align}
The off-shell free energy density $ F_{{\rm cri}}(c, \zeta,\Tilde{\mathcal{C}}) $ is in perfect agreement with the light-cone gauge result  (e.g., see equation (3.12) of \cite{Minwalla:2020ysu}). $F_{{\rm cri}}(c, \zeta,\Tilde{\mathcal{C}})$ matches with the previous results, modulo terms that are independent of both temperature and $\{\lambda_\alpha\}$.





\bibliographystyle{JHEP}
\bibliography{Reference_large_N.bib}

@article{minwalla202X:CSMonS2,
      title={Fermionic Chern--Simons theory on $S^2 \times S^1$ at large-$N$ in the `temporal' gauge}, 
      author={Shiraz Minwalla and Souparna Nath and Nikhil Tanwar and Vatsal},
      journal = "\textbf{To Appear}" 
}

@article{Ongoing, 
author = "Halder, Indranil and Janagal, Lavneet  and Minwalla, Shiraz and Patel, Chintan and Prabhakar, Naveen and Radicevic, Djordje",
    title = "{Three Dimensional Bose Fermi Dualities in the non relativistic limit}",
    journal = "\textbf{To Appear}",
    }

@article{tHooft:1974pnl,
    author = "'t Hooft, Gerard",
    title = "{A Two-Dimensional Model for Mesons}",
    reportNumber = "CERN-TH-1820",
    doi = "10.1016/0550-3213(74)90088-1",
    journal = "Nucl. Phys. B",
    volume = "75",
    pages = "461--470",
    year = "1974"
}

@article{Argurio:2018uup,
    author = "Argurio, Riccardo and Bertolini, Matteo and Bigazzi, Francesco and Cotrone, Aldo L. and Niro, Pierluigi",
    title = "{QCD domain walls, Chern-Simons theories and holography}",
    eprint = "1806.08292",
    archivePrefix = "arXiv",
    primaryClass = "hep-th",
    doi = "10.1007/JHEP09(2018)090",
    journal = "JHEP",
    volume = "09",
    pages = "090",
    year = "2018"
}

@article{Argurio:2019tvw,
    author = "Argurio, Riccardo and Bertolini, Matteo and Mignosa, Francesco and Niro, Pierluigi",
    title = "{Charting the phase diagram of QCD$_{3}$}",
    eprint = "1905.01460",
    archivePrefix = "arXiv",
    primaryClass = "hep-th",
    doi = "10.1007/JHEP08(2019)153",
    journal = "JHEP",
    volume = "08",
    pages = "153",
    year = "2019"
}

@article{Argurio:2020her,
    author = "Argurio, Riccardo and Armoni, Adi and Bertolini, Matteo and Mignosa, Francesco and Niro, Pierluigi",
    title = "{Vacuum structure of large $N$ $QCD_{3}$ from holography}",
    eprint = "2006.01755",
    archivePrefix = "arXiv",
    primaryClass = "hep-th",
    doi = "10.1007/JHEP07(2020)134",
    journal = "JHEP",
    volume = "07",
    pages = "134",
    year = "2020"
}

@article{Armoni:2019lgb,
    author = "Armoni, Adi and Dumitrescu, Thomas T. and Festuccia, Guido and Komargodski, Zohar",
    title = "{Metastable vacua in large-N QCD$_{3}$}",
    eprint = "1905.01797",
    archivePrefix = "arXiv",
    primaryClass = "hep-th",
    doi = "10.1007/JHEP01(2020)004",
    journal = "JHEP",
    volume = "01",
    pages = "004",
    year = "2020"
}

@article{Gabai:2022snc,
    author = "Gabai, Barak and Sandor, Joshua and Yin, Xi",
    title = "{Anyon Scattering from Lightcone Hamiltonian: the Singlet Channel}",
    eprint = "2205.09144",
    archivePrefix = "arXiv",
    primaryClass = "hep-th",
    month = "5",
    year = "2022"
}

@article{GurAri:2012is,
	Archiveprefix = {arXiv},
	Author = {Gur-Ari, Guy and Yacoby, Ran},
	Doi = {10.1007/JHEP02(2013)150},
	Eprint = {1211.1866},
	Journal = {JHEP},
	Pages = {150},
	Primaryclass = {hep-th},
	Slaccitation = {%%CITATION = ARXIV:1211.1866;%%},
	Title = {{Correlators of Large N Fermionic Chern-Simons Vector Models}},
	Volume = {1302},
	Year = {2013}}

@article{Gabai:2022vri,
    author = "Gabai, Barak and Sever, Amit and Zhong, De-liang",
    title = "{Line Operators in Chern-Simons\textendash{}Matter Theories and Bosonization in Three Dimensions}",
    eprint = "2204.05262",
    archivePrefix = "arXiv",
    primaryClass = "hep-th",
    doi = "10.1103/PhysRevLett.129.121604",
    journal = "Phys. Rev. Lett.",
    volume = "129",
    number = "12",
    pages = "121604",
    year = "2022"
}

@article{Maldacena:2011jn,
    author = "Maldacena, Juan and Zhiboedov, Alexander",
    title = "{Constraining Conformal Field Theories with A Higher Spin Symmetry}",
    eprint = "1112.1016",
    archivePrefix = "arXiv",
    primaryClass = "hep-th",
    doi = "10.1088/1751-8113/46/21/214011",
    journal = "J. Phys. A",
    volume = "46",
    pages = "214011",
    year = "2013"
}

@article{Maldacena:2012sf,
    author = "Maldacena, Juan and Zhiboedov, Alexander",
    title = "{Constraining conformal field theories with a slightly broken higher spin symmetry}",
    eprint = "1204.3882",
    archivePrefix = "arXiv",
    primaryClass = "hep-th",
    reportNumber = "PUPT-2410",
    doi = "10.1088/0264-9381/30/10/104003",
    journal = "Class. Quant. Grav.",
    volume = "30",
    pages = "104003",
    year = "2013"
}

@article{Aharony_2008,
	doi = {10.1088/1126-6708/2008/10/091},
  
	url = {https://doi.org/10.1088%2F1126-6708%2F2008%2F10%2F091},
  
	year = 2008,
	month = {oct},
  
	publisher = {Springer Science and Business Media {LLC}
},
  
	volume = {2008},
  
	number = {10},
  
	pages = {091--091},
  
	author = {Ofer Aharony and Oren Bergman and Daniel Louis Jafferis and Juan Maldacena},
  
	title = {$ \mathcal{N} = 6$ superconformal Chern-Simons-matter theories, M2-branes and their gravity duals},
  
	journal = {Journal of High Energy Physics}
}

@article{Jain:2022ajd,
    author = "Jain, Prabhav and Jain, Sachin and Sahoo, Bibhut and Dhruva, K. S. and Zade, Aashna",
    title = "{Mapping Slightly Broken Higher Spin (SBHS) theory correlators to Free theory correlators: A momentum space bootstrap using SBHS symmetry}",
    eprint = "2207.05101",
    archivePrefix = "arXiv",
    primaryClass = "hep-th",
    month = "7",
    year = "2022"
}

@article{Skvortsov:2018uru,
    author = "Skvortsov, Evgeny",
    title = "{Light-Front Bootstrap for Chern-Simons Matter Theories}",
    eprint = "1811.12333",
    archivePrefix = "arXiv",
    primaryClass = "hep-th",
    doi = "10.1007/JHEP06(2019)058",
    journal = "JHEP",
    volume = "06",
    pages = "058",
    year = "2019"
}

@article{Charan:2017jyc,
	Archiveprefix = {arXiv},
	Author = {Charan, V. Guru and Prakash, Shiroman},
	Doi = {10.1007/JHEP02(2018)094},
	Eprint = {1711.11300},
	Journal = {JHEP},
	Pages = {094},
	Primaryclass = {hep-th},
	Slaccitation = {%%CITATION = ARXIV:1711.11300;%%},
	Title = {{On the Higher Spin Spectrum of Chern-Simons Theory coupled to Fermions in the Large Flavour Limit}},
	Volume = {02},
	Year = {2018}}

@article{Gandhi:2021gwn,
    author = "Gandhi, Yatharth and Jain, Sachin and John, Renjan Rajan",
    title = "{Anyonic correlation functions in Chern-Simons matter theories}",
    eprint = "2106.09043",
    archivePrefix = "arXiv",
    primaryClass = "hep-th",
    doi = "10.1103/PhysRevD.106.046014",
    journal = "Phys. Rev. D",
    volume = "106",
    number = "4",
    pages = "046014",
    year = "2022"
}

@article{Jain:2021vrv,
    author = "Jain, Sachin and John, Renjan Rajan and Mehta, Abhishek and Nizami, Amin A. and Suresh, Adithya",
    title = "{Higher spin 3-point functions in 3d CFT using spinor-helicity variables}",
    eprint = "2106.00016",
    archivePrefix = "arXiv",
    primaryClass = "hep-th",
    doi = "10.1007/JHEP09(2021)041",
    journal = "JHEP",
    volume = "09",
    pages = "041",
    year = "2021"
}

@article{Jain:2021wyn,
    author = "Jain, Sachin and John, Renjan Rajan and Mehta, Abhishek and Nizami, Amin A. and Suresh, Adithya",
    title = "{Momentum space parity-odd CFT 3-point functions}",
    eprint = "2101.11635",
    archivePrefix = "arXiv",
    primaryClass = "hep-th",
    doi = "10.1007/JHEP08(2021)089",
    journal = "JHEP",
    volume = "08",
    pages = "089",
    year = "2021"
}

@article{Mishra:2020wos,
    author = "Mishra, Amiya",
    title = "{On thermal correlators and bosonization duality in Chern-Simons theories with massive fundamental matter}",
    eprint = "2010.03699",
    archivePrefix = "arXiv",
    primaryClass = "hep-th",
    reportNumber = "TIFR/TH/20-38",
    doi = "10.1007/JHEP01(2021)109",
    journal = "JHEP",
    volume = "01",
    pages = "109",
    year = "2021"
}

@article{Jain:2020puw,
    author = "Jain, Sachin and John, Renjan Rajan and Malvimat, Vinay",
    title = "{Constraining momentum space correlators using slightly broken higher spin symmetry}",
    eprint = "2008.08610",
    archivePrefix = "arXiv",
    primaryClass = "hep-th",
    doi = "10.1007/JHEP04(2021)231",
    journal = "JHEP",
    volume = "04",
    pages = "231",
    year = "2021"
}

@article{Minwalla:2020ysu,
    author = "Minwalla, Shiraz and Mishra, Amiya and Prabhakar, Naveen",
    title = "{Fermi seas from Bose condensates in Chern-Simons matter theories and a bosonic exclusion principle}",
    eprint = "2008.00024",
    archivePrefix = "arXiv",
    primaryClass = "hep-th",
    reportNumber = "TIFR/TH/20-22",
    doi = "10.1007/JHEP11(2020)171",
    journal = "JHEP",
    volume = "11",
    pages = "171",
    year = "2020"
}

@article{Jain:2020rmw,
    author = "Jain, Sachin and John, Renjan Rajan and Malvimat, Vinay",
    title = "{Momentum space spinning correlators and higher spin equations in three dimensions}",
    eprint = "2005.07212",
    archivePrefix = "arXiv",
    primaryClass = "hep-th",
    doi = "10.1007/JHEP11(2020)049",
    journal = "JHEP",
    volume = "11",
    pages = "049",
    year = "2020"
}

@article{Inbasekar:2020hla,
    author = "Inbasekar, Karthik and Janagal, Lavneet and Shukla, Ashish",
    title = "{Scattering Amplitudes in $\mathcal{N} = 3$ Supersymmetric $SU(N)$ Chern-Simons-Matter Theory at Large $N$}",
    eprint = "2001.02363",
    archivePrefix = "arXiv",
    primaryClass = "hep-th",
    doi = "10.1007/JHEP04(2020)101",
    journal = "JHEP",
    volume = "04",
    pages = "101",
    year = "2020"
}

@article{Ghosh:2019sqf,
    author = "Ghosh, Sudip and Mazumdar, Subhajit",
    title = "{Thermal correlators and bosonization dualities in large N Chern-Simons matter theories}",
    eprint = "1912.06589",
    archivePrefix = "arXiv",
    primaryClass = "hep-th",
    doi = "10.1007/JHEP02(2023)042",
    journal = "JHEP",
    volume = "02",
    pages = "042",
    year = "2023"
}

@article{Kalloor:2019xjb,
    author = "Kalloor, Rohit R.",
    title = "{Four-point functions in large $N$ Chern-Simons fermionic theories}",
    eprint = "1910.14617",
    archivePrefix = "arXiv",
    primaryClass = "hep-th",
    doi = "10.1007/JHEP10(2020)028",
    journal = "JHEP",
    volume = "10",
    pages = "028",
    year = "2020"
}

@article{Jensen:2019mga,
    author = "Jensen, Kristan and Patil, Priti",
    title = "{Chern-Simons dualities with multiple flavors at large $N$}",
    eprint = "1910.07484",
    archivePrefix = "arXiv",
    primaryClass = "hep-th",
    doi = "10.1007/JHEP12(2019)043",
    journal = "JHEP",
    volume = "12",
    pages = "043",
    year = "2019"
}

@article{Inbasekar:2019azv,
    author = "Inbasekar, Karthik and Janagal, Lavneet and Shukla, Ashish",
    title = "{Mass-deformed $N=3$ supersymmetric Chern-Simons-matter theory}",
    eprint = "1908.08119",
    archivePrefix = "arXiv",
    primaryClass = "hep-th",
    reportNumber = "TIFR/TH/19-29",
    doi = "10.1103/PhysRevD.100.085008",
    journal = "Phys. Rev. D",
    volume = "100",
    number = "8",
    pages = "085008",
    year = "2019"
}

@article{Inbasekar:2019wdw,
    author = "Inbasekar, Karthik and Jain, Sachin and Malvimat, Vinay and Mehta, Abhishek and Nayak, Pranjal and Sharma, Tarun",
    title = "{Correlation functions in ${\cal N}=2$ Supersymmetric vector matter Chern-Simons theory}",
    eprint = "1907.11722",
    archivePrefix = "arXiv",
    primaryClass = "hep-th",
    reportNumber = "TAUP-3036\textbackslash{}19, TAUP-3036-19",
    doi = "10.1007/JHEP04(2020)207",
    journal = "JHEP",
    volume = "04",
    pages = "207",
    year = "2020"
}

@article{Jain:2019fja,
    author = "Jain, Sachin and Malvimat, Vinay and Mehta, Abhishek and Prakash, Shiroman and Sudhir, Nidhi",
    title = "{All order exact result for the anomalous dimension of the scalar primary in Chern-Simons vector models}",
    eprint = "1906.06342",
    archivePrefix = "arXiv",
    primaryClass = "hep-th",
    doi = "10.1103/PhysRevD.101.126017",
    journal = "Phys. Rev. D",
    volume = "101",
    number = "12",
    pages = "126017",
    year = "2020"
}

@article{Li:2019twz,
    author = "Li, Zhijin",
    title = "{Bootstrapping conformal four-point correlators with slightly broken higher spin symmetry and $3D$ bosonization}",
    eprint = "1906.05834",
    archivePrefix = "arXiv",
    primaryClass = "hep-th",
    doi = "10.1007/JHEP10(2020)007",
    journal = "JHEP",
    volume = "10",
    pages = "007",
    year = "2020"
}

@article{Aharony:2019mbc,
    author = "Aharony, Ofer and Sharon, Adar",
    title = "{Large N renormalization group flows in 3d $ \mathcal{N} $ = 1 Chern-Simons-Matter theories}",
    eprint = "1905.07146",
    archivePrefix = "arXiv",
    primaryClass = "hep-th",
    doi = "10.1007/JHEP07(2019)160",
    journal = "JHEP",
    volume = "07",
    pages = "160",
    year = "2019"
}

@article{Halder:2019foo,
    author = "Halder, Indranil and Minwalla, Shiraz",
    title = "{Matter Chern Simons Theories in a Background Magnetic Field}",
    eprint = "1904.07885",
    archivePrefix = "arXiv",
    primaryClass = "hep-th",
    doi = "10.1007/JHEP11(2019)089",
    journal = "JHEP",
    volume = "11",
    pages = "089",
    year = "2019"
}

@article{Dey:2019ihe,
    author = "Dey, Anshuman and Halder, Indranil and Jain, Sachin and Minwalla, Shiraz and Prabhakar, Naveen",
    title = "{The large N phase diagram of $ \mathcal{N} $ = 2 SU(N) Chern-Simons theory with one fundamental chiral multiplet}",
    eprint = "1904.07286",
    archivePrefix = "arXiv",
    primaryClass = "hep-th",
    reportNumber = "TIFR/TH/19-10",
    doi = "10.1007/JHEP11(2019)113",
    journal = "JHEP",
    volume = "11",
    pages = "113",
    year = "2019"
}

@article{Chattopadhyay:2019lpr,
    author = "Chattopadhyay, Arghya and Suvankar, Dutta and Neetu",
    title = "{Chern-Simons Theory on Seifert Manifold and Matrix Model}",
    eprint = "1902.07538",
    archivePrefix = "arXiv",
    primaryClass = "hep-th",
    doi = "10.1103/PhysRevD.100.126009",
    journal = "Phys. Rev. D",
    volume = "100",
    number = "12",
    pages = "126009",
    year = "2019"
}

@article{Dey:2018ykx,
    author = "Dey, Anshuman and Halder, Indranil and Jain, Sachin and Janagal, Lavneet and Minwalla, Shiraz and Prabhakar, Naveen",
    title = "{Duality and an exact Landau-Ginzburg potential for quasi-bosonic Chern-Simons-Matter theories}",
    eprint = "1808.04415",
    archivePrefix = "arXiv",
    primaryClass = "hep-th",
    reportNumber = "TIFR/TH/18-26",
    doi = "10.1007/JHEP11(2018)020",
    journal = "JHEP",
    volume = "11",
    pages = "020",
    year = "2018"
}

@article{Aharony:2018pjn,
    author = "Aharony, Ofer and Jain, Sachin and Minwalla, Shiraz",
    title = "{Flows, Fixed Points and Duality in Chern-Simons-matter theories}",
    eprint = "1808.03317",
    archivePrefix = "arXiv",
    primaryClass = "hep-th",
    doi = "10.1007/JHEP12(2018)058",
    journal = "JHEP",
    volume = "12",
    pages = "058",
    year = "2018"
}

@article{Aitken:2018cvh,
    author = "Aitken, Kyle and Baumgartner, Andrew and Karch, Andreas",
    title = "{Novel 3d bosonic dualities from bosonization and holography}",
    eprint = "1807.01321",
    archivePrefix = "arXiv",
    primaryClass = "hep-th",
    doi = "10.1007/JHEP09(2018)003",
    journal = "JHEP",
    volume = "09",
    pages = "003",
    year = "2018"
}

@article{Yacoby:2018yvy,
    author = "Yacoby, Ran",
    title = "{Scalar Correlators in Bosonic Chern-Simons Vector Models}",
    eprint = "1805.11627",
    archivePrefix = "arXiv",
    primaryClass = "hep-th",
    month = "5",
    year = "2018"
}

@article{Aharony:2018npf,
    author = "Aharony, Ofer and Alday, Luis F. and Bissi, Agnese and Yacoby, Ran",
    title = "{The Analytic Bootstrap for Large $N$ Chern-Simons Vector Models}",
    eprint = "1805.04377",
    archivePrefix = "arXiv",
    primaryClass = "hep-th",
    doi = "10.1007/JHEP08(2018)166",
    journal = "JHEP",
    volume = "08",
    pages = "166",
    year = "2018"
}

@article{Karch:2018mer,
    author = "Karch, Andreas and Tong, David and Turner, Carl",
    title = "{Mirror Symmetry and Bosonization in 2d and 3d}",
    eprint = "1805.00941",
    archivePrefix = "arXiv",
    primaryClass = "hep-th",
    doi = "10.1007/JHEP07(2018)059",
    journal = "JHEP",
    volume = "07",
    pages = "059",
    year = "2018"
}

@article{Turiaci:2018nua,
    author = "Turiaci, Gustavo J. and Zhiboedov, Alexander",
    title = "{Veneziano Amplitude of Vasiliev Theory}",
    eprint = "1802.04390",
    archivePrefix = "arXiv",
    primaryClass = "hep-th",
    doi = "10.1007/JHEP10(2018)034",
    journal = "JHEP",
    volume = "10",
    pages = "034",
    year = "2018"
}

@article{Chattopadhyay:2018wkp,
    author = "Chattopadhyay, Arghya and Dutta, Parikshit and Dutta, Suvankar",
    title = "{From Phase Space to Integrable Representations and Level-Rank Duality}",
    eprint = "1801.07901",
    archivePrefix = "arXiv",
    primaryClass = "hep-th",
    doi = "10.1007/JHEP05(2018)117",
    journal = "JHEP",
    volume = "05",
    pages = "117",
    year = "2018"
}

@article{Jensen:2017bjo,
    author = "Jensen, Kristan",
    title = "{A master bosonization duality}",
    eprint = "1712.04933",
    archivePrefix = "arXiv",
    primaryClass = "hep-th",
    doi = "10.1007/JHEP01(2018)031",
    journal = "JHEP",
    volume = "01",
    pages = "031",
    year = "2018"
}

@article{Aitken:2017nfd,
    author = "Aitken, Kyle and Baumgartner, Andrew and Karch, Andreas and Robinson, Brandon",
    title = "{3d Abelian Dualities with Boundaries}",
    eprint = "1712.02801",
    archivePrefix = "arXiv",
    primaryClass = "hep-th",
    doi = "10.1007/JHEP03(2018)053",
    journal = "JHEP",
    volume = "03",
    pages = "053",
    year = "2018"
}

@article{Benini:2017aed,
    author = "Benini, Francesco",
    title = "{Three-dimensional dualities with bosons and fermions}",
    eprint = "1712.00020",
    archivePrefix = "arXiv",
    primaryClass = "hep-th",
    reportNumber = "SISSA-57-2017-FISI, SISSA 57/2017/FISI",
    doi = "10.1007/JHEP02(2018)068",
    journal = "JHEP",
    volume = "02",
    pages = "068",
    year = "2018"
}

@article{Cordova:2017vab,
    author = "Cordova, Clay and Hsin, Po-Shen and Seiberg, Nathan",
    title = "{Global Symmetries, Counterterms, and Duality in Chern-Simons Matter Theories with Orthogonal Gauge Groups}",
    eprint = "1711.10008",
    archivePrefix = "arXiv",
    primaryClass = "hep-th",
    doi = "10.21468/SciPostPhys.4.4.021",
    journal = "SciPost Phys.",
    volume = "4",
    number = "4",
    pages = "021",
    year = "2018"
}

@article{Inbasekar:2017sqp,
    author = "Inbasekar, Karthik and Jain, Sachin and Majumdar, Sucheta and Nayak, Pranjal and Neogi, Turmoli and Sharma, Tarun and Sinha, Ritam and Umesh, V.",
    title = "{Dual superconformal symmetry of $ \mathcal{N} $ = 2 Chern-Simons theory with fundamental matter at large N}",
    eprint = "1711.02672",
    archivePrefix = "arXiv",
    primaryClass = "hep-th",
    reportNumber = "TAUP-3027-17, TIFR-TH-17-42, TAUP-3027/17, TIFR/TH/17-42",
    doi = "10.1007/JHEP06(2019)016",
    journal = "JHEP",
    volume = "06",
    pages = "016",
    year = "2019"
}

@article{Inbasekar:2017ieo,
    author = "Inbasekar, Karthik and Jain, Sachin and Nayak, Pranjal and Umesh, V.",
    title = "{All tree level scattering amplitudes in Chern-Simons theories with fundamental matter}",
    eprint = "1710.04227",
    archivePrefix = "arXiv",
    primaryClass = "hep-th",
    reportNumber = "TAUP-3026/17, TIFR/TH/17-30",
    doi = "10.1103/PhysRevLett.121.161601",
    journal = "Phys. Rev. Lett.",
    volume = "121",
    number = "16",
    pages = "161601",
    year = "2018"
}

@article{Gomis:2017ixy,
    author = "Gomis, Jaume and Komargodski, Zohar and Seiberg, Nathan",
    title = "{Phases Of Adjoint QCD$_3$ And Dualities}",
    eprint = "1710.03258",
    archivePrefix = "arXiv",
    primaryClass = "hep-th",
    doi = "10.21468/SciPostPhys.5.1.007",
    journal = "SciPost Phys.",
    volume = "5",
    number = "1",
    pages = "007",
    year = "2018"
}

@article{Jensen:2017xbs,
    author = "Jensen, Kristan and Karch, Andreas",
    title = "{Embedding three-dimensional bosonization dualities into string theory}",
    eprint = "1709.07872",
    archivePrefix = "arXiv",
    primaryClass = "hep-th",
    doi = "10.1007/JHEP12(2017)031",
    journal = "JHEP",
    volume = "12",
    pages = "031",
    year = "2017"
}

@article{Jensen:2017dso,
    author = "Jensen, Kristan and Karch, Andreas",
    title = "{Bosonizing three-dimensional quiver gauge theories}",
    eprint = "1709.01083",
    archivePrefix = "arXiv",
    primaryClass = "hep-th",
    doi = "10.1007/JHEP11(2017)018",
    journal = "JHEP",
    volume = "11",
    pages = "018",
    year = "2017"
}

@article{Gaiotto:2017tne,
    author = "Gaiotto, Davide and Komargodski, Zohar and Seiberg, Nathan",
    title = "{Time-reversal breaking in QCD$_{4}$, walls, and dualities in 2 + 1 dimensions}",
    eprint = "1708.06806",
    archivePrefix = "arXiv",
    primaryClass = "hep-th",
    doi = "10.1007/JHEP01(2018)110",
    journal = "JHEP",
    volume = "01",
    pages = "110",
    year = "2018"
}

@article{Giombi:2017txg,
    author = "Giombi, Simone",
    title = "{Testing the Boson/Fermion Duality on the Three-Sphere}",
    eprint = "1707.06604",
    archivePrefix = "arXiv",
    primaryClass = "hep-th",
    reportNumber = "PUPT-2530",
    month = "7",
    year = "2017"
}

@article{Komargodski:2017keh,
    author = "Komargodski, Zohar and Seiberg, Nathan",
    title = "{A symmetry breaking scenario for QCD$_{3}$}",
    eprint = "1706.08755",
    archivePrefix = "arXiv",
    primaryClass = "hep-th",
    doi = "10.1007/JHEP01(2018)109",
    journal = "JHEP",
    volume = "01",
    pages = "109",
    year = "2018"
}

@article{Nosaka:2017ohr,
    author = "Nosaka, Tomoki and Yokoyama, Shuichi",
    title = "{Complete factorization in minimal $ \mathcal{N}=4 $ Chern-Simons-matter theory}",
    eprint = "1706.07234",
    archivePrefix = "arXiv",
    primaryClass = "hep-th",
    reportNumber = "KIAS-P17045, YITP-17-64",
    doi = "10.1007/JHEP01(2018)001",
    journal = "JHEP",
    volume = "01",
    pages = "001",
    year = "2018"
}

@article{Sezgin:2017jgm,
    author = "Sezgin, Ergin and Skvortsov, Evgeny D. and Zhu, Yaodong",
    title = "{Chern-Simons Matter Theories and Higher Spin Gravity}",
    eprint = "1705.03197",
    archivePrefix = "arXiv",
    primaryClass = "hep-th",
    doi = "10.1007/JHEP07(2017)133",
    journal = "JHEP",
    volume = "07",
    pages = "133",
    year = "2017"
}

@article{Benini:2017dus,
    author = "Benini, Francesco and Hsin, Po-Shen and Seiberg, Nathan",
    title = "{Comments on global symmetries, anomalies, and duality in (2 + 1)d}",
    eprint = "1702.07035",
    archivePrefix = "arXiv",
    primaryClass = "cond-mat.str-el",
    reportNumber = "SISSA-06-2017-FISI",
    doi = "10.1007/JHEP04(2017)135",
    journal = "JHEP",
    volume = "04",
    pages = "135",
    year = "2017"
}

@article{Giombi:2017rhm,
    author = "Giombi, S. and Kirilin, V. and Skvortsov, E.",
    title = "{Notes on Spinning Operators in Fermionic CFT}",
    eprint = "1701.06997",
    archivePrefix = "arXiv",
    primaryClass = "hep-th",
    reportNumber = "PUPT-2517, LMU-ASC-05-17",
    doi = "10.1007/JHEP05(2017)041",
    journal = "JHEP",
    volume = "05",
    pages = "041",
    year = "2017"
}

@article{Aharony:2016jvv,
    author = "Aharony, Ofer and Benini, Francesco and Hsin, Po-Shen and Seiberg, Nathan",
    title = "{Chern-Simons-matter dualities with $SO$ and $USp$ gauge groups}",
    eprint = "1611.07874",
    archivePrefix = "arXiv",
    primaryClass = "cond-mat.str-el",
    reportNumber = "SISSA-62-2016-FISI",
    doi = "10.1007/JHEP02(2017)072",
    journal = "JHEP",
    volume = "02",
    pages = "072",
    year = "2017"
}

@article{Wadia:2016zpd,
    author = "Wadia, Spenta R.",
    title = "{Chern\textendash{}Simons theories with fundamental matter: A brief review of large $N$ results including Fermi\textendash{}Bose duality and the S-matrix}",
    doi = "10.1142/S0217751X16300520",
    journal = "Int. J. Mod. Phys. A",
    volume = "31",
    number = "32",
    pages = "1630052",
    year = "2016"
}

@article{Giombi:2016zwa,
    author = "Giombi, S. and Gurucharan, V. and Kirilin, V. and Prakash, S. and Skvortsov, E.",
    title = "{On the Higher-Spin Spectrum in Large N Chern-Simons Vector Models}",
    eprint = "1610.08472",
    archivePrefix = "arXiv",
    primaryClass = "hep-th",
    reportNumber = "PUPT-2512, LMU-ASC-52-16",
    doi = "10.1007/JHEP01(2017)058",
    journal = "JHEP",
    volume = "01",
    pages = "058",
    year = "2017"
}

@article{Karch:2016aux,
    author = "Karch, Andreas and Robinson, Brandon and Tong, David",
    title = "{More Abelian Dualities in 2+1 Dimensions}",
    eprint = "1609.04012",
    archivePrefix = "arXiv",
    primaryClass = "hep-th",
    doi = "10.1007/JHEP01(2017)017",
    journal = "JHEP",
    volume = "01",
    pages = "017",
    year = "2017"
}

@article{Radicevic:2016wqn,
    author = "Radi\v{c}evi\'c, \DH{}or\dj{}e and Tong, David and Turner, Carl",
    title = "{Non-Abelian 3d Bosonization and Quantum Hall States}",
    eprint = "1608.04732",
    archivePrefix = "arXiv",
    primaryClass = "hep-th",
    doi = "10.1007/JHEP12(2016)067",
    journal = "JHEP",
    volume = "12",
    pages = "067",
    year = "2016"
}

@article{Hsin:2016blu,
    author = "Hsin, Po-Shen and Seiberg, Nathan",
    title = "{Level/rank Duality and Chern-Simons-Matter Theories}",
    eprint = "1607.07457",
    archivePrefix = "arXiv",
    primaryClass = "hep-th",
    doi = "10.1007/JHEP09(2016)095",
    journal = "JHEP",
    volume = "09",
    pages = "095",
    year = "2016"
}

@inproceedings{Giombi:2016ejx,
    author = "Giombi, Simone",
    title = "{Higher Spin \textemdash{} CFT Duality}",
    booktitle = "{Theoretical Advanced Study Institute in Elementary Particle Physics}: {New Frontiers in Fields and Strings}",
    eprint = "1607.02967",
    archivePrefix = "arXiv",
    primaryClass = "hep-th",
    doi = "10.1142/9789813149441_0003",
    pages = "137--214",
    year = "2017"
}

@article{Karch:2016sxi,
    author = "Karch, Andreas and Tong, David",
    title = "{Particle-Vortex Duality from 3d Bosonization}",
    eprint = "1606.01893",
    archivePrefix = "arXiv",
    primaryClass = "hep-th",
    doi = "10.1103/PhysRevX.6.031043",
    journal = "Phys. Rev. X",
    volume = "6",
    number = "3",
    pages = "031043",
    year = "2016"
}

@article{Murugan:2016zal,
    author = "Murugan, Jeff and Nastase, Horatiu",
    title = "{Particle-vortex duality in topological insulators and superconductors}",
    eprint = "1606.01912",
    archivePrefix = "arXiv",
    primaryClass = "hep-th",
    doi = "10.1007/JHEP05(2017)159",
    journal = "JHEP",
    volume = "05",
    pages = "159",
    year = "2017"
}

@article{Seiberg:2016gmd,
    author = "Seiberg, Nathan and Senthil, T. and Wang, Chong and Witten, Edward",
    title = "{A Duality Web in 2+1 Dimensions and Condensed Matter Physics}",
    eprint = "1606.01989",
    archivePrefix = "arXiv",
    primaryClass = "hep-th",
    doi = "10.1016/j.aop.2016.08.007",
    journal = "Annals Phys.",
    volume = "374",
    pages = "395--433",
    year = "2016"
}

@article{Gur-Ari:2016xff,
    author = "Gur-Ari, Guy and Hartnoll, Sean A. and Mahajan, Raghu",
    title = "{Transport in Chern-Simons-Matter Theories}",
    eprint = "1605.01122",
    archivePrefix = "arXiv",
    primaryClass = "hep-th",
    reportNumber = "SU-ITP-16-09",
    doi = "10.1007/JHEP07(2016)090",
    journal = "JHEP",
    volume = "07",
    pages = "090",
    year = "2016"
}

@article{Yokoyama:2016sbx,
    author = "Yokoyama, Shuichi",
    title = "{Scattering Amplitude and Bosonization Duality in General Chern-Simons Vector Models}",
    eprint = "1604.01897",
    archivePrefix = "arXiv",
    primaryClass = "hep-th",
    reportNumber = "YITP-16-49",
    doi = "10.1007/JHEP09(2016)105",
    journal = "JHEP",
    volume = "09",
    pages = "105",
    year = "2016"
}

@article{Aharony:2015mjs,
    author = "Aharony, Ofer",
    title = "{Baryons, monopoles and dualities in Chern-Simons-matter theories}",
    eprint = "1512.00161",
    archivePrefix = "arXiv",
    primaryClass = "hep-th",
    reportNumber = "WIS-12-15-NOV-DPPA",
    doi = "10.1007/JHEP02(2016)093",
    journal = "JHEP",
    volume = "02",
    pages = "093",
    year = "2016"
}

@article{Geracie:2015drf,
    author = "Geracie, Michael and Goykhman, Mikhail and Son, Dam T.",
    title = "{Dense Chern-Simons Matter with Fermions at Large N}",
    eprint = "1511.04772",
    archivePrefix = "arXiv",
    primaryClass = "hep-th",
    doi = "10.1007/JHEP04(2016)103",
    journal = "JHEP",
    volume = "04",
    pages = "103",
    year = "2016"
}

@article{Radicevic:2015yla,
    author = "Radi\v{c}evi\'c, \DJ{}or\dj{}e",
    title = "{Disorder Operators in Chern-Simons-Fermion Theories}",
    eprint = "1511.01902",
    archivePrefix = "arXiv",
    primaryClass = "hep-th",
    reportNumber = "SU-ITP-15-16",
    doi = "10.1007/JHEP03(2016)131",
    journal = "JHEP",
    volume = "03",
    pages = "131",
    year = "2016"
}

@article{Minwalla:2015sca,
    author = "Minwalla, Shiraz and Yokoyama, Shuichi",
    title = "{Chern Simons Bosonization along RG Flows}",
    eprint = "1507.04546",
    archivePrefix = "arXiv",
    primaryClass = "hep-th",
    reportNumber = "TIFR-TH-15-19",
    doi = "10.1007/JHEP02(2016)103",
    journal = "JHEP",
    volume = "02",
    pages = "103",
    year = "2016"
}

@article{Gur-Ari:2015pca,
    author = "Gur-Ari, Guy and Yacoby, Ran",
    title = "{Three Dimensional Bosonization From Supersymmetry}",
    eprint = "1507.04378",
    archivePrefix = "arXiv",
    primaryClass = "hep-th",
    reportNumber = "PUPT-2482",
    doi = "10.1007/JHEP11(2015)013",
    journal = "JHEP",
    volume = "11",
    pages = "013",
    year = "2015"
}

@article{Bedhotiya:2015uga,
    author = "Bedhotiya, Akshay and Prakash, Shiroman",
    title = "{A test of bosonization at the level of four-point functions in Chern-Simons vector models}",
    eprint = "1506.05412",
    archivePrefix = "arXiv",
    primaryClass = "hep-th",
    doi = "10.1007/JHEP12(2015)032",
    journal = "JHEP",
    volume = "12",
    pages = "032",
    year = "2015"
}

@article{Inbasekar:2015tsa,
    author = "Inbasekar, Karthik and Jain, Sachin and Mazumdar, Subhajit and Minwalla, Shiraz and Umesh, V. and Yokoyama, Shuichi",
    title = "{Unitarity, crossing symmetry and duality in the scattering of $ \mathcal{N}=1 $ susy matter Chern-Simons theories}",
    eprint = "1505.06571",
    archivePrefix = "arXiv",
    primaryClass = "hep-th",
    reportNumber = "TIFR-TH-15-15",
    doi = "10.1007/JHEP10(2015)176",
    journal = "JHEP",
    volume = "10",
    pages = "176",
    year = "2015"
}

@article{Aharony:2015pla,
    author = "Aharony, Ofer and Narayan, Prithvi and Sharma, Tarun",
    title = "{On monopole operators in supersymmetric Chern-Simons-matter theories}",
    eprint = "1502.00945",
    archivePrefix = "arXiv",
    primaryClass = "hep-th",
    reportNumber = "WIS-10-14-DEC-DPPA",
    doi = "10.1007/JHEP05(2015)117",
    journal = "JHEP",
    volume = "05",
    pages = "117",
    year = "2015"
}

@article{Moshe:2014bja,
    author = "Moshe, Moshe and Zinn-Justin, Jean",
    title = "{3D Field Theories with Chern--Simons Term for Large $N$ in the Weyl Gauge}",
    eprint = "1410.0558",
    archivePrefix = "arXiv",
    primaryClass = "hep-th",
    doi = "10.1007/JHEP01(2015)054",
    journal = "JHEP",
    volume = "01",
    pages = "054",
    year = "2015"
}

@article{Frishman:2014cma,
    author = "Frishman, Yitzhak and Sonnenschein, Jacob",
    title = "{Large N Chern-Simons with massive fundamental fermions - A model with no bound states}",
    eprint = "1409.6083",
    archivePrefix = "arXiv",
    primaryClass = "hep-th",
    doi = "10.1007/JHEP12(2014)165",
    journal = "JHEP",
    volume = "12",
    pages = "165",
    year = "2014"
}

@article{Dandekar:2014era,
    author = "Dandekar, Yogesh and Mandlik, Mangesh and Minwalla, Shiraz",
    title = "{Poles in the $S$-Matrix of Relativistic Chern-Simons Matter theories from Quantum Mechanics}",
    eprint = "1407.1322",
    archivePrefix = "arXiv",
    primaryClass = "hep-th",
    doi = "10.1007/JHEP04(2015)102",
    journal = "JHEP",
    volume = "04",
    pages = "102",
    year = "2015"
}

@article{Gurucharan:2014cva,
    author = "Gurucharan, V. and Prakash, Shiroman",
    title = "{Anomalous dimensions in non-supersymmetric bifundamental Chern-Simons theories}",
    eprint = "1404.7849",
    archivePrefix = "arXiv",
    primaryClass = "hep-th",
    doi = "10.1007/JHEP09(2014)009",
    journal = "JHEP",
    volume = "09",
    pages = "009",
    year = "2014",
    note = "[Erratum: JHEP 11, 045 (2017)]"
}

@article{Bardeen:2014qua,
    author = "Bardeen, William A.",
    title = "{The Massive Fermion Phase for the U(N) Chern-Simons Gauge Theory in D=3 at Large N}",
    eprint = "1404.7477",
    archivePrefix = "arXiv",
    primaryClass = "hep-th",
    reportNumber = "FERMILAB-PUB-14-113-T",
    doi = "10.1007/JHEP10(2014)039",
    journal = "JHEP",
    volume = "10",
    pages = "039",
    year = "2014"
}

@article{Jain:2014nza,
    author = "Jain, Sachin and Mandlik, Mangesh and Minwalla, Shiraz and Takimi, Tomohisa and Wadia, Spenta R. and Yokoyama, Shuichi",
    title = "{Unitarity, Crossing Symmetry and Duality of the S-matrix in large N Chern-Simons theories with fundamental matter}",
    eprint = "1404.6373",
    archivePrefix = "arXiv",
    primaryClass = "hep-th",
    reportNumber = "TIFR-TH-14-12, HRI-ST-1405, ICTS-2014-04",
    doi = "10.1007/JHEP04(2015)129",
    journal = "JHEP",
    volume = "04",
    pages = "129",
    year = "2015"
}

@article{Bardeen:2014paa,
    author = "Bardeen, William A. and Moshe, Moshe",
    title = "{Spontaneous Breaking of Scale Invariance in a D=3 U(N ) Model with Chern-Simons Gauge Fields}",
    eprint = "1402.4196",
    archivePrefix = "arXiv",
    primaryClass = "hep-th",
    reportNumber = "FERMILAB-PUB-13-552-T, CERN-PH-TH-2013-299",
    doi = "10.1007/JHEP06(2014)113",
    journal = "JHEP",
    volume = "06",
    pages = "113",
    year = "2014"
}

@article{Fujitsuka:2013fga,
    title={Higgs branch localization of 3d N = 2 theories},
   volume={2014},
   ISSN={2050-3911},
   url={http://dx.doi.org/10.1093/ptep/ptu158},
   DOI={10.1093/ptep/ptu158},
   number={12},
   journal={Progress of Theoretical and Experimental Physics},
   publisher={Oxford University Press (OUP)},
   author={Fujitsuka, M. and Honda, M. and Yoshida, Y.},
   year={2014},
   month=dec, pages={123B02-123B02}
}

@article{Yokoyama:2013pxa,
    author = "Yokoyama, Shuichi",
    title = "{A Note on Large N Thermal Free Energy in Supersymmetric Chern-Simons Vector Models}",
    eprint = "1310.0902",
    archivePrefix = "arXiv",
    primaryClass = "hep-th",
    reportNumber = "TIFR-TH-13-26",
    doi = "10.1007/JHEP01(2014)148",
    journal = "JHEP",
    volume = "01",
    pages = "148",
    year = "2014"
}

@article{Jain:2013gza,
    author = "Jain, Sachin and Minwalla, Shiraz and Yokoyama, Shuichi",
    title = "{Chern Simons duality with a fundamental boson and fermion}",
    eprint = "1305.7235",
    archivePrefix = "arXiv",
    primaryClass = "hep-th",
    reportNumber = "TIFR-TH-13-17",
    doi = "10.1007/JHEP11(2013)037",
    journal = "JHEP",
    volume = "11",
    pages = "037",
    year = "2013"
}

@article{Aharony:2013dha,
    author = "Aharony, Ofer and Razamat, Shlomo S. and Seiberg, Nathan and Willett, Brian",
    title = "{3d dualities from 4d dualities}",
    eprint = "1305.3924",
    archivePrefix = "arXiv",
    primaryClass = "hep-th",
    reportNumber = "WIS-04-13-APR-DPPA",
    doi = "10.1007/JHEP07(2013)149",
    journal = "JHEP",
    volume = "07",
    pages = "149",
    year = "2013"
}

@article{Takimi:2013zca,
    author = "Takimi, Tomohisa",
    title = "{Duality and higher temperature phases of large N Chern-Simons matter theories on $S^2$ x $S^1$}",
    eprint = "1304.3725",
    archivePrefix = "arXiv",
    primaryClass = "hep-th",
    doi = "10.1007/JHEP07(2013)177",
    journal = "JHEP",
    volume = "07",
    pages = "177",
    year = "2013"
}

@article{Jain:2013py,
    author = "Jain, Sachin and Minwalla, Shiraz and Sharma, Tarun and Takimi, Tomohisa and Wadia, Spenta R. and Yokoyama, Shuichi",
    title = "{Phases of large $N$ vector Chern-Simons theories on $S^2 \times S^1$}",
    eprint = "1301.6169",
    archivePrefix = "arXiv",
    primaryClass = "hep-th",
    reportNumber = "TIFR-TH-13-02, ICTS-2012-14",
    doi = "10.1007/JHEP09(2013)009",
    journal = "JHEP",
    volume = "09",
    pages = "009",
    year = "2013"
}

@article{Aharony:2012ns,
    author = "Aharony, Ofer and Giombi, Simone and Gur-Ari, Guy and Maldacena, Juan and Yacoby, Ran",
    title = "{The Thermal Free Energy in Large N Chern-Simons-Matter Theories}",
    eprint = "1211.4843",
    archivePrefix = "arXiv",
    primaryClass = "hep-th",
    reportNumber = "WIS-18-12-NOV-DPPA",
    doi = "10.1007/JHEP03(2013)121",
    journal = "JHEP",
    volume = "03",
    pages = "121",
    year = "2013"
}

@article{Hwang:2012jh,
    author = "Hwang, Chiung and Kim, Hee-Cheol and Park, Jaemo",
    title = "{Factorization of the 3d superconformal index}",
    eprint = "1211.6023",
    archivePrefix = "arXiv",
    primaryClass = "hep-th",
    doi = "10.1007/JHEP08(2014)018",
    journal = "JHEP",
    volume = "08",
    pages = "018",
    year = "2014"
}

@article{Yokoyama:2012fa,
    author = "Yokoyama, Shuichi",
    title = "{Chern-Simons-Fermion Vector Model with Chemical Potential}",
    eprint = "1210.4109",
    archivePrefix = "arXiv",
    primaryClass = "hep-th",
    reportNumber = "TIFR-TH-12-34",
    doi = "10.1007/JHEP01(2013)052",
    journal = "JHEP",
    volume = "01",
    pages = "052",
    year = "2013"
}

@article{Chang:2012kt,
    author = "Chang, Chi-Ming and Minwalla, Shiraz and Sharma, Tarun and Yin, Xi",
    title = "{ABJ Triality: from Higher Spin Fields to Strings}",
    eprint = "1207.4485",
    archivePrefix = "arXiv",
    primaryClass = "hep-th",
    reportNumber = "TIFR-TH-12-29",
    doi = "10.1088/1751-8113/46/21/214009",
    journal = "J. Phys. A",
    volume = "46",
    pages = "214009",
    year = "2013"
}

@article{Aharony:2012nh,
    author = "Aharony, Ofer and Gur-Ari, Guy and Yacoby, Ran",
    title = "{Correlation Functions of Large N Chern-Simons-Matter Theories and Bosonization in Three Dimensions}",
    eprint = "1207.4593",
    archivePrefix = "arXiv",
    primaryClass = "hep-th",
    reportNumber = "WIS-13-12-JUL-DPPA",
    doi = "10.1007/JHEP12(2012)028",
    journal = "JHEP",
    volume = "12",
    pages = "028",
    year = "2012"
}

@article{Jain:2012qi,
    author = "Jain, Sachin and Trivedi, Sandip P. and Wadia, Spenta R. and Yokoyama, Shuichi",
    title = "{Supersymmetric Chern-Simons Theories with Vector Matter}",
    eprint = "1207.4750",
    archivePrefix = "arXiv",
    primaryClass = "hep-th",
    reportNumber = "TIFR-TH-12-30",
    doi = "10.1007/JHEP10(2012)194",
    journal = "JHEP",
    volume = "10",
    pages = "194",
    year = "2012"
}

@article{Dimofte:2011py,
    author = "Dimofte, Tudor and Gaiotto, Davide and Gukov, Sergei",
    title = "{3-Manifolds and 3d Indices}",
    eprint = "1112.5179",
    archivePrefix = "arXiv",
    primaryClass = "hep-th",
    doi = "10.4310/ATMP.2013.v17.n5.a3",
    journal = "Adv. Theor. Math. Phys.",
    volume = "17",
    number = "5",
    pages = "975--1076",
    year = "2013"
}

@article{Aharony:2011jz,
    author = "Aharony, Ofer and Gur-Ari, Guy and Yacoby, Ran",
    title = "{d=3 Bosonic Vector Models Coupled to Chern-Simons Gauge Theories}",
    eprint = "1110.4382",
    archivePrefix = "arXiv",
    primaryClass = "hep-th",
    doi = "10.1007/JHEP03(2012)037",
    journal = "JHEP",
    volume = "03",
    pages = "037",
    year = "2012"
}

@article{Giombi:2011kc,
    author = "Giombi, Simone and Minwalla, Shiraz and Prakash, Shiroman and Trivedi, Sandip P. and Wadia, Spenta R. and Yin, Xi",
    title = "{Chern-Simons Theory with Vector Fermion Matter}",
    eprint = "1110.4386",
    archivePrefix = "arXiv",
    primaryClass = "hep-th",
    doi = "10.1140/epjc/s10052-012-2112-0",
    journal = "Eur. Phys. J. C",
    volume = "72",
    pages = "2112",
    year = "2012"
}

@article{Benini:2011mf,
    author = "Benini, Francesco and Closset, Cyril and Cremonesi, Stefano",
    title = "{Comments on 3d Seiberg-like dualities}",
    eprint = "1108.5373",
    archivePrefix = "arXiv",
    primaryClass = "hep-th",
    reportNumber = "PU-2391, WIS-07-11-AUG-DPPA, TAUP-2931-11",
    doi = "10.1007/JHEP10(2011)075",
    journal = "JHEP",
    volume = "10",
    pages = "075",
    year = "2011"
}

@article{Imamura:2011su,
    author = "Imamura, Yosuke and Yokoyama, Shuichi",
    title = "{Index for three dimensional superconformal field theories with general R-charge assignments}",
    eprint = "1101.0557",
    archivePrefix = "arXiv",
    primaryClass = "hep-th",
    reportNumber = "UT-11-01, TIT-HEP-607",
    doi = "10.1007/JHEP04(2011)007",
    journal = "JHEP",
    volume = "04",
    pages = "007",
    year = "2011"
}

@article{Giombi:2009wh,
    author = "Giombi, Simone and Yin, Xi",
    title = "{Higher Spin Gauge Theory and Holography: The Three-Point Functions}",
    eprint = "0912.3462",
    archivePrefix = "arXiv",
    primaryClass = "hep-th",
    doi = "10.1007/JHEP09(2010)115",
    journal = "JHEP",
    volume = "09",
    pages = "115",
    year = "2010"
}

@article{Kim:2009wb,
    author = "Kim, Seok",
    title = "{The Complete superconformal index for N=6 Chern-Simons theory}",
    eprint = "0903.4172",
    archivePrefix = "arXiv",
    primaryClass = "hep-th",
    reportNumber = "IMPERIAL-TP-09-SK-01",
    doi = "10.1016/j.nuclphysb.2009.06.025",
    journal = "Nucl. Phys. B",
    volume = "821",
    pages = "241--284",
    year = "2009",
    note = "[Erratum: Nucl.Phys.B 864, 884 (2012)]"
}

@article{Klebanov:2002ja,
    author = "Klebanov, I. R. and Polyakov, A. M.",
    title = "{AdS dual of the critical O(N) vector model}",
    eprint = "hep-th/0210114",
    archivePrefix = "arXiv",
    reportNumber = "PUPT-2053",
    doi = "10.1016/S0370-2693(02)02980-5",
    journal = "Phys. Lett. B",
    volume = "550",
    pages = "213--219",
    year = "2002"
}

@article{Sezgin:2002rt,
    author = "Sezgin, E. and Sundell, P.",
    title = "{Massless higher spins and holography}",
    eprint = "hep-th/0205131",
    archivePrefix = "arXiv",
    reportNumber = "CTP-TAMU-08-02, UU-01-10",
    doi = "10.1016/S0550-3213(02)00739-3",
    journal = "Nucl. Phys. B",
    volume = "644",
    pages = "303--370",
    year = "2002",
    note = "[Erratum: Nucl.Phys.B 660, 403--403 (2003)]"
}

@article{Blau:1993tv,
    author = "Blau, Matthias and Thompson, George",
    title = "{Derivation of the Verlinde formula from Chern-Simons theory and the G/G model}",
    eprint = "hep-th/9305010",
    archivePrefix = "arXiv",
    reportNumber = "IC-93-83",
    doi = "10.1016/0550-3213(93)90538-Z",
    journal = "Nucl. Phys. B",
    volume = "408",
    pages = "345--390",
    year = "1993"
}

@article{Choudhury:2018iwf,
    author = "Choudhury, Sayantan and Dey, Anshuman and Halder, Indranil and Jain, Sachin and Janagal, Lavneet and Minwalla, Shiraz and Prabhakar, Naveen",
    title = "{Bose-Fermi Chern-Simons Dualities in the Higgsed Phase}",
    eprint = "1804.08635",
    archivePrefix = "arXiv",
    primaryClass = "hep-th",
    reportNumber = "TIFR/TH/18-10, TIFR-TH-18-10",
    doi = "10.1007/JHEP11(2018)177",
    journal = "JHEP",
    volume = "11",
    pages = "177",
    year = "2018"
}

@article{Minwalla:2022sef,
    author = "Minwalla, Shiraz and Mishra, Amiya and Prabhakar, Naveen and Sharma, Tarun",
    title = "{The Hilbert space of large N Chern-Simons matter theories}",
    eprint = "2201.08410",
    archivePrefix = "arXiv",
    primaryClass = "hep-th",
    reportNumber = "TIFR/TH/21-13",
    doi = "10.1007/JHEP07(2022)025",
    journal = "JHEP",
    volume = "07",
    pages = "025",
    year = "2022"
}

@article{Mehta:2022lgq,
    author = "Mehta, Umang and Minwalla, Shiraz and Patel, Chintan and Prakash, Shiroman and Sharma, Kartik",
    title = "{Crossing Symmetry in Matter Chern-Simons Theories at finite $N$ and $k$}",
    eprint = "2210.07272",
    archivePrefix = "arXiv",
    primaryClass = "hep-th",
    reportNumber = "TIFR/TH/22-41",
    month = "10",
    year = "2022"
}

@article{Wadia:1980rb,
    author = "Wadia, Spenta R.",
    title = "{On the Dyson-schwinger Equations Approach to the Large $N$ Limit: Model Systems and String Representation of {Yang-Mills} Theory}",
    reportNumber = "EFI-80/47-CHICAGO",
    doi = "10.1103/PhysRevD.24.970",
    journal = "Phys. Rev. D",
    volume = "24",
    pages = "970",
    year = "1981"
}

@article{Li:1987hx,
    author = "Li, M. and Wilets, L. and Birse, M. C.",
    title = "{{QCD} in Two-dimensions in the Axial Gauge}",
    doi = "10.1088/0305-4616/13/7/005",
    journal = "J. Phys. G",
    volume = "13",
    pages = "915--923",
    year = "1987"
}

@article{Bars:1977ud,
    author = "Bars, I. and Green, Michael B.",
    title = "{Poincare and Gauge Invariant Two-Dimensional QCD}",
    reportNumber = "COO-3075-179",
    doi = "10.1103/PhysRevD.17.537",
    journal = "Phys. Rev. D",
    volume = "17",
    pages = "537",
    year = "1978"
}

\end{document}